\documentclass{article}
\usepackage{url}
\usepackage{amsmath,amsthm,amssymb}
\usepackage{amsfonts}
\usepackage{natbib}
\usepackage{hyperref}
\usepackage{cleveref}
\usepackage{graphicx}
\usepackage{caption}
\usepackage{subcaption}
\usepackage{booktabs}
\usepackage{longtable}
\usepackage{xcolor}
\usepackage{footnote}
\usepackage{lineno}
\usepackage{layout}
\usepackage[disable]{todonotes}
\usepackage{algorithm}
\usepackage{algpseudocode}

\newcommand{\footremember}[2]{%
    \footnote{#2}
    \newcounter{#1}
    \setcounter{#1}{\value{footnote}}%
}
\newcommand{\footrecall}[1]{%
    \footnotemark[\value{#1}]%
}

\newenvironment{delayedproof}[1]
 {\begin{proof}[\raisedtarget{#1}Proof of \Cref{#1}]}
 {\end{proof}}
\newcommand{\raisedtarget}[1]{%
  \raisebox{\fontcharht\font`P}[0pt][0pt]{\hypertarget{#1}{}}%
}
\newcommand{\proofref}[1]{\hyperlink{#1}{proof}}

\newcommand{\Rd}[1]{\mathbb{R}^{#1}}

\newtheorem{mytheorem}{Theorem}
\newtheorem{mylemma}{Lemma}
\newtheorem{mycorollary}{Corollary}
\newtheorem{assumption}{Assumption}
\crefname{assumption}{assumption}{assumptions}
\Crefname{assumption}{Assumption}{Assumptions}
\newtheorem{definition}{Definition}
\numberwithin{mytheorem}{section} %
\numberwithin{mylemma}{section} %
\numberwithin{mycorollary}{section} %
\numberwithin{assumption}{section} %
\numberwithin{definition}{section}

\newcommand{\E}{\mathbb{E}}
\newcommand{\Var}{\text{Var}}
\newcommand{\Cov}{\text{Cov}}

\newcommand{\obs}{d}
\newcommand{\obsn}{\obs^{(n)}}
\newcommand{\dataset}{\{\obs^{(n)}\}_{n=1}^{\nobs}}
\newcommand{\covar}{x}
\newcommand{\covarn}{\covar^{(n)}}
\newcommand{\response}{y}
\newcommand{\responsen}{\response^{(n)}}

\newcommand{\param}{\beta}
\newcommand{\funcname}{g}
\newcommand{\func}[1]{g(#1)}
\newcommand{\coord}[1]{{#1}_v}
\newcommand{\coordsym}{v}
\newcommand{\coef}{\theta}
\newcommand{\noise}{\sigma}
\newcommand{\paramdim}{V}
\newcommand{\prior}{p}
\newcommand{\nobs}{N}
\newcommand{\allObsIndices}{\{1,2,\ldots,\nobs\}}
\newcommand{\loglik}[2]{L(\obs^{({#1})} \mid #2)} %
\newcommand{\normalposterior}{p(\param \mid \dataset ) }

\newcommand{\weight}{w}
\newcommand{\allzeros}{\mathbf{0}_{\nobs}}
\newcommand{\allexceptzero}{[0,1]^{\nobs} \setminus \{ \allzeros \}}
\newcommand{\allones}{\mathbf{1}_{\nobs}}
\newcommand{\normalizer}{Z(\weight)}

\newcommand{\weightedlogprob}{\prior(\param) \exp\left(\sum_{n=1}^{\nobs} \weight_n \loglik{n}{\param}\right)}
\newcommand{\posterior}[1]{p(\param \mid #1, \dataset )}
\newcommand{\Eposterior}[1]{\E_{#1}}
\newcommand{\indexset}{I}
\newcommand{\indexFromW}{q}
\newcommand{\wFromIndex}{q^{-1}}

\newcommand{\qoinoinput}{\phi}
\newcommand{\qoi}[1]{\qoinoinput(#1)}

\newcommand{\bound}[1]{f_{#1}}
\newcommand{\lowerbound}{\delta}
\newcommand{\diffdomain}{\{ \weight \in [0,1]^{\nobs}: \max_n \weight_n \geq \lowerbound \}}

\newcommand{\dropout}{\alpha}
\newcommand{\feasibleweight}{W_{\dropout}}

\newcommand{\amip}{\Delta(\dropout)} %
\newcommand{\amiw}{w(\dropout)}  %
\newcommand{\amis}{U(\dropout)} %
\newcommand{\numdrop}{\lfloor \nobs \dropout \rfloor}
\newcommand{\indict}[1]{\mathbb{I}\{#1\}}

\newcommand{\infl}{\psi}
\newcommand{\infln}{\infl_{n}}
\newcommand{\limitVarMat}{\Sigma}
\newcommand{\meaninfl}{f}
\newcommand{\sdinfl}{h}
\newcommand{\mcmcmeaninfl}{\hat{\meaninfl}}
\newcommand{\mcmcsdinfl}{\hat{\sdinfl}}

\newcommand{\firstOrderError}{\text{Err}_{\text{1st}}(\indexset)}
\newcommand{\zerothOrderError}{\text{Err}_{\text{0th}}(\indexset)}

\newcommand{\mcmcdraws}[1]{\param^{(#1)}} %
\newcommand{\mcmccoord}[1]{\param^{(#1)}_v}
\newcommand{\markovchain}{(\param^{(1)},\ldots,\param^{(\samplesize)})}

\newcommand{\sampleidx}{s}

\newcommand{\numchains}{J}
\newcommand{\samplesize}{S}

\newcommand{\mcmcinfl}{\hat{\psi}}
\newcommand{\mcmcinfln}{\hat{\psi}_n}
\newcommand{\mcmcamis}{\widehat{U}}
\newcommand{\mcmcamip}{\widehat{\Delta}}

\newcommand{\groundtruthmcmcinfl}[1]{\psi^*_{#1}}
\newcommand{\groundtruthamis}[1]{U^*({#1})}
\newcommand{\groundtruthamip}[1]{\Delta^*({#1})}

\newcommand{\empiricaldist}[2]{\frac{1}{#1} \sum_{i=1}^{#1} \delta_{\{{#2}^{(i)}\}}(\cdot) }
\newcommand{\blocklength}{L}
\newcommand{\bootstrapsize}{B}
\newcommand{\btdraws}[1]{{\param^*}^{(#1)}} %
\newcommand{\btsample}{(\btdraws{1},\btdraws{2},\ldots,\btdraws{\samplesize})}
\newcommand{\btmcmcamip}{\widehat{\Delta}^*}
\newcommand{\numblocks}{M}

\newcommand{\nominal}{\eta}

\newcommand{\amipub}{\Delta^{ub}(\dropout)}
\newcommand{\amiplb}{\Delta^{lb}(\dropout)}
\newcommand{\amipci}{[\amiplb, \amipub]}

\newcommand{\meanOfIndexSet}{\bar{\covar}_{\indexset}}

\newcommand{\groupn}{g^{(n)}}
\newcommand{\numgroup}{G}
\newcommand{\groupmean}{\theta}

\newcommand{\dispersion}{\tau}
\newcommand{\nobsj}[1]{N_{#1}}
\newcommand{\fullnobsj}[1]{N_{#1}^*}
\newcommand{\meanj}[1]{M_{#1}}
\newcommand{\fullmeanj}[1]{M_{#1}^*}
\newcommand{\precj}[1]{\Lambda_{#1}}
\newcommand{\fullprecj}[1]{\Lambda_{#1}^*}
\newcommand{\sumprec}{\Lambda}
\newcommand{\fullmu}{\mu^*}
\newcommand{\fullprec}{\Lambda^*}
\newcommand{\shrmeanj}[1]{\tilde{\globalmean}_{#1}}

\newcommand{\globalmean}{\mu}

\newcommand{\const}{c}

\newcommand{\soici}{[V^{lb}, V^{ub}]}

\newcommand{\intercept}{\mu}
\newcommand{\slope}{\theta}

\newcommand{\gControlLocation}{\mu}
\newcommand{\gTreatmentLocation}{\tau}
\newcommand{\lControlLocation}{\mu^{(\text{country})}}
\newcommand{\lTreatmentLocation}{\tau^{(\text{country})}}
\newcommand{\gControlScale}{\xi}
\newcommand{\gTreatmentScale}{\theta}
\newcommand{\lControlScale}{\xi^{(\text{country})}}
\newcommand{\lTreatmentScale}{\theta^{(\text{country})}}
\newcommand{\sdControl}{\sigma_{(\text{control})}}
\newcommand{\sdTreatment}{\sigma_{(\text{treatment})}}
\newcommand{\sdControlScale}{\psi_{(\text{control})}}
\newcommand{\sdTreatmentScale}{\psi_{(\text{treatment})}}

\newcommand{\locationVar}{l}
\newcommand{\locationVarn}{\locationVar^{(n)}}
\newcommand{\timeVar}{t}
\newcommand{\timeVarn}{\timeVar^{(n)}}
\newcommand{\lIntercept}{\intercept^{(\text{region})}}
\newcommand{\tIntercept}{\intercept^{(\text{time})}}
\newcommand{\lSlope}{\slope^{(\text{region})}}
\newcommand{\tSlope}{\slope^{(\text{time})}}

\newcommand{\Adraw}[1]{A^{(#1)}}
\newcommand{\Bdraw}[1]{B^{(#1)}}
\newcommand{\Cdraw}[1]{C^{(#1)}}

\newcommand{\gap}{\omega}

\newcommand{\jacob}[1]{\mathbf{J}_{#1}} 
\newcommand{\LL}[1]{L_{#1}}

\newcommand{\normquant}[1]{z_{#1}}
\newcommand{\defaultquant}{\normquant{0.975}}

\newcommand{\defaultsamplesize}{4000}
\newcommand{\defaultblocklength}{10}
\newcommand{\defaultbootstrapsize}{200}

\newcommand{\defaultnumchains}{960}
\newcommand{\defaultdropout}{0.05}
\newcommand{\defaultdropoutaspercentage}{5\%}
\newcommand{\defaultnominal}{0.95}

\begin{document}
\title{Sensitivity of MCMC-based analyses to small-data removal}

\author{%
Tin D.\ Nguyen\footremember{MIT}{Massachusetts Institute of Technology}\footremember{MIT_IBM}{MIT-IBM Watson AI Lab}%
\and Ryan Giordano\footnote{University of California, Berkeley}%
\and Rachael Meager\footnote{University of New South Wales}%
\and Tamara Broderick\footrecall{MIT}\hspace{4pt}\footrecall{MIT_IBM}%
}

\maketitle              %

\begin{abstract}
If the conclusion of a data analysis is sensitive to dropping very few data points, that conclusion might hinge on the particular data at hand 
rather than representing a more broadly applicable truth. 
How could we check whether this sensitivity holds? One idea is to consider every small subset of data, drop it from the dataset, and re-run our analysis. 
But running MCMC to approximate a Bayesian posterior is already very expensive; running multiple times is prohibitive, and the number of re-runs 
needed here is combinatorially large. 
Recent work proposes a fast and accurate approximation to find the worst-case dropped data subset, but that work was developed for problems based on estimating equations --- and does not directly handle Bayesian posterior approximations using MCMC. 
We make two principal contributions in the present work. 
We adapt the existing data-dropping approximation to estimators computed via MCMC. 
Observing that Monte Carlo errors induce variability in the approximation, we use a variant of the bootstrap to quantify this uncertainty.
We demonstrate how to use our approximation in practice to determine whether there is non-robustness in a problem.
Empirically, our method is accurate in simple models, such as linear regression.
In models with complicated structure, such as hierarchical models, the performance of our method is mixed.
\end{abstract}

\section{Introduction}

Consider this motivating example.
\citet{Angelucci2015} conducted a randomized controlled trial (RCT) in Mexico to study whether microcredit improves business profits. 
One might choose to analyze the data from this RCT using a simple (but non-conjugate) Bayesian model and Markov chain Monte Carlo (MCMC).
Based on an MCMC estimate of the posterior mean effect of microcredit, an analyst might conclude that microcredit actually reduces profit.
So, microcredit might be viewed as detrimental to the businesses in this RCT.

Next, if a policymaker wants to advocate against microcredit deployment outside of Mexico, they need to know if microcredit remains detrimental beyond the data gathered in \citet{Angelucci2015}.
More broadly, many researchers analyze data with Bayesian models and MCMC \citep{Senf2020,Meager2022,Jones2021,Porter2022} and want to know if their conclusions generalize beyond their data.

Standard tools to assess generalization do not answer this question entirely.
An analyst might use frequentist tools (confidence interval, p-values) to predict whether their inferences hold in the broader population.
The validity of these methods technically depends on the assumption that the gathered data is an independent and identically distributed (i.i.d.)\ sample from a broader population.
In practice, we have reason to suspect that this assumption is not met; for instance, it might not be reasonable to assume that data collected in Mexico and data collected in a separate country are i.i.d.\ from the same distribution.

As pointed out by \citet{Shiffman2023}, an analyst might hope that deviations from the i.i.d.\ assumption are small enough that (a) their conclusions remain the same in the broader population and (b) standard tools accurately assess generalization.
On the other hand, the analyst might worry that this hope is misplaced if small, realistic deviations from i.i.d.-ness could affect the substantive conclusions of an analysis.
An often-realistic kind of deviation is the missingness of a small fraction of data; for instance, some percentage of the population might not respond to a survey.
So, if it were possible to remove a small fraction of data and change conclusions, the analyst might worry about generalization.

\citet{Broderick2023} were the first to formulate sensitivity to dropping a small fraction of data as a check on generalization.
Along with the formulation, one contribution of that work is a fast approximation to detect sensitivity when the analysis in question is based on estimating equations \citep{Kosorok2008}[Chapter 13].
Regardless of how estimators are constructed, in general, the brute-force approach to finding an influential small fraction of data is computationally intractable.
One would need to enumerate all possible data subsets of a given cardinality and re-analyze on each subset. Even when the fraction of data removed is small and each analysis takes little time, there are too many such subsets to consider; see the discussion at the end of \cref{sec:methods}.
For estimating equations, \citet{Broderick2023} approximate the effect of dropping data with a first-order Taylor series approximation; this approximation can be optimized very efficiently, while the brute-force approach is not at all practical.

Neither \citet{Broderick2023} nor subsequent existing work on small-data removals \citep{Kuschnig2021,Moitra2022,Shiffman2023,Freund2023} can be immediately applied to determine sensitivity in MCMC.
Since MCMC cannot be cast as the root of an estimating equation or the solution to an optimization problem, neither \citet{Broderick2023} nor \citet{Shiffman2023} apply to our situation.
As \citet{Kuschnig2021,Moitra2022,Freund2023} focus on ordinary least squares (OLS), their work does not address our problem, either. 

\paragraph{Our contributions.}
We extend \citet{Broderick2023} to handle analyses based on MCMC.
In \cref{sec:background}, we introduce necessary concepts in Bayesian decision-making, and we describe sensitivity to small-data dropping in more detail.
In \cref{sec:taylor-series}, we form a first-order approximation to the effect of removing observations; to do so, we use known results on how much a posterior expectation locally changes under small perturbations to the total log likelihood \citep{Diaconis1986,Ruggeri1993,Gustafson1996,Giordano2023}.
As this approximation involves posterior covariances, in \cref{sec:estimate-infl}, we re-use the MCMC draws that an analyst would have already generated to estimate what happens when data is removed.
Recognizing that Monte Carlo errors induce variability in our approximation, in \cref{sec:ci}, we use a variant of the bootstrap \citet{Efron1979} to quantify this uncertainty.
For more discussion on how our methodology relates to existing work, see \cref{sec:related_work}.
In \cref{sec:theory}, we provide some theoretical bounds on the quality of our approximation.

Experimentally, we apply our method to three Bayesian analyses.
In \cref{sec:experiments}, we can detect non-robustness in econometric and ecological studies. 
However, while our approximation performs well in simple models such as linear regression, it is less reliable in complex models, such as ones with many random effects.

\subsection{Related work} \label{sec:related_work}

Our work arguably fits into the intersection of three lines of work.
We have already mentioned the first: papers on detecting sensitivity to small-data removal.

The second line of work estimates the changes that happen to a posterior expectation because of small perturbations to the total log likelihood.
There are two conceptually distinct approaches to this sensitivity analysis.
\begin{itemize}
    \item One approach \citep[e.g.][]{Arya2022,Arya2023,Seyer2023} applies to when the posterior is approximated with a Metropolis-Hastings algorithm. In particular, this approach computes the gradient of the Metropolis-Hastings sampler to small perturbations in the total log likelihood. More broadly, there is a literature on estimating gradients for random processes with discrete components \citep{Kleijnen1996,Fu2012,Heidergott2008}.
    \item The other approach does not compute the gradient of the MCMC algorithm or steps within it. Instead, it directly computes (and then estimates) the gradient of the posterior expectation. Recent works in this literature include \citet{Giordano2018,Mohamed2020,Giordano2023,Giordano2023b}, while foundational works include \citet{Diaconis1986,Ruggeri1993,Gustafson1996}. 
\end{itemize}
In our work, we take the second approach. 
A priori, it is not clear which approach is superior. 
Two reasons to prefer the second approach over the first approach are the following.
While the discrete operations in Metropolis-Hastings, e.g.\ the accept/reject steps, pose a key challenge in the first approach, they do not cause any issues in the second approach; the second approach is ``oblivious'' to details regarding how the posterior is approximated.
In addition, suppose that an analyst wishes to compute gradients of multiple quantities of interest.
If they follow the first approach, for each quantity of interest, they would need to re-run the sampling algorithm to estimate the gradient.
Taking the second approach, the analyst needs to run the sampling algorithm only once; the analyst may then use the resulting draws to simultaneously estimate the gradient of multiple quantities of interest.
On the other hand, the first approach might be better than the second approach in the following way.
Our experiments later show that gradient estimates coming from the second approach can be noisy.
The first approach, with the promise of variance reduction through a good choice of Markov chain coupling, might produce more accurate gradient estimates. 
It is an interesting direction for future work to apply the first approach to our problem and compare the performance of the two approaches.

While papers taking the second approach have already mentioned how to estimate the effect of dropping an individual observation, these estimates have not been used to assess whether conclusions based on MCMC are sensitive to the removal of a small data fraction.
Some works \citep[e.g.][]{Gustafson1996,Giordano2018,Giordano2023b} generate perturbations by varying prior or likelihood choice. 
\citet{Giordano2023} estimate the frequentist variability of Bayesian procedures, a task that can be seen as equivalent to the goal of bootstrap resampling. No existing work aims to find a small fraction of data that, if dropped, would change conclusions.

The third set of works, in the Bayesian case influence literature, quantifies the importance of individual observations to a Bayesian analysis.
As we will explain, existing works do not tackle our problem.
Early works in this area include \citet{Johnson1983,Mcculloch1989,Lavine1992,Carlin1991}, while recent works include \citet{Marshall2007,Millar2007,vanderlinde2007,Thomas2018,Pratola2023}.
Such papers focus on the identification of outliers, rather than predictions about whether the conclusion changes after removing a small amount of data.
Generally, this literature defines an observation to be an outlier if the Kullback--Leibler (KL) divergence between the posterior after removing the observation and the original posterior is large.
For conclusions based on posterior functionals, such as the mean, we are not aware of how to systematically connect the KL divergence to the sensitivity of the decision-making process; in fact, recent work \citep{Huggins2020} has shown that comparing probability distributions based on the KL divergence can be misleading if an analyst really cared about the comparison between the distributions' means or variances.

\section{Background} \label{sec:background}
We introduce notation in two parts. First, we cover relevant concepts from Bayesian data analysis. Second, we extend the notation to dropping data.

\subsection{Bayesian data analysis} 

Suppose we have a dataset $\dataset$.
For instance, in regression, each observation is a vector of covariates $\covarn$ and a response $\responsen$; in this case, we write $\obsn = (\covarn, \responsen)$.
Consider a parameter $\param \in \Rd{\paramdim}$ of interest. 
To estimate the latent $\param$, one option is to take a Bayesian approach.
First, we probabilistically model the link between $\param$ and the data through a likelihood function $\loglik{n}{\param}$.
As an example, in linear regression, $\param$ consists of the coefficients $\coef$ and the noise $\noise$, with the likelihood equaling
$
    \loglik{n}{\param} = -\frac{1}{2\sigma^2} (\responsen - \coef^T \covarn)^2 - \frac{1}{2}\log(2\pi \sigma^2) . 
$ 
Secondly, we specify a prior distribution over the latent parameters, and use $\prior(\param)$ to denote the prior density. 
Then, the density of the posterior distribution of $\param$ given the data is 
\begin{equation*}
    \normalposterior \propto \prior(\param) \prod_{n=1}^{\nobs} \exp(\loglik{n}{\param}).
\end{equation*}

In practice, an analyst uses a functional of the posterior to make conclusions.
One prominent functional is the posterior mean $\E \func{\param}$, where $\funcname$ is a mapping from $\Rd{\paramdim}$ to $\mathbb{R}$. 
As an example, in linear regression, commonly a practitioner will
make a decision based on the sign of the posterior mean of a
particular regression coefficient.
Other decisions are made with credible intervals.
An econometrician might declare that an intervention is helping some
population if the vast majority of the posterior mass for a particular
coefficient lies above zero. 
That is, the practitioner checks if the lower bound of a credible interval lies above zero. This decision might be considered to reflect a Bayesian notion of
\emph{significance}.
Decisions might also be made with approximate credible intervals; while exact intervals are based on posterior quantiles, an approximate interval is often based on the sum between the posterior mean and a multiple of the posterior standard deviation.

Computationally, in general, the functionals needed to make a conclusion are not available in closed form.
To approximate posterior functionals, practitioners frequently use Markov chain Monte Carlo (MCMC) methods.
Let $\markovchain$ denote a set of MCMC draws that target the posterior distribution; a draw refers to a single $\mcmcdraws{s}$, and $\samplesize$ is the number of draws. 
In practice, we estimate expectations using $\markovchain$, and make a decision based on such estimates.  

\subsection{Sensitivity to small-data removal} \label{sec:non-robustness}

With notation for Bayesian data analyses in place, we introduce the problem of sensitivity to small-data removal.

A Bayesian analyst might be worried if the substantive decision
arising from their data analysis changed after removing some small
fraction $\dropout$ of the data. 
For instance,
\begin{itemize}
    \item If their decision were based on the sign of the posterior mean, they
    would be worried if that sign changed.
    \item If their decision were based on zero falling outside a credible interval, they would be worried if we can make the credible interval contain zero. 
    \item If their decision was based on both the sign and the significance, they would be worried if we can both change the posterior mean's sign and put a majority of the posterior mass on the opposite side of zero.
\end{itemize}
In general, we expect an analyst to be worried if we could remove a
small fraction $\dropout$ of the data and change their decision.

To describe non-robustness precisely and to develop our approximation, we need notation to indicate the dependence of posterior functionals on the presence of data points.
We introduce a vector of data weights $\weight = (\weight_1,\weight_2,\ldots,\weight_{\nobs})$, where $\weight_n$ is the weight for the $n$-th observation. Each $\weight_n$ is constrained to be in the interval $[0,1]$.
The whole vector $\weight$ defines the so-called \emph{weighted} posterior distribution. 
\begin{definition}
    Let 
    $
        \normalizer := \int \prior(\param) \prod_{n=1}^{\nobs} \exp(\weight_n \loglik{n}{\param}) d\param. 
    $
    If $\normalizer < \infty$, the weighted posterior distribution associated with $\weight$ has density 
    \begin{equation*}
        \posterior{\weight} := \frac{1}{\normalizer} \weightedlogprob.
    \end{equation*}
\end{definition}
Note that $\weight_n$ encodes the inclusion of $\obsn$ in the analysis.
If $\weight_n = 0$, the $n$-th observation is ignored; if $\weight_n = 1$, the $n$-th observation is fully included.
If it exists, we recover the standard posterior density by setting all weights to $1$: $\weight = \allones = (1,1,\ldots,1)$.  
It is possible that $\weightedlogprob$ is not integrable for some $\weight$.
For instance, consider the case when the prior $\prior(\param)$ is improper and all weights have been set to zero: $\weight = \allzeros = (0,0,\ldots,0)$.
In the following, we assume that any contribution of the likelihood is enough to define a proper posterior.
\begin{assumption}\label{assume:feasible-has-finite-normalization}
    $\forall \weight \in \allexceptzero$, $\normalizer < \infty$.
\end{assumption}
This assumption is immediate in the case of a proper prior and standard likelihood.

The notation $\posterior{\weight}$ emphasizes the dependence on $\weight$, and will supersede the $\normalposterior$ notation.
To indicate expectations under the weighted posterior, we use the subscript $\weight$; $\Eposterior{\weight}$ is the expectation taken with respect to the randomness $\param \sim \posterior{\weight}$.

With the weighted posterior notation, we extend concepts from the standard analysis to the new analysis involving weights.
The value of a posterior functional depends on $\weight$. 
For instance, the posterior mean under the weighted posterior is $\Eposterior{\weight} \func{\param}$, and we recover the standard posterior mean by setting $\weight = \allones$.

The Bayesian analyst's non-robustness concern can be formalized as follows.
For $\dropout \in (0, 1)$, let $\feasibleweight$ denote the set of all weight vectors that correspond to dropping no more than $100\dropout\%$ of the data, i.e.\
\begin{equation*}
    \feasibleweight := \left\{\weight \in \{0,1\}^{\nobs} : \frac{1}{\nobs} \sum_{n=1}^{\nobs} (1 - \weight_n) \leq \dropout \right\},
\end{equation*}
We say the analysis is non-robust if there exists a weight $\weight$ that a) corresponds to dropping a small amount of data ($\weight \in \feasibleweight)$ and b) changes the conclusion.

We focus on decision problems that satisfy the following simplifying assumption: there exists a posterior functional, which we denote by $\qoi{\weight}$, such that $\qoi{\allones} < 0$ and the conclusion changes if and only if $\qoi{\weight} > 0$. If we are interested in other decision boundaries or the other direction of change, we can add a constant to $\qoinoinput$ or multiply it by -1, respectively; so the preceding assumption is made without loss of generality. 
We call the functional $\qoinoinput$ a ``quantity of interest'' (QoI).

We next show how the changes described at the start of this section fit this framework.
First consider a conclusion based on the sign of the posterior mean of a parameter. If the full-data posterior mean ($\Eposterior{\allones} \func{\param}$) were positive, we would take 
\begin{equation*}
    \qoi{\weight} = - \Eposterior{\weight} \func{\param}.
\end{equation*}
Since the full-data posterior mean is positive, $\qoi{\allones} < 0$.
And $\qoi{\weight} > 0$ is equivalent to the posterior mean (after removing the data) being negative.
Next consider a conclusion based on whether zero falls in a standard approximate credible interval; we will abbreviate this situation as a conclusion based on ``significance.'' If the approximate credible interval's left endpoint\footnote{Our approximate credible interval multiplies the posterior standard deviation by $\defaultquant$, which is the $97.5\%$ quantile of the standard normal, but we can replace this value with other scaling without undue effort.} ($\Eposterior{\allones} \func{\param} - \defaultquant \sqrt{\Var_{\allones}\func{\param}}$) were positive, we would take 
\begin{equation*} 
    \qoi{\weight} = - (\Eposterior{\weight} \func{\param} - \defaultquant \sqrt{\Var_{\weight}\func{\param}}).
\end{equation*}
$\qoi{\weight} > 0$ is equivalent to moving the left endpoint below zero, thus changing from a significant result to a non-significant one.
Finally, consider a case where our conclusion is different if we can change to a significant result of the opposite sign. If the approximate credible interval's left endpoint were positive, we would take 
\begin{equation*} 
    \qoi{\weight} = - (\Eposterior{\weight} \func{\param} + \defaultquant \sqrt{\Var_{\weight}\func{\param}}).
\end{equation*}
On the full data, the right endpoint is above zero.
$\qoi{\weight} > 0$ is equivalent to moving the right endpoint below zero. In this case, the conclusion has changed from a positive result to a significant negative result.

Under our assumptions so far, checking for non-robustness is equivalent to a) finding the maximum value of $\qoi{\weight}$ subject to $\weight \in \feasibleweight$ and b) checking its sign. 
The outcome of this comparison remains the same if we retain the feasible set, maximize the objective function $\qoi{\weight} - c$, and compare the optimal value with $-c$, for $c$ being any constant that does not depend on weight.
Out of later convenience, we set $c = \qoi{\allones}$.
As in \citet[Section 2]{Broderick2023}, we define the Maximum Influence Perturbation to be the largest change, induced in a quantity of interest, by dropping no more than $100\dropout\%$ of the data.
In our notation, it is the optimal value of the following optimization problem:
\begin{equation} \label{eq:mip}
    \max_{\weight \in \feasibleweight} \left( \qoi{\weight} - \qoi{\allones} \right).
\end{equation}
If the Maximum Influence Perturbation is more than $-\qoi{\allones}$, then the conclusion is non-robust.
The set of observations that achieve the Maximum Influence Perturbation is called the Most Influential Set; to report it, we compute the optimal solution of \cref{eq:mip} and find its zero indices.

In general, the brute force approach to solve \cref{eq:mip} takes a prohibitively long time. 
We need to enumerate every data subset that drops no more than $100\dropout\%$ of the original data.
And, for each subset, we would need to re-run MCMC to re-estimate the quantity of interest. 
There are more than $\binom{\nobs}{\numdrop}$ elements in $\feasibleweight$.
One of our later numerical studies involves $\nobs = 16{,}560$ observations; 
even for $\dropout = 0.001$, there are more than $10^{54}$ subsets to consider.
Each Markov chain already takes a noticeable amount of time to construct; in this analysis, to generate $\samplesize = 4{,}000$ draws, we need to run the chain for $1$ minute.
The total time to compute the Maximum Influence Perturbation would be on the order of $10^{48}$ years.

\section{Methods} \label{sec:methods}
As the brute force solution to \cref{eq:mip} is computationally prohibitive, we turn to approximation methods.
In this section, we provide a series of approximations to the Maximum Influence Perturbation problem.

\subsection{Taylor series} \label{sec:taylor-series}

Our first approximation relies on the first-order Taylor series of the quantity of interest $\qoi{\weight}$.
This idea of approximating the Maximum Influence Perturbation with Taylor series was first proposed by \citet{Broderick2023}, in the context of Z-estimators.
Our work extends this idea to conclusions based on MCMC. 

To be able to form a Taylor series, we require that the quantity of interest $\qoi{\weight}$ is differentiable with respect to the weight $\weight$.
We are not aware of a complete theory (necessary and sufficient conditions) for this differentiability.
However, through \cref{assume:mean-and-sd,assume:finiteness}, we state a set of sufficient conditions. 

\begin{assumption}\label{assume:mean-and-sd}
    Let $\funcname$ be a function from $\Rd{\paramdim}$ to the real line.
    Let $\qoi{\weight}$ be a linear combination of a posterior mean of $g(\beta)$ and its corresponding posterior standard deviation. In particular, there exist constants $\const_1$ and $\const_2$, with no dependence on $\weight$, such that
    \begin{equation*}
        \qoi{\weight} = \const_1 \Eposterior{\weight} \func{\param} +  \const_2 \sqrt{\Var_{\weight}\func{\param}}.
    \end{equation*}
\end{assumption}
A typical choice of $\funcname$ is the function that returns the $\coordsym$-th coordinate of a $\paramdim$-dimensional vector.

It might appear that constraining $\qoi{\weight}$ to be a linear combination of the posterior mean and standard deviation is overly restrictive.
However, this choice encompasses many cases of practical interest; recall from \cref{sec:non-robustness} that the quantities of interest for changing sign, changing significance, and producing a significant result of the opposite sign all take the form in \cref{assume:mean-and-sd}.
Furthermore, the choice of constraining $\qoi{\weight}$ to be a linear combination of the posterior mean and standard deviation in \cref{assume:mean-and-sd} is done out of convenience.
Our framework can also handle quantities of interest that involve higher moments of the posterior distribution, and the function that combines these moments need not be linear, but we omit these cases for brevity.
However, we note that posterior quantiles in general do not satisfy \cref{assume:mean-and-sd} and leave to future work the question of how to diagnose the sensitivity of such quantities of interest. 

\begin{assumption}\label{assume:finiteness}
    For any $\weight \in [0,1]^{\nobs} \setminus \{ \allzeros \}$, the following functions have finite expectations under the weighted posterior: $|\func{\param}|$, $\func{\param}^2$, $|\loglik{n}{\param}|$ (for all $n$), $|\func{\param}\loglik{n}{\param}|$ (for all $n$) and $|\func{\param}^2\loglik{n}{\param}|$ (for all $n$).
\end{assumption}
The assumption is mild.
It is satisfied by for instance, linear regression under Gaussian likelihood and $\func{\param} = \coord{\param}$.

Under \cref{assume:feasible-has-finite-normalization,assume:mean-and-sd,assume:finiteness}, $\qoi{\weight}$ is continuously differentiable with respect to $\weight$.
\begin{mytheorem} \label{thm:linear-map-exists}
    Take \cref{assume:feasible-has-finite-normalization,assume:mean-and-sd,assume:finiteness}.
    For any $\lowerbound \in (0,1)$, $\qoi{\weight}$ is continuously differentiable with respect to $\weight$ on $\diffdomain$.
    The $n$-th partial derivative\footnote{If $\weight_n$ lies on the boundary, the partial derivative is understood to be one-sided.} at $\weight$ is equal to $\const_1 \meaninfl + \const_2 \sdinfl $ where 
    \begin{equation} \label{eq:mean-sensitivity}
       \meaninfl = \Cov_{\weight} \left( \func{\param}, \loglik{n}{\param}  \right), 
    \end{equation} and 
    \begin{equation} \label{eq:sd-sensitivity}
        \sdinfl = \frac{\Cov_{\weight} \left( \func{\param}^2, \loglik{n}{\param}  \right)  - 2\mathbb{E}_{\weight} \func{\param} \times \Cov_{\weight} \left( \func{\param}, \loglik{n}{\param}  \right)    }{\sqrt{\Var_{\weight} \func{\param}}}.
    \end{equation}
\end{mytheorem}
See the \proofref{thm:linear-map-exists}.
This theorem is a specific instance of the sensitivity of posterior expectations with respect to log likelihood perturbations; for further reading, see \citet{Diaconis1986,Basu1996,Gustafson1996}. 
\Cref{thm:linear-map-exists} establishes both the existence of the partial derivatives and their formula. 
\Cref{eq:mean-sensitivity} is the partial derivative of the posterior mean with respect to the weights, while \cref{eq:sd-sensitivity} is that for the posterior standard deviation.

Based on \cref{thm:linear-map-exists}, we define the $n$-th \emph{influence} as the partial derivative of $\qoi{\weight}$ at $\weight = \allones$:
\begin{equation*}
    \infln := \frac{\partial \qoi{\weight}}{\partial \weight_n} \big|_{\weight = \allones}.
\end{equation*}
Then, the first-order Taylor series approximation of $\qoi{\weight} - \qoi{\allones}$ is 
\begin{equation} \label{eq:qoi-taylor-series}
    \qoi{\weight} - \qoi{\allones} \approx \sum_{n=1}^{\nobs} \infln (\weight_n - 1).
\end{equation}
This approximation predicts that leaving out the $n$-th observation ($\weight_n = 0$) changes the quantity of interest by $-\infln$. 
Using \cref{eq:qoi-taylor-series}, we approximately solve \cref{eq:mip} by replacing its objective function but keeping its feasible set: 
\begin{equation} \label{eq:first-order}
    \begin{aligned}
    \max_{\weight} \quad & \sum_{n=1}^{\nobs}  (w_n-1) \infln    \\
    \textrm{s.t.} \quad & \weight_n \in \{0,1\}, \hspace{10pt} \frac{1}{\nobs} \sum_{n=1}^{\nobs}  (1-\weight_n) \leq \dropout.
    \end{aligned}
\end{equation}
Solving \cref{eq:first-order} is straightforward.
For any $\weight \in \feasibleweight$, the objective function is equal to $\sum_{n: \weight_n = 0}(-\infln)$.
Let $\amiw$ be the optimal solution and $\amip$ be the optimal value of \cref{eq:first-order}.
We denote $\amis$ to be the set of observations omitted according to $\amiw$: $\amis := \{\obs_{n}: \amiw_{n} = 0 \}$.
Let $r_1,r_2,\ldots,r_{\nobs}$ be indices of the $\infln$ sorted in increasing order: $\infl_{r_1} \leq \infl_{r_2} \leq \ldots \leq \infl_{r_{\nobs}}$.
Let $m$ be the smallest index such that $\infl_{r_{m+1}} \geq 0$; if none exists, set $m$ to $\nobs$.
If $m \geq 1$, $\amiw$ assigns weight $0$ to the observations $r_1,r_2,\ldots,r_{\min(m,  \numdrop)}$, and $1$ to the remaining ones.
Otherwise, $m = 0$ and $\amiw$ assigns weight $1$ to all observations.
Following \citet{Broderick2023}, we call the optimal objective value of \cref{eq:first-order} by the name Approximate Maximum Influence Perturbation (AMIP), and denote it by $\amip$.
It is equal to the negative of $\sum_{n=1}^{\numdrop} \infl_{r_n} \indict{\infl_{r_n}  < 0} $, where $\indict{\cdot}$ equals one if its argument is true and 0 otherwise.

\subsection{Estimating the influence} \label{sec:estimate-infl}

To solve \cref{eq:first-order}, we need to compute each influence $\infln$.
In this section, we use MCMC to estimate $\infln$.

Because of \cref{thm:linear-map-exists} and the fact that $\infln$ is the partial derivative at $\weight = \allones$, we know that $\infln$ is a function of certain expectations and covariances under the full-data posterior. 
Therefore, the MCMC draws from the full-data posterior, which are already used to estimate $\qoi{\allones}$, can be used to estimate $\infln$; see 
\cref{alg:EI} for the details.
In a nutshell, we replace all population expectations with sample averages.
The estimate of $\infln$ will be called $\mcmcinfln$.

\begin{algorithm}[t]
    \caption{Estimate of Influence (EI)} \label{alg:EI}
    \textbf{Inputs:} \\
    $\const_1,\const_2$ \Comment{$\qoi{\weight}$-defining constants} \\
    $\markovchain$ \Comment{Markov chain Monte Carlo draws}
    \begin{algorithmic}[1]
        \Procedure{EI}{$\const_1,\const_2,\markovchain$}
            \State $m \gets \frac{1}{\samplesize} \sum_{\sampleidx=1}^{\samplesize} \func{\mcmcdraws{s}}$, $k \gets \frac{1}{\samplesize} \sum_{\sampleidx=1}^{\samplesize} \func{\mcmcdraws{s}}^2$ 
            \State $v \gets k - m^2$
            \For{$n \gets 1, \ldots, \nobs$}
                \State  $a \gets \frac{1}{\samplesize} \sum_{\sampleidx=1}^{\samplesize}\func{\mcmcdraws{s}}\loglik{n}{\mcmcdraws{s}}$ 
                \State  $b \gets \frac{1}{\samplesize} \sum_{\sampleidx=1}^{\samplesize}\func{\mcmcdraws{s}}^2\loglik{n}{\mcmcdraws{s}}$
                \State  $u \gets \frac{1}{\samplesize} \sum_{\sampleidx=1}^{\samplesize}\loglik{n}{\mcmcdraws{s}}$
                \State  $\mcmcmeaninfl \gets a - m u$ \Comment{Estimate of \cref{eq:mean-sensitivity}}
                \State $g \gets b - k u$
                \State $\mcmcsdinfl \gets (g - 2 m \mcmcmeaninfl)/(\sqrt{v})$ \Comment{Estimate of \cref{eq:sd-sensitivity}}
                \State $\mcmcinfln \gets c_1 \mcmcmeaninfl + c_2 \mcmcsdinfl$ \Comment{Estimate of $\infln$ }
            \EndFor
        \State \textbf{return} $(\mcmcinfl_1, \mcmcinfl_2, \ldots, \mcmcinfl_{\nobs})$
        \EndProcedure 
    \end{algorithmic}
\end{algorithm}

Since $\mcmcinfln$ is only an approximation of $\infln$, we are not able to solve \cref{eq:first-order} exactly; rather, we solve only an approximation of it.
\Cref{alg:sosie} details the procedure.
We denote the outputs of \cref{alg:sosie} by $\mcmcamip$ and $\mcmcamis$:
\begin{equation} \label{eq:sosie-output}
    (\mcmcamip, \mcmcamis) := -\text{SoSIE}(\const_1, \const_2, \markovchain, \dropout).
\end{equation}
While $\mcmcamip$ is a point estimate of $\amip$, $\mcmcamis$ is a point estimate of $\amis$.

\begin{algorithm}[t]
    \caption{Sum of Sorted Influence Estimate (SoSIE)}\label{alg:sosie}
    \textbf{Inputs:} \\
    $\const_1,\const_2$ \Comment{$\qoi{\weight}$-defining constants} \\
    $\markovchain$ \Comment{Markov chain Monte Carlo draws} \\
    $\dropout$ \Comment{Fraction of data to drop}
    \begin{algorithmic}[1]
        \Procedure{SoSIE}{$\const_1$, $\const_2$, $\markovchain$, $\dropout$}
            \State $\hat{\psi} \gets \text{EI}(\const_1, \const_2,\markovchain)$
            \State Find ranks $v_1, v_2,\ldots,v_N$ such that $\mcmcinfl_{v_1} \leq \mcmcinfl_{v_2} \leq  \ldots \leq \mcmcinfl_{v_N}$ 
            \State Find the smallest $p$ such that $\mcmcinfl_{v_{p+1}} \geq 0$. If none exists, set $p$ to $\nobs$.
            \State If $p \geq 1$, $\mcmcamis \gets  \{\obs_{v_1},\ldots,\obs_{v_{\min(p, \numdrop)}} \}$. Otherwise, $\mcmcamis \gets \emptyset$
            \State $\mcmcamip \gets -\sum_{m = 1}^{\numdrop} \mcmcinfl_{v_m} \indict{\mcmcinfl_{v_m} < 0}$
            \State \textbf{return} $\mcmcamip,\mcmcamis$
        \EndProcedure 
    \end{algorithmic}
\end{algorithm}

\subsection{Confidence intervals for AMIP} \label{sec:ci}
$\mcmcamip$ from \cref{eq:sosie-output} is a noisy point estimate.
One concern regarding the quality of $\mcmcamip$ is noise due to sampling variability of $\markovchain$.
In this section, we design confidence intervals for $\amip$. 
We begin by considering the special case when the samples $\markovchain$ come from exact sampling.
Then, we relax the exact sampling assumption, and consider general Markov chain Monte Carlo samples.

\subsubsection{Exact sampling} \label{sec:exact_sampling}

For certain prior and likelihoods, we are able to draw exact Monte Carlo samples from the posterior distribution; for instance, consider conjugate models \citep{Diaconis1979} or models in which convenient augmentation schemes have been discovered, such as Bayesian logistic regression with Polya-Gamma augmentation \citep{Polson2013}.
In these cases, we can assume $\markovchain$ is an i.i.d.\ sample of size $\samplesize$ drawn from the full-data posterior distribution. 
And $\mcmcamip$ from \cref{eq:sosie-output} can be thought of as an estimator constructed from an i.i.d.\ sample, though we emphasize that the sample in question is not the data $\dataset$, but $\markovchain$. 
To highlight the dependence between $\mcmcamip$ and $\markovchain$, we will use the notation $\mcmcamip\markovchain$.  
The estimator $\mcmcamip$ is a complex, non-smooth function of the sample; the act of taking the minimum across the estimated influences $\mcmcinfln$ is non-smooth.
We do not attempt to prove distributional results for this estimator or use such results to quantify uncertainty.
Instead, we appeal to the bootstrap \citep{Efron1979}, a general-purpose technique to quantify the sampling variability of estimators.

Our confidence interval construction proceeds in three steps. 
First, we define the \emph{bootstrap distribution} of $\mcmcamip$. 
Second, we approximate this distribution with an empirical distribution based on Monte Carlo draws. 
Finally, we use the range spanned by quantiles of this empirical distribution as our confidence interval for $\amip$.
    
To define the bootstrap distribution, consider the empirical distribution of the sample $\markovchain$: 
\begin{equation*}
    \empiricaldist{\samplesize}{\param}. 
\end{equation*}
We denote one draw from this empirical distribution by $\btdraws{s}$.
A \emph{bootstrap sample} is a set of $\samplesize$ draws: $\btsample$. 
The bootstrap distribution of $\mcmcamip$ is the distribution of $\mcmcamip\btsample$, where the randomness is taken over the bootstrap sample but is conditional on the original sample $\markovchain$.

Clearly, the bootstrap distribution is discrete with finite support. 
If we chose to, we could enumerate its support and compute its probability mass function, by enumerating all possible values a bootstrap sample can take.
However, this is time consuming.
It suffices to approximate the bootstrap distribution with Monte Carlo draws.
We will abbreviate the draw $\mcmcamip\btsample$ as $\btmcmcamip$. We generate a total number of $\bootstrapsize$ such draws. 
As $\bootstrapsize$ increases, the empirical distribution of  
$
    (\btmcmcamip_1, \btmcmcamip_2, \ldots, \btmcmcamip_{\bootstrapsize})
$ 
becomes a better approximation of the bootstrap distribution.
However, the computational cost scales up with $\bootstrapsize$.
In practice, $\bootstrapsize$ in the hundreds is commonplace. Our numerical work uses $\bootstrapsize = \defaultbootstrapsize$.

We now define confidence intervals for $\amip$. 
Each interval is parametrized by $\nominal$, the nominal coverage level, which is valued in $(0,1)$. 
We compute two quantiles of the empirical distribution over 
$
    (\btmcmcamip_1, \btmcmcamip_2, \ldots, \btmcmcamip_{\bootstrapsize}),
$
the $(1-\nominal)/2$ and $(1+\nominal)/2$ quantiles.\footnote{We use R's quantile() to compute the sample quantiles. When $(1+\nominal)/2 \times \bootstrapsize$ is not an integer, the $(1+\nominal)/2$ quantile is defined by linearly interpolating the order statistics.} We define the interval spanned by these two values as our confidence interval.
By default, we set $\nominal = \defaultnominal$.

One limitation of our current work is that we do not make theoretical claims regarding the actual coverage of such confidence intervals. 
Although bootstrap confidence intervals can always be computed, whether the actual coverage matches the nominal coverage $\nominal$ depends on structural properties of the estimator and regularity conditions on the sample.  
To verify the quality of these confidence intervals, we turn to empirics.
We leave to future work the task of formulating reasonable assumptions and theoretically analyzing the actual coverage. 

\subsubsection{General MCMC}

In \cref{sec:exact_sampling}, we made the simplifying assumption that exact sampling was possible.
We now lift this assumption and handle the case in which $\markovchain$ arose from a Markov chain Monte Carlo algorithm (e.g., Hamiltonian Monte Carlo).
This case is much more common in practice than the exact sampling case.

To construct confidence intervals, one idea is to use the previous section's construction without modification.
In other words, one could apply the bootstrap to a non-i.i.d.\ sample. But recall that the Markov chain states are not independent of each other. 
Theoretically, it is known that the bootstrap struggles on non-i.i.d.\ samples, for even simple estimators. 
For example, if the estimator in question is the sample mean and the draws exhibit positive autocorrelation, under mild regularity conditions, the bootstrap variance estimate seriously underestimates the true sampling variance, even in the limit of infinite sample size \citep[Theorem 2.2]{Lahiri2003}.
In our case, the bootstrap likely struggles on the sample means that are involved in the definition of $\mcmcamip$ from \cref{eq:sosie-output}; for instance, it is very common for some coordinate $\coordsym$ that $(\mcmccoord{1}, \mcmccoord{2}, \ldots, \mcmccoord{\samplesize})$ exhibits positive autocorrelation in practice. 
Therefore, we have reason to be pessimistic about the ability of bootstrap confidence intervals to adequately cover $\amip$. 

Fundamentally, the bootstrap fails in the non-i.i.d.\ case because the draws that form the bootstrap sample do not have any dependence, while the draws that form the original sample do.
To improve upon the bootstrap, one option is to resample in a way that respects the original sample's dependence structure. 
We recognize that the sample in question, $\markovchain$, is a (multivariate) time series. So we focus on methods that perform well under time series dependence. 
One such scheme is the \emph{non-overlapping block bootstrap} \citep{Lahiri2003,Carlstein1986}.\footnote{The original paper \citep{Carlstein1986} did not use the term ``non-overlapping block bootstrap'' to describe the technique. The name comes from \citet{Lahiri2003}.} 
The sample $\markovchain$ is divided up into a number of blocks, where each block is a vector of contiguous draws.
Let $\blocklength$ be the number of elements in a block, and let $\numblocks :=  \lfloor \samplesize/\blocklength \rfloor$ denote the number of blocks.\footnote{All samples from the non-overlapping block bootstrap distribution will have length $\numblocks * \blocklength$. By construction, it may be the case that $\numblocks * \blocklength < \samplesize$. In all of our experiments, we choose $\numblocks, \blocklength, \samplesize$ so that $\numblocks * \blocklength = \samplesize$ exactly.}
The $m$-th block is defined as
\begin{equation*}
    B_m := \left( \mcmcdraws{(m-1)\blocklength+1},\ldots,\mcmcdraws{mL} \right).
\end{equation*}
To generate one sample from the non-overlapping block bootstrap distribution, we first draw a set of $\numblocks$ blocks; in particular, we draw them with replacement from the original set of $\numblocks$ blocks. We call the draws $B_1^*,\ldots,B_{\numblocks}^*$.
Then, we write the elements of these drawn blocks in a contiguous series.
For example, when $\markovchain = (\mcmcdraws{1}, \mcmcdraws{2}, \mcmcdraws{3}, \mcmcdraws{4})$ and $\blocklength = 2$, the two original blocks are $(\mcmcdraws{1}, \mcmcdraws{2})$, and $(\mcmcdraws{3}, \mcmcdraws{4})$.
The set of possible samples from resampling include $(\mcmcdraws{1}, \mcmcdraws{2}, \mcmcdraws{1}, \mcmcdraws{2})$ and $(\mcmcdraws{3}, \mcmcdraws{4}, \mcmcdraws{3}, \mcmcdraws{4})$ but not $(\mcmcdraws{1}, \mcmcdraws{3}, \mcmcdraws{1}, \mcmcdraws{3})$.

The name ``non-overlapping block bootstrap'' comes from the fact that these blocks, viewed as sets, are disjoint from each other.
The name is needed in \citet{Lahiri2003} to distinguish from other blocking rules. However, we consider only the above blocking rule, so moving forward we will refer to the procedure as simply the \emph{block bootstrap}. 
Intuitively, the block bootstrap sample is a good approximation of the original sample if the latter has short-term dependence; in such a case, the original sample itself can be thought of as the concatenation of smaller, i.i.d.\ subsamples, and the generation of a block bootstrap sample mimics that construction.
In well-behaved probabilistic models with well-tuned algorithms, the MCMC draws can be expected to have only short-term dependence, and the block bootstrap is a good choice.

The block bootstrap has one hyperparameter: the block length $\blocklength$.
We would like both $\blocklength$ and $\numblocks$ to be large; large $\blocklength$ captures time series dependence at larger lags, and large $\numblocks$ is close to having many i.i.d.\ subsamples.
However, since their product is constrained to be $\samplesize$, the choice of $\blocklength$ is a trade-off.
In numerical studies, we set $\blocklength = \defaultblocklength$.

Our construction of confidence intervals for general MCMC proceeds identically to the previous section's construction, except for the step of generating the bootstrap sample: instead of drawing from the vanilla bootstrap, we draw from the block bootstrap. 
We will denote the endpoints of such an interval by $\amiplb$ (lower endpoint) and $\amipub$ (upper endpoint). 

Similar to the previous section, we do not make theoretical claims on the actual coverage of our block bootstrap confidence intervals; instead, we verify the quality of the intervals through later numerical studies.

\subsection{Putting everything together} \label{sec:final-output}
Now, we chain together the intermediate approximations from the previous sections to form our final estimate of \cref{eq:mip}.
We then explain how to use it to determine non-robustness. 

Rather than a single point estimate of the Maximum Influence Perturbation, we provide the interval $\amipci$ constructed in \cref{sec:ci}.
This approximation is the result of combining \cref{sec:ci}, where $\amipci$ is designed to cover $\amip$ with high probability, with \cref{sec:taylor-series}, where $\amip$ approximates the Maximum Influence Perturbation.
Our final estimate of the Most Influential Set is $\mcmcamis$ from \cref{eq:sosie-output}.
This approximation is the result of combining \cref{sec:estimate-infl}, where $\mcmcamis$ approximates $\amis$, with \cref{sec:taylor-series}, where $\amis$ approximates the Most Influential Set.

To determine non-robustness, we use $\amipci$ as follows. 
Recall that we have assumed for simplicity that the decision threshold is zero, and that $\qoi{\allones} < 0$.
We believe that the interval $[\qoi{\allones} + \amiplb, \qoi{\allones} + \amipub]$ contains the quantity of interest after removing the most extreme observations.
Therefore, our assessment of non-robustness depends on the relationship between this interval and the threshold zero in the following way:
\begin{itemize}
    \item $\qoi{\allones} + \amiplb > 0$. Hence, $[\qoi{\allones} + \amiplb, \qoi{\allones} + \amipub ]$ is entirely on the opposite side of $0$ compared to $\qoi{\allones}$. We declare the analysis to be non-robust. 
    \item $\qoi{\allones} + \amipub < 0$. Hence, $[\qoi{\allones} + \amiplb, \qoi{\allones} + \amipub ]$ is entirely on the same side of $0$ compared to $\qoi{\allones}$. We do not declare non-robustness.
    \item $\qoi{\allones} + \amiplb \leq 0 \leq \qoi{\allones} + \amipub$. The interval contains $0$, and we abstain from making an assessment about non-robustness. We recommend practitioners run more MCMC draws to reduce the width of the confidence interval.
\end{itemize}

While $\amipci$ plays the main role in determining non-robustness, $\mcmcamis$ plays a supporting role.
For problems in which drawing a second MCMC sample is not prohibitively expensive, we can refit the analysis without the data points in $\mcmcamis$.
Performing the refit is one way of verifying the quality of our assessment (of non-robustness); if $\amipci$ declares that the conclusion is non-robust, and the conclusion truly changes after removing $\mcmcamis$ and refitting, then we conclusively know that our assessment is correct.

\section{Theory} \label{sec:theory}
In this section, we theoretically quantify the errors incurred by our approximations.
First, in \cref{sec:first-order-accuracy}, we analyze the error made by approximating $\qoi{\weight}$ with a first-order Taylor series.
Although our analysis is limited to a simple probablistic model, we conclusively show that this error is always small relative to a natural notion of scale.
Second, in \cref{sec:est-properties}, we analyze the error made by using MCMC to estimate influences $\infln$.
Under more stringent assumptions than those needed to apply our procedure,
we show that our estimator possesses a number of desirable properties.
For one, our estimator of $\amip$ ($\mcmcamip$ from \cref{eq:sosie-output}) is consistent in the limit $\samplesize \to \infty$.
\subsection{Accuracy of first-order approximation} \label{sec:first-order-accuracy}

In this section, we investigate the error incurred by replacing $\qoi{\weight}$ with the Taylor series from \cref{sec:taylor-series}.
While the Taylor series approximation applies to any model that satisfies \cref{assume:feasible-has-finite-normalization,assume:mean-and-sd,assume:finiteness}, our error analysis is limited to a normal model.
To ground our analysis, we first need a notion of scale. 
A baseline approximation to dropping data is to do nothing, i.e.\ approximate $\qoi{\weight}$ with $\qoi{\allones}$. 
We use the error of this ``zeroth-order'' Taylor series as the scale.
In this section, we show that the first-order error is small in this natural scale.
In \cref{sec:more-theory}, we calculate the errors for a hierarchical extension of the normal model.
For such a model, although we can articulate conditions under which the first-order error is smaller than the zeroth-order error, such conditions are not immediately interpretable; we leave to future work to provide a more intuitive understanding of these conditions.

We begin by detailing the data, prior, and likelihood for the normal model.
We will also specify a quantity of interest.
The $n$-th observed data point is $\obsn = \covarn$.
The parameter of interest is the population mean $\globalmean$.
The likelihood of an observation is Gaussian with a known standard deviation $\noise$.
In other words, the $n$-th log-likelihood evaluated at $\globalmean$ is 
$
    \loglik{n}{\globalmean} = 
       \frac{1}{2}  \log \left( \frac{1}{2\pi \noise^2} \right) - \frac{1}{2\noise^2}[ (\covarn)^2 - 2 \covarn \globalmean + \globalmean^2].
$
We choose the uniform distribution over the real line as the prior for $\globalmean$.
The quantity of interest is the posterior mean of $\globalmean$.

We next pin down the two notions of error.
We define the first-order error to be the (signed) difference between $\qoi{\weight}$ and $\qoi{\allones} + \sum_n (\weight_n - 1) \infln$.
We mainly care when $\weight$ encodes the full removal of certain observations and full inclusion of the remaining ones; i.e.\ $\weight \in \{0,1\}^{\nobs}$.
If we let $\indexFromW$ be the function that returns the zero indices of such a weight ($\indexFromW(\weight) = \{n: \weight_n = 0\}$), then its inverse $\wFromIndex$ takes a set of observation indices $\indexset$ and produces a weight valued in $\{0,1\}^{\nobs}$. In what follows, we take $\indexset \subsetneq \allObsIndices$ and $\indexset \neq \emptyset$.
We reformulate the error as a function of $\indexset$ instead of $\weight$ by replacing $\weight$ with $\wFromIndex(\indexset)$ in the definition of error. 
After reformulation, we can write the error as follows.
\begin{equation*}
    \firstOrderError := \qoi{\wFromIndex(\indexset)} - \qoi{\allones} + \sum_{n \in I} \infln.
\end{equation*}
For the zeroth-order approximation, i.e.\ approximating $\qoi{\weight}$ with $\qoi{\allones}$, the error is 
\begin{equation*}
    \zerothOrderError := \qoi{\wFromIndex(\indexset)} - \qoi{\allones}.
\end{equation*}

To display the error formulas, it is convenient to introduce the following notation.
We define the sample average of observations as a function of $\indexset$:
$
    \meanOfIndexSet := (1/|\indexset|) \sum_{n \in \indexset} \covarn.  
$
We denote the sample average of the whole dataset by $\bar{\covar}$.

In this model, expectations under the weighted posterior have closed forms. 
We can derive an explicit expression for the first-order error. 

\begin{mylemma}\label{lem:normal-error} 
    For the normal model, $\firstOrderError$ is equal to
    \begin{equation*} 
        \frac{|\indexset|^2 (\bar{\covar}  - \meanOfIndexSet )}
             { \nobs (\nobs - |\indexset|) }
    \end{equation*}
\end{mylemma} 
We prove \cref{lem:normal-error} in the \proofref{lem:normal-error}.
The error is a function of $\indexset$ through the a) the cardinality of the set $|\indexset|$ and b) the difference between the whole dataset's sample mean, $\bar{\covar}$, and the sample mean for elements in $\indexset$.
Because $\indexset$ is a strict subset of $\allObsIndices$, $|\indexset| < \nobs$. 
So, the denominator is always non-zero, and the error is always well-defined.

We also have an explicit expression for the zeroth-order error.

\begin{mylemma} \label{lem:normal-zeroth-order-error}
    For the normal model, $\zerothOrderError$ is equal to
    \begin{equation*} 
        \frac{|\indexset| (\bar{\covar}  - \meanOfIndexSet )}
             { (\nobs - |\indexset|) }.
    \end{equation*}
\end{mylemma}

We prove \cref{lem:normal-zeroth-order-error} in the \proofref{lem:normal-zeroth-order-error}.
Comparing the expression in \cref{lem:normal-zeroth-order-error} with the expression in \cref{lem:normal-error}, we see that the first-order error is equal to the zeroth-order error times $|\indexset|/\nobs$, which is $\dropout$, the fraction of data removed.
We are interested in $\dropout$ close to $0$.
So, the first-order error is substantively smaller than the zeroth-order error for $\dropout$ of interest. 

\subsection{Desirable properties of MCMC estimators}  \label{sec:est-properties}

Recall from \cref{sec:ci} that one concern regarding the quality of $\mcmcamip$ is the $\markovchain$-induced sampling uncertainty. 
Theoretically analyzing this uncertainty is difficult, with one obstacle being  that $\mcmcamip$ is a non-smooth function of $\markovchain$.
In this section, we settle for the easier goal of analyzing the sampling uncertainty of the influence estimates $\mcmcinfln$.
We expect such theoretical characterizations to play a role in the eventual theoretical characterizations of $\mcmcamip$, but we leave this step to future work.

In this analysis, we make more restrictive assumptions than those needed for \cref{thm:linear-map-exists} to hold.
We assume that the sample $\markovchain$ comes from exact sampling; the independence across draws makes it easier to analyze sampling uncertainty.
We focus on the quantity of interest equaling the posterior mean ($\const_1 = 1, \const_2 = 0$ in the sense of \cref{assume:mean-and-sd}). The choice $\const_1 = 1$ is made out of convenience. A similar analysis can be conducted when $\const_2 \neq 0$, but we omit it for brevity. 
Finally, we need more stringent moment conditions than \cref{assume:finiteness}.

\begin{assumption} \label{assume:more-moments-and-iid}
    The functions $|\func{\param}^2 \loglik{i}{\param} \loglik{j}{\param}|$ (across $i,j$) have finite expectation under the full-data posterior.
\end{assumption}
This moment condition guarantees that the sample covariance between $\func{\param}$ and $\loglik{i}{\param}$ has finite variance under the full-data posterior.
When proving desirable properties about the sample variance, such as consistency, one typical moment condition is that the population kurtosis is finite.
Here, the assumed finite variance plays a similar role (in our analysis of sample covariance consistency) as that played by finite kurtosis (in an analysis of sample variance consistency).

With the assumptions in place, we begin by showing that the sampling uncertainty of $\mcmcinfln$ goes to zero in the limit of $\samplesize \to \infty$.

\begin{mylemma} \label{lem:mcmcinfln-variance}
    Take \cref{assume:feasible-has-finite-normalization,assume:mean-and-sd,assume:finiteness,assume:more-moments-and-iid}.
    Take $\markovchain$ to be an i.i.d.\ sample.
    Let $\mcmcinfl$ be the output of \cref{alg:EI} for $\const_1 = 1$, $\const_2 = 0$, and $\markovchain$.
    Then, there exists a constant $C$ such that for all $n$ and for all $\samplesize$, $\Var(\mcmcinfln) \leq  C/\samplesize$.
\end{mylemma}

We prove \cref{lem:mcmcinfln-variance} in the \proofref{lem:mcmcinfln-variance}.
That the variance of individual $\mcmcinfln$ goes to zero at the rate of $1/\samplesize$ is not surprising; $\mcmcinfln$ is a sample covariance, after all.

We use \cref{lem:mcmcinfln-variance} to show consistency of different estimators.

\begin{mytheorem} \label{thm:mcmcinfl-mcmcamip-consistent}
    Take \cref{assume:feasible-has-finite-normalization,assume:mean-and-sd,assume:finiteness,assume:more-moments-and-iid}.
    Take $\markovchain$ to be an i.i.d.\ sample.
    Let $\mcmcinfl$ be the output of \cref{alg:EI} for $\const_1 = 1$, $\const_2 = 0$, and $\markovchain$.
    Then $\max_{n=1}^{\nobs} | \mcmcinfln - \infln |$ converges in probability to $0$ in the limit $\samplesize \to \infty$, and $\mcmcamip$ converges in probability to $\amip$ in the limit $\samplesize \to \infty$.
\end{mytheorem}
We prove \cref{thm:mcmcinfl-mcmcamip-consistent} in the \proofref{thm:mcmcinfl-mcmcamip-consistent}. 
Our theorem states that the vector $\mcmcinfl$ is a consistent estimator for the vector $\infl$ and $\mcmcamip$ is a consistent estimator for $\amip$.

Not only is $\mcmcinfl$ consistent in estimating $\infl$, it is also asymptotically normal.
\begin{mytheorem} \label{thm:mcmcinfl-asymptotic-normal}
    Take \cref{assume:feasible-has-finite-normalization,assume:mean-and-sd,assume:finiteness,assume:more-moments-and-iid}.
    Take $\markovchain$ to be an i.i.d.\ sample.
    Let $\mcmcinfl$ be the output of \cref{alg:EI} for $\const_1 = 1$, $\const_2 = 0$, and $\markovchain$.
    Then 
    $
        \sqrt{\samplesize} ( \mcmcinfl - \infl )
    $ 
    converges in distribution to $N(\allzeros, \limitVarMat)$ where $\limitVarMat$ is the $\nobs \times \nobs$ matrix whose $(i,j)$ entry, $\limitVarMat_{i,j}$, is the covariance between
    $$
    \left( 
        \func{\param}  - \Eposterior{\allones} \func{\param}
    \right) 
    \left( 
        \loglik{i}{\param}  -  \Eposterior{\allones} \loglik{i}{\param} 
    \right)
    $$
    and
    $$
    \left( 
        \func{\param}  - \Eposterior{\allones} \func{\param}
    \right) 
    \left( 
        \loglik{j}{\param}  -  \Eposterior{\allones} \loglik{j}{\param} 
    \right),$$
    taken under the full-data posterior.
\end{mytheorem}
We prove \cref{thm:mcmcinfl-asymptotic-normal} in the \proofref{thm:mcmcinfl-asymptotic-normal}. 
Heuristically, for each $n$, the distribution of $\mcmcinfln$ is the Gaussian centered at $\infln$, with standard deviation $\sqrt{\limitVarMat_{n,n}}/\sqrt{\samplesize}$.

\subsubsection{Normal model with unknown precision.}

The quantity $\sqrt{\limitVarMat_{n,n}}/\sqrt{\samplesize}$ eventually goes to zero as $\samplesize \rightarrow \infty$. But for finite $\samplesize$, this standard deviation can be large, and $\mcmcinfln$ can be an imprecise estimate of $\infln$.
To illustrate this phenomenon, we will derive $\limitVarMat_{n,n}$ in the context of a simple probabilistic model: a normal model with unknown precision.

We first introduce the model and the associated quantity of interest.
The data is a set of $\nobs$ real values: $\obsn = \covarn$, where $\covarn \in \mathbb{R}$.
The parameters of interest are the mean $\mu$ and the precision $\tau$ of the population.
The log-likelihood of an observation based on $\mu$ and $\tau$ is Gaussian:  
$ 
    \frac{1}{2} \log \left( \frac{\tau}{2\pi} \right) - \frac{1}{2}\tau [ (\covarn)^2 - 2 \covarn \mu + \mu^2].
$
The priors are chosen as follows.
$\mu$ is distributed uniformly over the real line, and $\tau$ is distributed according to a gamma distribution.
The quantity of interest is the posterior mean of $\mu$.

For this probabilistic model, the assumptions of \cref{thm:mcmcinfl-asymptotic-normal} are satisfied. 
We show that the variance $\limitVarMat_{n,n}$ behaves like a quartic function of the observation $\covarn$. 

\begin{mylemma} \label{lem:normal-gamma-limitVarMat}
    In the normal-gamma model, there exists constants $D_1$, $D_2$, and $D_3$, where $D_1 > 0$, such that for all $n$, $\limitVarMat_{n,n}$ is equal to $D_1 (\covarn - \bar{\covar})^4 + D_2 (\covarn - \bar{\covar})^2 + D_3$.
\end{mylemma} 
We prove \cref{lem:normal-gamma-limitVarMat} in the \proofref{lem:normal-gamma-limitVarMat}. 
$D_1, D_2, D_3$ are based on posterior expectations.
For instance, the proof shows that $D_1 = \frac{ \Eposterior{\allones} [\tau^{-1}(\tau - \Eposterior{\allones} \tau)^2]}{4\nobs}$.
It is easy to show that for the normal-gamma model,
\begin{equation*}
    \Cov_{\allones}(\mu, \loglik{n}{\mu,\tau}) = \frac{\covarn - \bar{\covar}}{\nobs}.
\end{equation*}
Hence, while the mean of $\mcmcinfln$ behaves like a linear function of $\covarn - \bar{\covar}$, its standard deviation behaves like a quadratic function of $\covarn - \bar{\covar}$.
In other words, the more influence an observation has, the harder it is to accurately determine its influence!

\section{Experimental Setup} \label{sec:setup}
For the rest of the paper, we check the quality of our approximations empirically on real data analyses.
In this section, we only describe the checks; for the actual results, see \cref{sec:experiments}.

A practitioner with a particular definition of ``small data'' can set $\dropout$ to reflect their concern.
We consider a number of $\dropout$ values.
We set the maximum value of $\dropout$ to be $0.01$.
This choice is motivated by \citet{Broderick2023}.
Many analyses are non-robust to removing $1\%$ of the data, and we a priori think that $\dropout > 1\%$ is a large amount of data to remove.
We vary $\log_{10}(\dropout)$ in an equidistant grid of length $10$ from $-3$ to $-2$.
The ten values are $0.10\%$, $0.13\%$, $0.17\%$, $0.22\%$, $0.28\%$, $0.36\%$, $0.46\%$, $0.60\%$, $0.77\%$ and $1.00\%$.
In addition to these 10 values, we also consider $\alpha$ that correponds to removing only one observation from the data: in all, there are 11 values of $\dropout$ under consideration.

For the range of dropout fractions specified above and across three common quantities of interest corresponding to sign, significance, and significant result of opposite sign changes, we walk through what a practitioner would do in practice (although they would choose only one $\dropout$ and one decision).
Our method proposes an influential data subset and a change in the quantity of interest, represented by a confidence interval. 

Ideally, we want to check if our interval includes the result of the worst-case data to leave out.
We are unable to do so, since we do not know how to compute the worst-case result in a reasonable amount of time.
We settle for the following checks.

In the first check, for a particular MCMC run, we plot how the change from re-running minus the proposed data compares to the confidence interval.
We recommend the user run this check if re-running MCMC a second time is not too computationally expensive.

Unfortunately, such refitting does not paint a complete picture of approximation quality.
For instance, the MCMC run might be unlucky since MCMC is random.
To be more comprehensive, we run additional checks.
We do not expect users to run these tests, as their computational costs are high.
The central question is how frequently (under MCMC randomness) the confidence interval includes the result after removing the worst-case data.
To assess this frequency, we recall the approximations made in constructing the confidence interval, and check the quality of each approximation separately.
In one approximation, we estimate dropping data with a linear approximation; in the other approximation, we construct a confidence interval around the result of the linear approximation.
So, we have two checks.
The first (\cref{sec:amip-coverage-exp}) checks how frequently the confidence interval includes the result of the linear approximation, i.e.\ the AMIP.
The second (\cref{sec:rerun-interpolation}) checks whether the AMIP is a good approximation of dropping data. 
To understand \emph{why} we observe the coverage in \cref{sec:amip-coverage-exp}, in \cref{sec:soi-coverage-exp} we isolate the impact of the sorting step in the construction of our confidence interval.

\subsection{Estimating coverage of confidence intervals for AMIP} \label{sec:amip-coverage-exp}

We estimate how frequently $\amipci$ covers the AMIP by using another level of Monte Carlo.
Recall that $\amipci$ is intended to be a confidence interval covering $\amip$  a fraction $\nominal$ of the time.
If the estimated coverage is far from $\nominal$, we have evidence that $\amipci$ does not achieve the desired nominal coverage.

We draw $\numchains$ Markov chains; we set $\numchains = \defaultnumchains$.
On each chain, we estimate the influences and construct the confidence interval $\amipci$.
Observe that, for each $n$, we have $\numchains$ estimates of $\infln$.
We take the sample mean across chains and denote this quantity by $\groundtruthmcmcinfl{n}$. Because of variance reduction through averaging, $\groundtruthmcmcinfl{n}$ is a much better estimate of $\infln$ than individual $\mcmcinfln$.
We denote the indices of the $\numdrop$ most negative $\groundtruthmcmcinfl{n}$ by $\groundtruthamis{\dropout}$.
We sort $\groundtruthmcmcinfl{n}$ across $n$ and sum the $\numdrop$ most negative $\groundtruthmcmcinfl{n}$.
This sum is denoted by $\groundtruthamip{\dropout}$; we use it in place of the ground truth $\amip$. 
We use the sample mean of the indicators $\indict{\groundtruthamip{\dropout} \in \amipci}$ as the point estimate of the coverage.
We also report a $95\%$ confidence interval for the coverage.
This interval is computed using binomial tests designed in \citet{Clopper1934} and implemented as R's binom.test() function.

\subsection{Estimating coverage of confidence intervals for sum-of-influence} \label{sec:soi-coverage-exp}
It is possible that the estimated coverage of $\amipci$ is far from the nominal $\nominal$.
We suspect that such a discrepancy comes from the sorting of $\mcmcinfln$ to construct $\amip$.
To modularize out the sorting, we consider a target of inference that is simpler than $\amip$.
At a high level, we fix an index set $\indexset$, and define the target to be the sum of influences in $\indexset$: $\sum_{n \in I} \infln$.
On each sample $\markovchain$, our point estimate is $\sum_{n \in I} \mcmcinfln$: this estimate does not involve any sorting, while $\mcmcamip$ does. 
We construct the confidence interval, $\soici$, from the block bootstrap distribution of $\sum_{n \in I} \mcmcinfln$. 
The difference between $\soici$ and $\amipci$, which is constructed from the block bootstrap distribution of $\mcmcamip$, is that the former is not based on sorting the influence estimates.
If the actual coverage of $\soici$ is close to the nominal value, we have evidence that any miscoverage of $\amipci$ is due to this sorting.

From \cref{sec:amip-coverage-exp} we use $\groundtruthmcmcinfl{n}$ and the associated $\groundtruthamip{\dropout}$ and $\groundtruthamis{\dropout}$ as replacement for ground truths.
We set $\indexset$ to be $\groundtruthamis{\dropout}$.
We run another set of $\numchains$ Markov chains: for each chain, we construct the confidence interval $\soici$ by sampling from the block bootstrap distribution of the estimator $\sum_{n \in I} \mcmcinfln$. 
We report the sample mean of the indicators $\indict{\sum_{n \in I} \groundtruthmcmcinfl{n} \in \soici}$ as our point estimate of the coverage.
We also report a $95\%$ confidence interval for the coverage.
This interval is computed using binomial tests designed in \citet{Clopper1934} and implemented as R's binom.test() function.

\subsection{Re-running MCMC on interpolation path} \label{sec:rerun-interpolation}
Ideally, we want to know the difference between the Maximum Influence Perturbation and the AMIP.
As we have established, we do not know how to compute the former efficiently.
We settle for checking the linearity approximation made in \cref{sec:taylor-series}; recall that this approximation estimates $\qoi{\weight} - \qoi{\allones}$ with $\sum_{n}(\weight_n - 1) \infln$.
In particular, we expect the first-order Taylor series approximation to be arbitrarily good for $\weight$ arbitrarily close to $\allones$. 
By necessity, we are interested in some $\weight^*$ that has a non-trivial distance from $\allones$.
Plotting the quantity of interest $\qoi{\weight}$ on an interpolation path between $\allones$ and $\weight^*$, we get a sense of how much we have diverged from linearity by that point.

From \cref{sec:amip-coverage-exp}, we have $\groundtruthmcmcinfl{n}$ as our replacement for the ground truth $\infln$.
We focus on $\dropout = \defaultdropout$: $\defaultdropoutaspercentage$ is a large amount of data to remove, and a priori we expect the linear approximation to be poor.
Recall that $\groundtruthamis{\defaultdropout}$ is the set of $\lfloor \defaultdropout \nobs \rfloor$ observations that are most influential according to sorted $\groundtruthmcmcinfl{n}$.
Let $\weight^*$ be the $\nobs$-dimensional weight vector that is $1$ for observations in $\groundtruthamis{\defaultdropout}$ and $0$ otherwise.
For $\zeta \in [0,1]$, the linear approximation of $\qoi{\zeta w^* + (1-\zeta)\allones}$ is $\qoi{\allones} + \zeta \groundtruthamip{\defaultdropout}$.
In the extreme $\zeta = 0$, we do not leave out any data.
In the extreme $\zeta = 1$, we leave out the entirety of $\groundtruthamis{\defaultdropout}$ i.e.\ $\defaultdropoutaspercentage$ of the data.
An intermediate value $\zeta$ roughly corresponds\footnote{This correspondence is not exact, since for $\zeta  < 1$, all observations in $\groundtruthamis{\defaultdropout}$ are included in the analysis, only with downplayed contributions.} to removing $(\zeta 5) \%$ of the data. 
We discretize $[0,1]$ with $15$ values: $0$, $0.0010$, $0.0016$, $0.0027$, $0.0044$, $0.0072$, $0.0118$, $0.0193$, $0.0316$,  $0.0518$, $0.0848$, $0.1389$, $0.2276$, $0.3728$, $0.6105$, $1$.
For each value on this grid, we run MCMC to estimate $\qoi{\zeta w^* + (1-\zeta)\allones}$, and compare it to the linear approximation.

\section{Experiments} \label{sec:experiments}
In our experiments, we find that our approximation works well for a simple linear model. 
But we find that it can struggle in hierarchical models with more complex structure.

\subsection{Linear model} \label{sec:linear_model}
We consider a slight variation of a microcredit analysis from \citet{Meager2019}.
In \citet{Meager2019}, conclusions regarding microcredit efficacy were based on ordinary least squares (OLS).
We refer the reader to \citet[Section 4.3.2]{Broderick2023} for investigations of such conclusions' non-robustness.
Here, we instead consider an analogous Bayesian analysis using MCMC,
and we examine the robustness of conclusions from this analysis.
Even for this very simple Bayesian analysis, it is possible to change substantive conclusions by removing a small fraction of the data.

Our quality checks suggest that our approximation is accurate. 
Our confidence interval contains the refit after removing the proposed data.
The actual coverage of the confidence interval for AMIP is close to the nominal coverage.
The actual coverage of the confidence interval for sum-of-influence is also close to the nominal coverage. 
Even for dropping $\defaultdropoutaspercentage$ of the data, the linear approximation is still adequate. 

\subsubsection{Background and full-data fit}

\citet{Meager2019} studies the microcredit data from \citet{Angelucci2015}, which was an RCT conducted in Mexico. 
There are $\nobs = 16{,}560$ households in the RCT.
Each observation is $\obsn = (\covarn, \responsen)$, where $\covarn$ is the treatment status and $\responsen$ is the profit measured.
The log-likelihood for the $n$-th observation is 
$ \loglik{n}{\intercept,\slope,\sigma} = 
-\frac{1}{2\sigma^2} (\responsen - \slope \covarn - \intercept)^2 - \frac{1}{2}\log(2\pi \sigma^2).
$
Here, the model parameters are baseline profit $\intercept$, treatment effect $\slope$, and noise scale $\sigma$.
The most interesting parameter is $\slope$; as $\covarn$ is binary, $\slope$ compares the means in the treatment and control groups.
\citet{Meager2019} estimates the model parameters with OLS.
Our variation of the above analysis is as follows.
We put \textit{t} location-scale distribution priors on the model parameters, with the additional constraint that the noise scale $\sigma$ is positive; for exact values of the prior hyperparameters, see \cref{subsec:linear_model_details}.
We use Hamiltonian Monte Carlo (HMC) as implemented in Stan \citep{Carpenter2017} to approximate the full-data posterior.
We draw $\samplesize = \defaultsamplesize$ samples.

\begin{figure}[t]
    \centering
    \includegraphics{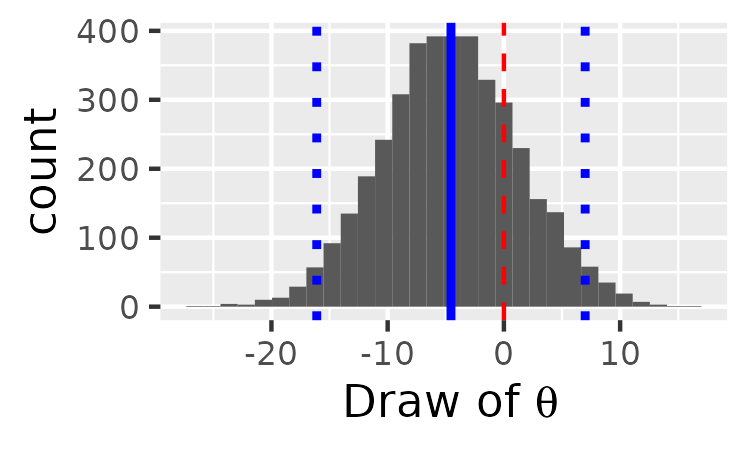}
    \caption{(Linear model) Histogram of treatment effect MCMC draws. The blue line indicates the sample mean. The dashed red line is the zero threshold. The dotted blue lines indicate estimates of approximate credible interval's endpoints.}
    \label{fig:mexico_hist}
\end{figure}

\Cref{fig:mexico_hist} plots the histogram of the treatment effect draws as well as key sample summaries. 
The sample mean is equal\footnote{We round to two decimal places when reporting results of our numerical studies.} to $-4.55$.
The sample standard deviation is $5.79$.
These values are close to the point estimate and the standard error from OLS \citep{Meager2019}.
Our estimate of the approximate credible interval's left endpoint is $-16.10$; our estimate of the right endpoint is $6.99$.
Based on these summaries, an analyst would likely conclude that while the posterior mean of the effect of microcredit is negative, the uncertainty interval covers zero, so they cannot confidently conclude that microcredit either helps or hurts.
These conclusions are in line with \citet{Meager2019}.

\subsubsection{Sensitivity results}

The running of our approximation takes very little time compared to the running of the original analysis.
Generating the draws in \cref{fig:mexico_hist} took 3 minutes on MIT Supercloud \citep{Reuther2018}.
For one $\dropout$ and one quantity of interest, it took less than $5$ seconds to make a confidence interval for what happens if we remove the most extreme data subset.
A user might check approximation quality by dropping a proposed subset and re-running MCMC; each such check took us around 3 minutes, the runtime of the original analysis.

\begin{figure}[t]
    \centering
    \includegraphics{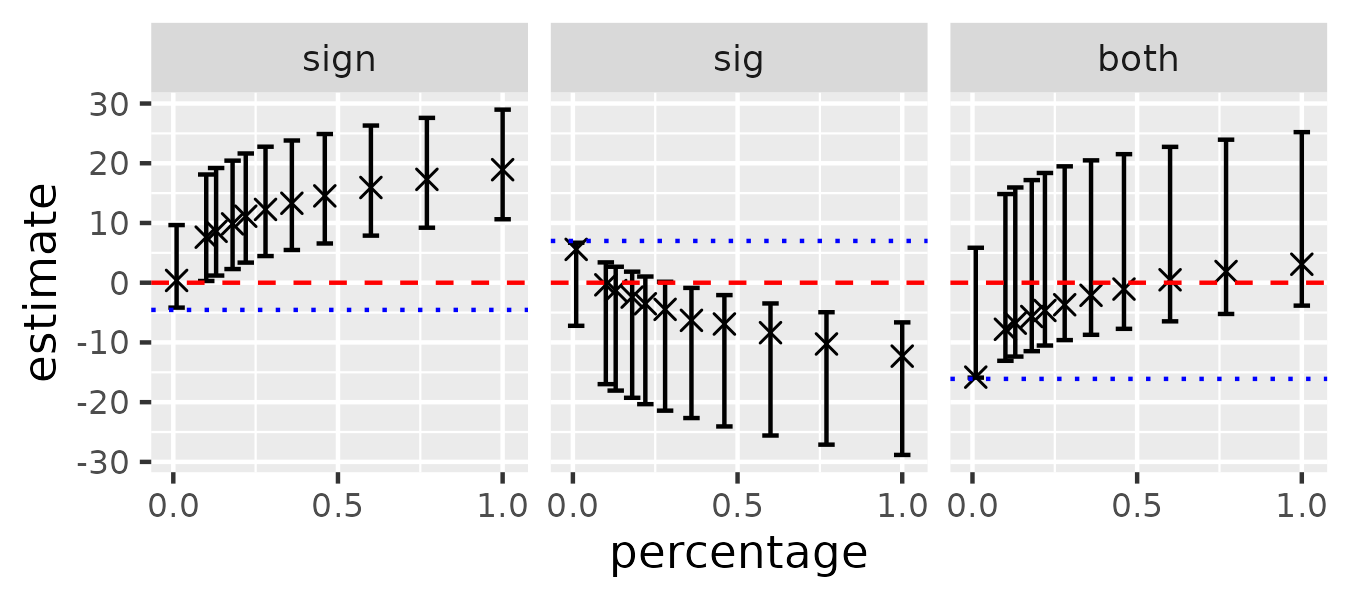}
    \caption{(Linear model) 
    Confidence interval and refit.
    At maximum, we remove $1\%$ of the data.
    Each panel corresponds to a target conclusion change: `sign' is the change in sign, `sig' is change in significance, and `both' is the change to a significant effect of the opposite sign.
    Error bars are confidence interval for refit after removing the most extreme data subset. Each `x' is the refit after removing the proposed data and re-running MCMC.
    The dotted blue line is the fit on the full data.
    }
    \label{fig:Mexico_refit}
\end{figure}

In \cref{fig:Mexico_refit}, we plot our confidence intervals and the result after removing the proposed data.
Although the confidence intervals are wide, they are still useful.
Across quantities of interest and removal fractions, our intervals contain the refit after removing the proposed data. 
For changing sign, our method predicts there exists a data subset of relative size at most $0.1\%$ such that if we remove it, we change the posterior mean's sign.
Refitting after removing the proposed data confirms this prediction.
For changing significance, our method predicts there exists a data subset of relative size at most $0.36\%$ such that if we remove it, we change the sign of the approximate credible interval's right endpoint; refitting confirms this prediction.
Our method is not able to predict whether the result can be changed to significant effect of the opposite sign for these $\dropout$ values and this number of samples; we recommend a larger number of MCMC samples. 

\subsubsection{Additional quality checks}

\begin{figure}[t]
    \centering
    \includegraphics{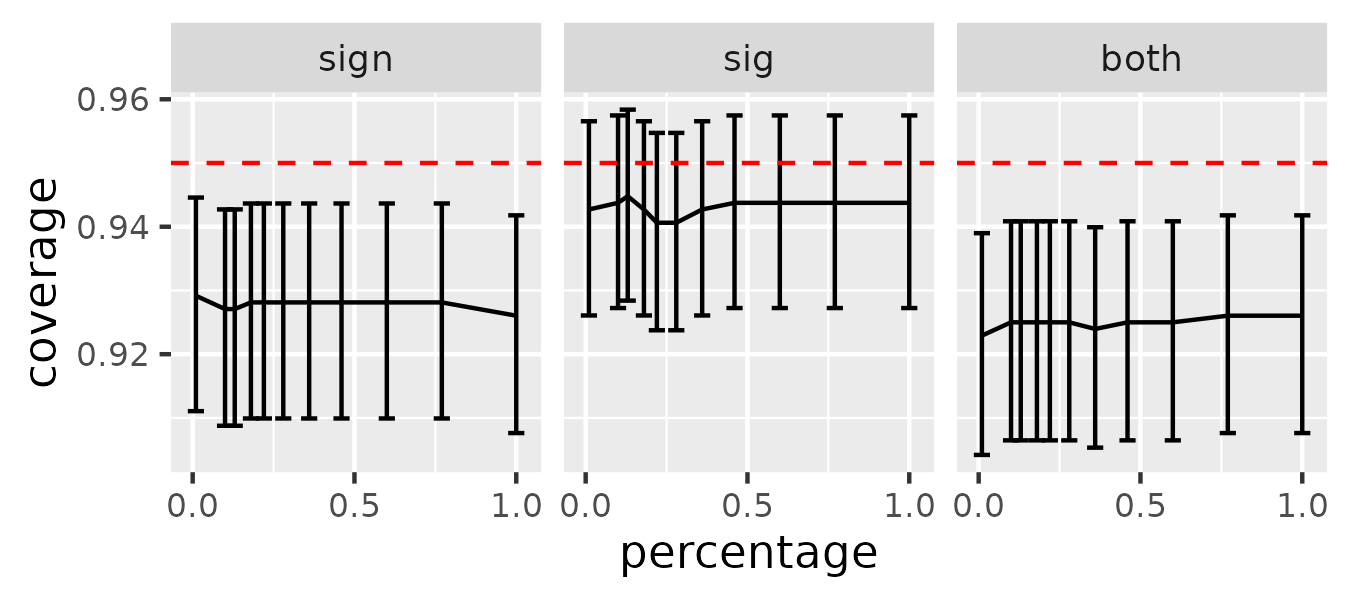}
    \caption{(Linear model) Monte Carlo estimate of AMIP confidence interval's coverage.
    Each panel corresponds to a target conclusion change.
    The dashed line is the nominal level $\nominal = \defaultnominal$. 
    The solid line is the sample mean of the indicator variable for the event that ground truth is contained in the confidence interval.
    The error bars are confidence intervals for the population mean of these indicators.}
    \label{fig:Mexico_amip_coverage}
\end{figure}

\Cref{fig:Mexico_amip_coverage} shows that the actual coverage of the confidence interval for the AMIP is close to the nominal one, across $\dropout$.
As the half-width of each error bar is small (only $0.02$), we believe that the difference between the  true coverage and our point estimate of it is small. 
For either `sign' or `both' QoI, the error bars do not contain the nominal $\nominal$.
However, the difference between the point estimate and the nominal $\eta$ is only $0.03$ at worst, which is small. 
For the `sig' QoI, the point estimate is within $0.005$ of the nominal value, and the error bars contain the nominal $\nominal$.

\begin{figure}[t]
    \centering
    \includegraphics{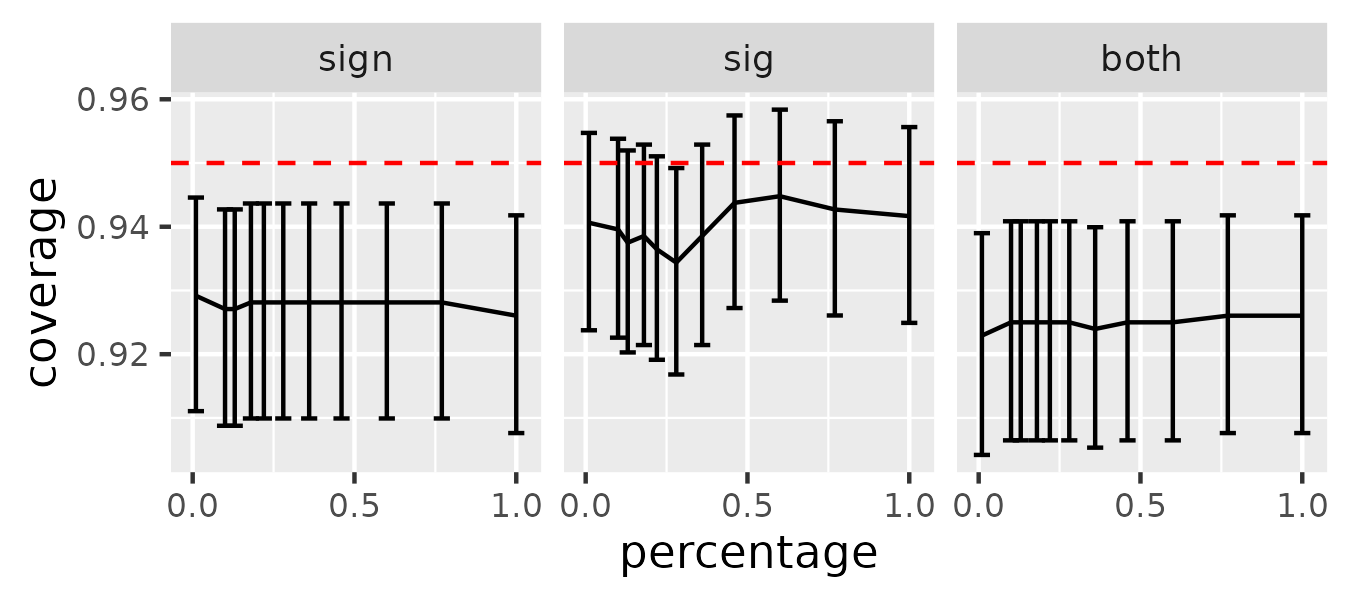}
    \caption{(Linear model) Monte Carlo estimate of sum-of-influence confidence interval's coverage.
    Each panel corresponds to a target conclusion change.
    The dashed line is the nominal level $\nominal = \defaultnominal$. 
    The solid line is the sample mean of the indicator variable for the event that ground truth is contained in the confidence interval, and error bars are confidence intervals for the population mean of these indicators.
    }
    \label{fig:Mexico_soi_coverage}
\end{figure}

\Cref{fig:Mexico_soi_coverage} shows that the actual coverage of the confidence interval for the sum-of-influence is close to the nominal one across $\dropout$.
The absolute errors between our estimate of coverage and the nominal $\nominal$ are similar to those seen in \cref{fig:Mexico_amip_coverage}.
This success suggests that the default block length, $\blocklength = \defaultblocklength$, is appropriate for this problem.

\begin{figure}[t]
    \centering
    \includegraphics{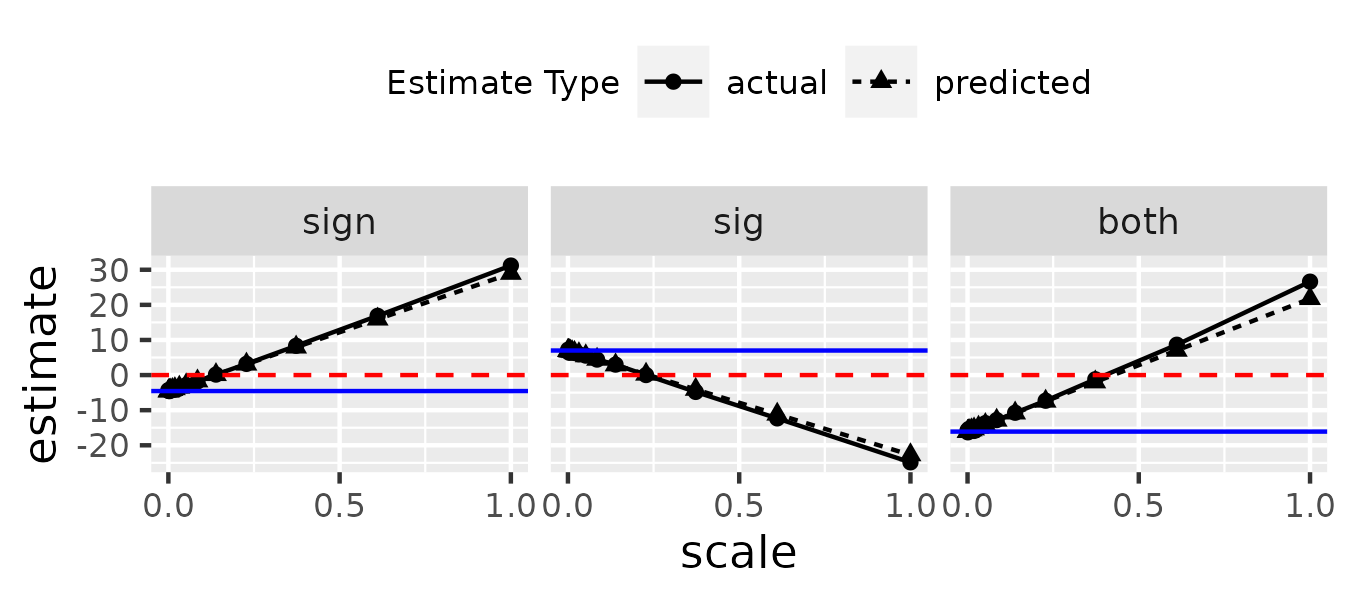}
    \caption{(Linear model) Quality of the linear approximation. 
    Each panel corresponds to a target conclusion change.
    The solid blue line is the full-data fit.
    The horizontal axis is the distance from the weight that represents the full data.
    We plot both the refit from rerunning MCMC and the linear approximation of the refit.}
    \label{fig:Mexico_linear_approx}
\end{figure}

\Cref{fig:Mexico_linear_approx} shows that the linear approximation works very well.
It is somewhat remarkable that the linear approximation is this good even after dropping $\defaultdropoutaspercentage$, which we consider to be a large fraction of data.
The horizontal axis (`scale') is the same as $\zeta$ in \cref{sec:rerun-interpolation}.
For all quantities of interest, the linear approximation and the refit lie mostly on top of each other; towards the right end of each panel, the approximation slightly underestimates the refit.

\subsection{Hierarchical model on microcredit data} \label{sec:microcredit_mixture}

We consider a part of the analysis of microcredit done in \citet{Meager2022}.
Originally, \citeauthor{Meager2022} studied a number of impacts made by microcredit, using data from seven separate RCTs analyzed under a hierarchical model fitted with MCMC.
In \citet{Broderick2023}, the authors fit this hierarchical model using variational inference \citep{Blei2017} and investigate the non-robustness of the conclusions based on that fit.
Here, we focus on only a component of the hierarchical model.
We fit this component, which is still a hierarchical model in itself, using MCMC, and examine the fit's non-robustness.

Our approximation does not work as well as it did for the linear model.
For the particular MCMC run, our confidence interval does not contain the refit after removing proposed data. 
The confidence interval for AMIP undercovers: the relative error between estimated coverage and nominal coverage is at most $9.1\%$.
The confidence interval for the sum-of-influence also undercovers: at worst, the relative error is $14.7\%$.
The linear approximation is adequate for the posterior mean even after removing $\defaultdropoutaspercentage$.
For the credible endpoints, the approximation is good until removing roughly $1.8\%$ of the data, and breaks down after that.

As articulated in \cref{sec:setup}, a priori, we think that $\dropout > 1\%$ is a \emph{large} data fraction to remove, and we are not worried about the Maximum Influence Perturbation for such $\dropout$.
So, that the linear approximation stops working after $1.8\%$ is not a cause for concern.
It is more pressing to improve the confidence intervals.
It is likely that a problem-dependent block length $\blocklength$ will outperform the default $\blocklength = \defaultblocklength$.

\subsubsection{Background and full-data fit}
To study the relationship between microcredit and profit, \citet{Meager2022} combines the data from \citet{Angelucci2015} with that from \citet{Attanasio2015,Augsburg2015,Banerjee2015,Crepon2015,Karlan2011,Tarozzi2015}.
In the aggregated data, each observation is a household, with $\obsn = (\covarn, \responsen, \groupn)$ where $\covarn$ is the treatment status, $\responsen$ is the profit measured, and $\groupn$ indicates the household's country.
\citet{Meager2022} uses a tailored hierarchical model that simultaneously estimates a number of effects.
This model separates the dataset into three parts: households with negative profit, households with zero profit, and households with positive profit.
Microcredit is modeled to have an impact on the proportion of data assigned to each part: for households with non-zero profit, microcredit is modeled to have an impact on the location and spread of the log of absolute profit.

For our experiment, we will not look at all the impacts estimated by \citet{Meager2022}'s model.
We focus only on how microcredit impacts the households with negative realizations of profit.
\citet{Meager2022}'s model is such that to study this impact, it suffices to a) filter out observations with non-negative profit from the aggregated data and b) use only a model component rather than the entire model.

The dataset on households with negative profits has $3{,}493$ observations.
The relevant model component from \citet{Meager2022} is the following.
They model all households in a given country as exchangeable, and ``share strength'' across countries.
The absolute value of the profit is modeled as coming from a log-normal distribution.
If the household is in country $k$, this distribution has mean 
$
  \lControlLocation_{k} + \lTreatmentLocation_{k} \covarn,
$
and variance  
$
  \exp \left( \lControlScale_{k}  + \lTreatmentScale_{k} \covarn \right)
$,
where $(\lControlLocation_{k}, \lTreatmentLocation_{k}, \lControlScale_{k}, \lTreatmentScale_{k})$ are latent parameters to be learned.
In other words, the access to microcredit has country-specific impacts on the location and scale of the log of absolute profit.
To borrow strength, the above country-specific parameters are modeled as coming from a common distribution.
For instance, there exists a \emph{global} parameter, $\gTreatmentLocation$, such that the $\lTreatmentLocation_{k}$'s are a priori independent Gaussian centered at $\gTreatmentLocation$. 
For complete specification of the model i.e.\ the list of all global parameters and the prior choice, see \cref{subsec:hierarchical_model_microcredit_details}.

\begin{figure}[t]
    \centering
    \includegraphics{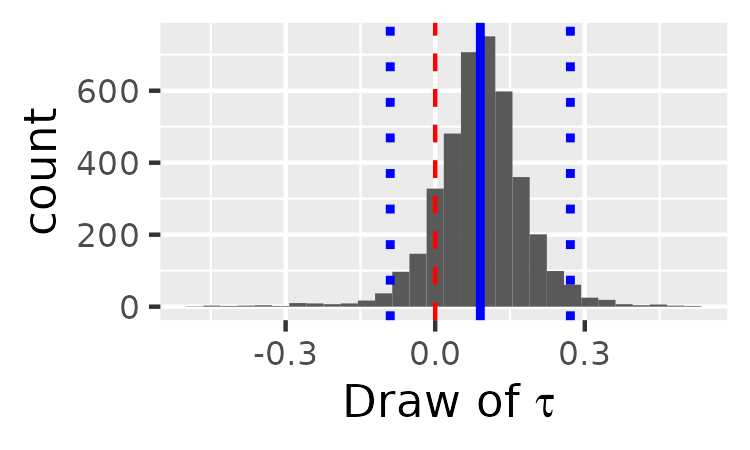}
    \caption{(Hierarchical model for microcredit) Histogram of treatment effect MCMC draws. 
    See the caption of \cref{fig:mexico_hist} for the meaning of the distinguished vertical lines.}
    \label{fig:negative_tail_hist}
  \end{figure}

Roughly speaking, $\gTreatmentLocation$ is an average \emph{treatment effect} across countries.
We use $\samplesize = \defaultsamplesize$ HMC draws to approximate the posterior.
\Cref{fig:negative_tail_hist} plots the histogram of the treatment effect draws and sample summaries.
The sample mean is equal to $0.09$.
The sample standard deviation is $0.09$.
These values are in agreement with the mean and standard deviation estimates obtained from fitting on the original model and data \citep{Meager2022}.
Our estimate of the approximate credible interval's left endpoint is $-0.09$; our estimate of the right endpoint is $0.27$.

Using the summaries in \cref{fig:negative_tail_hist}, an analyst might come to a
decision based on either 
(1) the observation that the posterior mean is positive, or 
(2) the observation that the uncertainty interval covers zero and therefore they cannot be confident of the sign of the unknown parameter.

\subsubsection{Sensitivity results}

Running our approximation takes very little time compared to running the original analysis.
Generating the draws in \cref{fig:negative_tail_hist} took 8 minutes.
For one $\dropout$ and one quantity of interest, it took less than $15$ seconds to make a confidence interval for what happens if we remove the most extreme data subset.
A user might check approximation quality by dropping a proposed subset and re-running MCMC; each such check took us around 8 minutes, the runtime of the original analysis.

\begin{figure}[t]
  \centering 
  \includegraphics{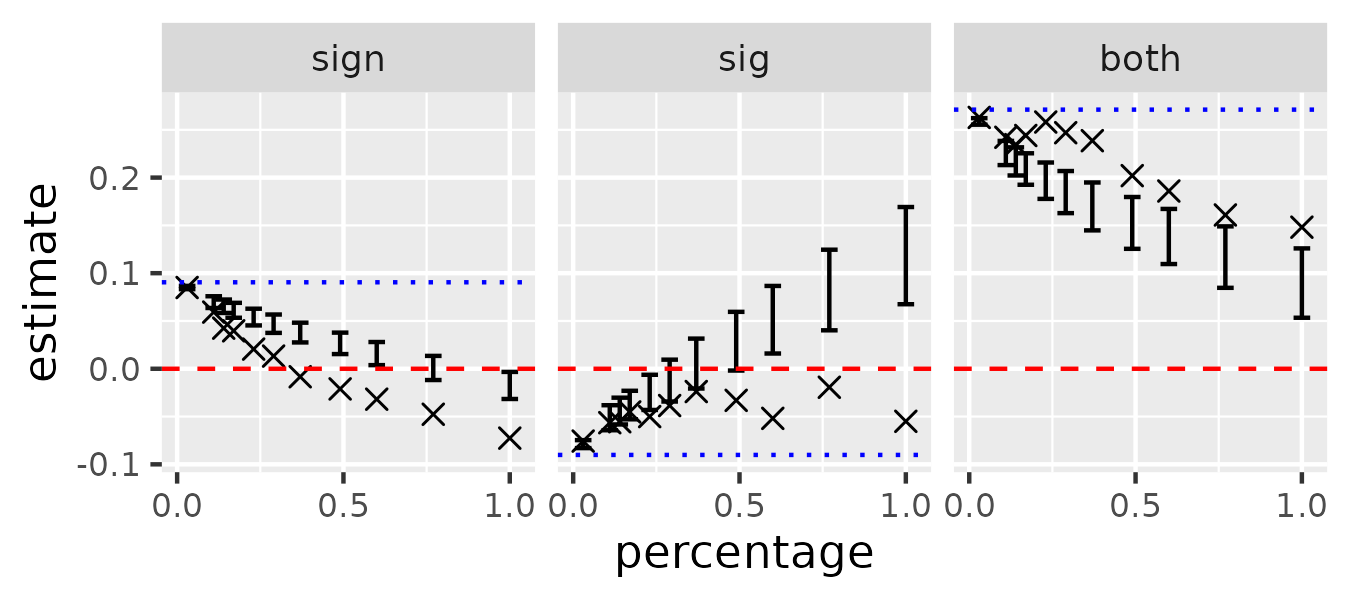}
  \caption{(Hierarchical model for microcredit) Confidence interval and refit. See the caption of \cref{fig:Mexico_refit} for meaning of annotated lines.}
  \label{fig:negative_tail_refit}
\end{figure}

\Cref{fig:negative_tail_refit} plots our confidence intervals and the result after removing the proposed data.
In general, our confidence interval predicts a more extreme change than the actual refit achieves.
The interval is therefore not conservative: if it predicts that a change is achievable, we cannot always trust that such a change is possible. 
The refit is not a monotone function of the proposed data's size in the case of `both' and `sig'.
The non-monotonicity indicates that not all observations in the proposed data induce the correct direction of change (upon their removal).
For instance, in the case of `sig', we aim to increase the credible left endpoint, but actually, the endpoint decreases between $\dropout = 0.46\%$ and $\dropout = 0.60\%$.
Since the proposed data is $\mcmcamis$ from \cref{alg:sosie}, it is apparent that the proposed data for $\dropout = 0.46\%$ is nested in the proposed data for $\dropout = 0.60\%$.
This means that some observations in the difference between these subsets actually decrease the left endpoint upon removal, rather than increase it.

Our method is not able to predict whether the posterior mean can change sign for these $\dropout$ values and this number of samples; 
likewise, our method is not able to predict whether the result can be changed to a significant effect of the opposite sign.
In either case, we recommend a larger number of MCMC samples. 
For changing significance, our method predicts there exists a data subset of relative size at most $0.60\%$ such that if we remove it, we change the sign of the approximate credible interval's left endpoint.
However, refitting does not confirm this prediction.

\subsubsection{Additional quality checks} 

\begin{figure}[t]
  \centering 
  \includegraphics{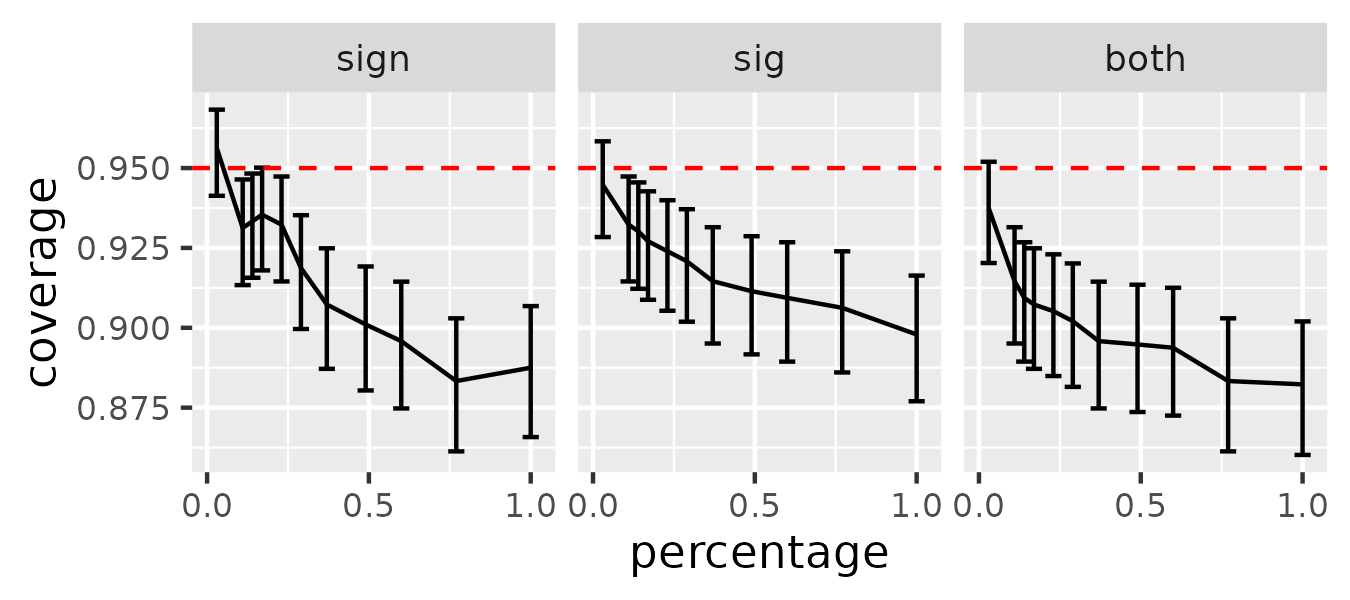}
  \caption{(Hierarchical model for microcredit) Monte Carlo estimate of AMIP confidence interval's coverage. See the caption of \cref{fig:Mexico_amip_coverage} for the meaning of the error bars and the distinguished lines.}
  \label{fig:negative_tail_amip_coverage}
\end{figure}

\Cref{fig:negative_tail_amip_coverage} shows that the confidence interval for the $\amip$ undercovers, but the degree of undercoverage is arguably mild. 
Our confidence interval for the true coverage does not contain the nominal $\nominal$ except for the smallest $\dropout$. 
As $\dropout$ increases, our point estimate of the coverage generally decreases: for the largest $\dropout$, the difference between our point estimate and the nominal $\nominal$ is $0.08$, which translates to a relative error of $8.4\%$.
If we compare $\nominal$ with the lower endpoint of our confidence interval for the true coverage, the worst relative error is $9.1\%$. 

\Cref{fig:negative_tail_amip_coverage} shows that the confidence interval for sum-of-influence has the right coverage for sign change, but undercovers for significance change and generating a significant result of the opposite sign.
At worst, in the case of `sig', the relative error between the nominal $\nominal$ and our estimate of true coverage is $14.7\%$.

Intuitively, the block bootstrap underestimates uncertainty if the block length is not large enough to overcome the time series dependence in the MCMC samples.
The miscoverage suggests that the default block length, $\blocklength = \defaultblocklength$, is too small for this problem.
One potential reason for the difference in coverage between `sign' and `sig' is that, the estimate of influence for `sign' involves a smaller number of objects than that for `sig'.
While an estimate of influence for `sign' involves $\func{\param}$ and $\loglik{n}{\param}$, an estimate of influence for `sig' involves $\func{\param}$, $\loglik{n}{\param}$, and $\func{\param}^2$.
It is possible that the default block length is enough to capture time series dependence for $\func{\param}$ and $\loglik{n}{\param}$, but is inadequate for $\func{\param}^2$.

\begin{figure}[t]
  \centering 
  \includegraphics{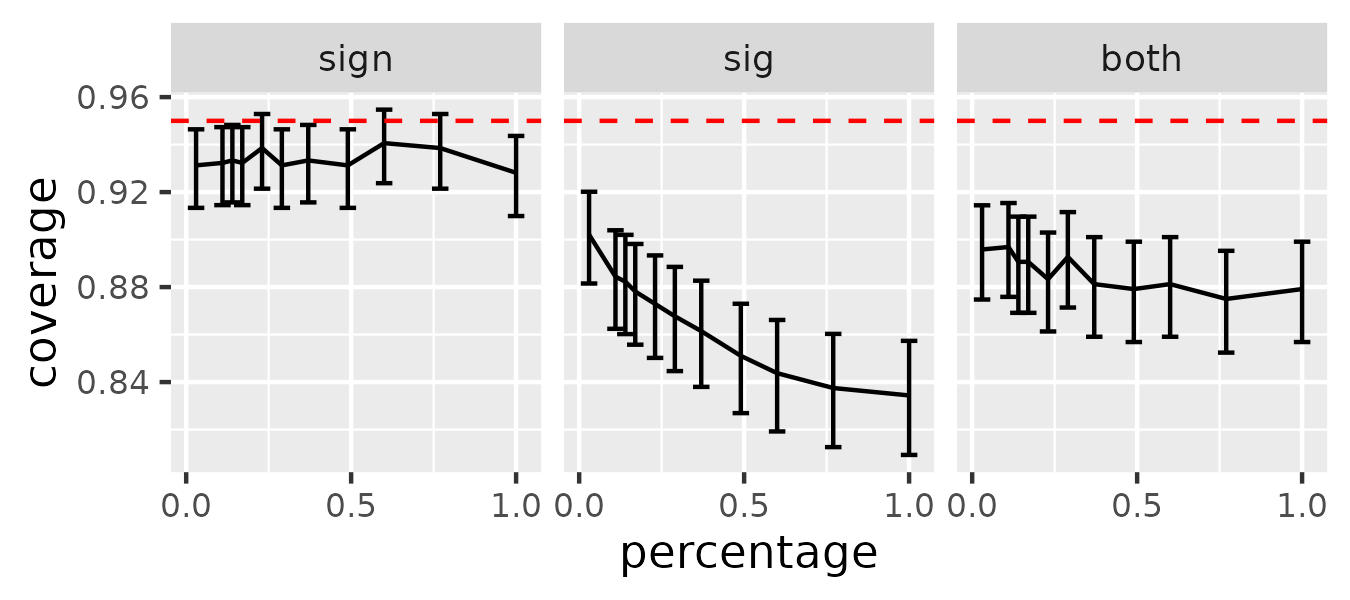}
  \caption{(Hierarchical model for microcredit) Monte Carlo estimate of sum-of-influence confidence interval's coverage. See the caption of \cref{fig:Mexico_soi_coverage} for the meaning of the panels and the distinguished lines.}
  \label{fig:negative_tail_soi_coverage}
\end{figure}

\begin{figure}[t]
  \centering 
  \includegraphics{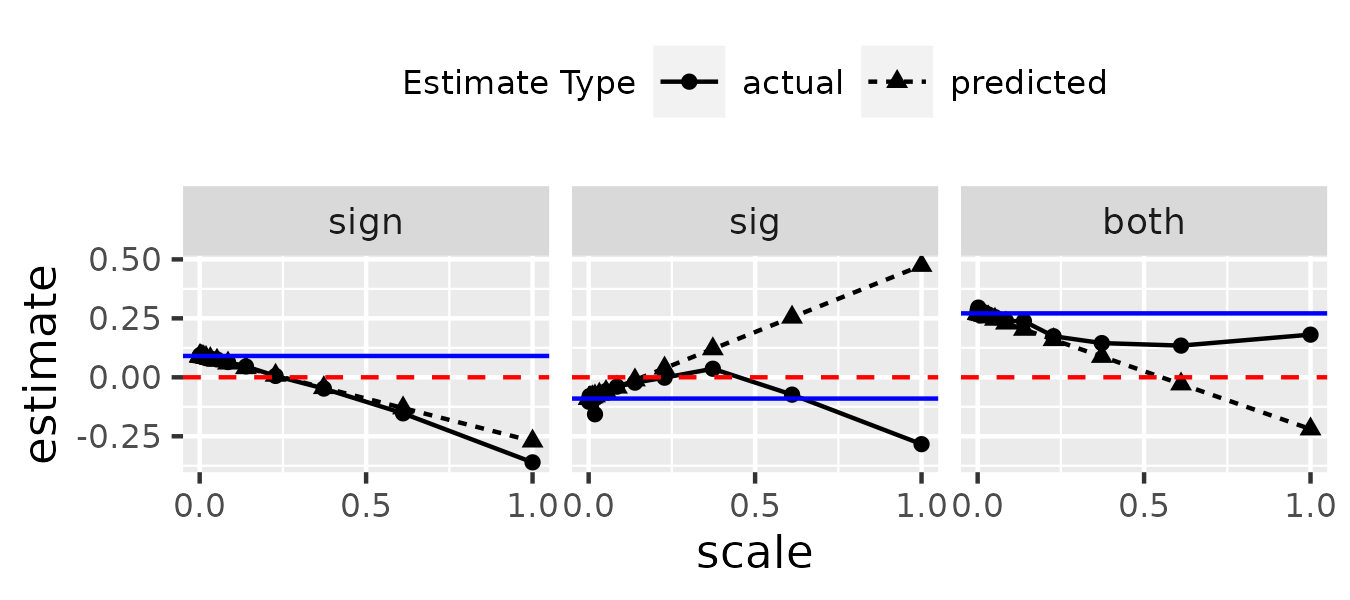}
  \caption{(Hierarchical model for microcredit) Quality of linear approximation. See the caption for \cref{fig:Mexico_linear_approx} for the meaning of the panels and the distinguished lines.}
  \label{fig:negative_tail_linear_approx}
\end{figure}

\Cref{fig:negative_tail_linear_approx} provides evidence that the linear approximation is adequate for $\zeta$ less than $0.3728$ for `both' QoI and `sig', but is grossly wrong for larger $\zeta$. 
Using the rough correspondence between $\zeta$ and amount of data dropped, we say that the linear approximation is adequate until dropping $1.8\%$ of the data.
For `both' QoI, the refit plateaus after dropping $1.8\%$, while the linear approximation continues to decrease.
For `sig', the refit decreases after dropping $1.8\%$, while the linear approximation continues to increase.
The approximation is good for `sign' even after removing $\defaultdropoutaspercentage$ of the data:
the refit and the prediction lie on top of each other for `sign'.

\subsection{Hierarchical model on tree mortality data} \label{sec:tree_mortality}

In the final experiment, we break from microcredit and look at ecological data.
In particular, we consider a slight tweak of the analysis of European tree mortality from \citet{Senf2020}.
\citeauthor{Senf2020} are acutely aware of generalization concerns.
While previous work on tree death had been limited in both time and space, \citet{Senf2020} designs a large study that stretches across Europe and over 30 years, in hopes of making a broad-scale assessment.
Our work shows that, even after an expansive study with generalization in mind, one might still worry about applying the findings at large, because of small-data sensitivity.

Our approximation also struggles in this case.
For the particular MCMC run used to estimate the full-data posterior, our confidence interval does not contain the refit after removing the proposed data.
As each MCMC run is already highly time-consuming, we do not run quality checks on the whole dataset.
We settle for running quality checks on a subsample of the data.
On the subsampled data, the confidence interval for AMIP undercovers: the undercoverage is severe for one of the quantities of interest. 
However, the confidence interval for sum-of-influence is close to achieving the nominal coverage.
For all three quantities of interest, the linear approximation is good up to removing roughly $1.1\%$ of the data.
For two of the three, it breaks down afterwards; for the remaining one, it continues to be good up to $3\%$, then falters.

As articulated in \cref{sec:setup}, we think that dropping more than $1\%$ of the data is already removing a large fraction.
We are not worried about the Maximum Influence Perturbation for such $\dropout$.
So, that the linear approximation stops working after $1.1\%$ is not a cause for concern.

\subsubsection{Background and full-data fit}
\citet{Senf2020} studies the relationship between drought and tree death in Europe.
To identify the association, they have compiled a dataset with $\nobs = 87{,}390$ observations.
Europe is divided into $2{,}913$ regions, and the data spans 30 years. Each observation is a set of measurements made in a particular region, which we denote as $\locationVarn$, and at a particular year, which we denote as $\timeVarn$.
For our purposes, it suffices to know that the measurement of (the opposite of) drought is called climatic water balance, and we denote it as $\covarn$. Larger values of $\covarn$ indicate that more water is available; i.e.\ there is less drought.
The response of interest, $\responsen$, is excess death of tree canopy. 

In our experiment, we mostly replicate \citep{Senf2020}'s probabilistic model: we use the same likelihood and make only an immaterial modification in the choice of priors.
For the likelihood, \citep{Senf2020} models each $\responsen$ as a realization from an exponentially modified Gaussian distribution.
Recall that such a distribution has three parameters, ($\mu, \sigma, \lambda$), and a random variate can be expressed as the sum between a normal variate $N(\mu, \sigma^2)$ and an exponential variate with rate $\lambda$.
When modelling $\{\responsen\}_{n=1}^{\nobs}$, the model uses the same $\sigma$ and $\lambda$ for all observations.
However, the mean $\mu$ is a function of $n$.
It is the sum of three components.
The first is an affine function of $\covarn$; i.e.\ $\intercept + \slope \covarn$ for some latent parameters $\intercept$ and $\slope$.
The second is a smoothing spline of $\covarn$; it is included to capture non-linear relationships, but we do not go into details here.
The third contains the random effects for the region $\locationVarn$ and the time $\timeVarn$; if the observation is located at $l$ and took place during $t$, this term is $(\tIntercept_t + \lIntercept_l) + (\tSlope_t + \lSlope_l) \covarn$.

At a high level, both \citet{Senf2020}'s prior and our prior share strength across regions and times by modeling the random effects as coming from a some common \emph{global} distributions. 
However, while \citet{Senf2020} uses an improper prior, we use a proper one.
Numerically, there is no perceptible difference between the two.
Theoretically, we prefer working with proper priors to avoid the integrability issue mentioned around \cref{assume:feasible-has-finite-normalization}. 
For complete specification of our model, see \cref{subsec:hierarchical_model_ecological_details}.

\begin{figure}[t]
  \centering
  \includegraphics{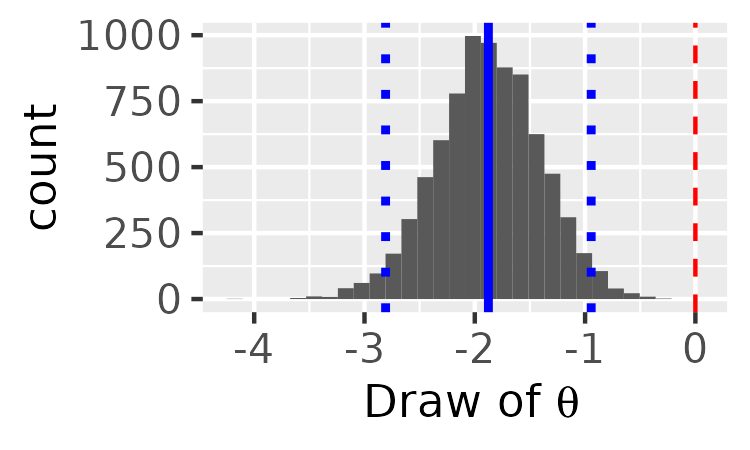}
  \caption{(Hierarchical model for tree mortality) Histogram of slope MCMC draws. 
  See the caption of \cref{fig:mexico_hist} for the meaning of the distinguished vertical lines.}
  \label{fig:tree_mortality_hist}
\end{figure}

Following \citet{Senf2020}, we make conclusions based on posterior functionals of $\slope$.
Roughly speaking, $\slope$ is the average (across time and space) \emph{association effect} that water balance has on excess tree death.
We use $\samplesize = 8000$ HMC draws to approximate the posterior.
\Cref{fig:tree_mortality_hist} plots the histogram of the association effect draws and sample summaries.
The sample mean is equal to $-1.88$.
The sample standard deviation is $0.48$.
These estimates are very close to those reported in \citet[Table 1]{Senf2020}.
Our estimate of the approximate credible interval's left endpoint is $-2.81$; our estimate of the right endpoint is $-0.94$.

In our parametrization, if $\slope$ were estimated to be negative, it would indicate that the availability of water is negatively associated with tree death.
In other words, drought is positively associated with tree death. 
Based on the sample summaries, a forest ecologist might decide that drought has a positive relationship with canopy mortality, since the posterior mean is negative, and this relationship is significant, since the approximate credible interval does not contain zero.

\subsubsection{Sensitivity results}

Running our approximation takes very little time compared to running the original analysis.
Generating the draws in \cref{fig:tree_mortality_hist} took 12 hours.
For one $\dropout$ and one quantity of interest, it took less than $2$ minutes to make a confidence interval for what happens if we remove the most extreme data subset.
A user might check approximation quality by dropping a proposed subset and re-running MCMC; each such check took us around $12$ hours, which is the runtime of the original analysis.

\begin{figure}[t]
    \centering 
    \includegraphics{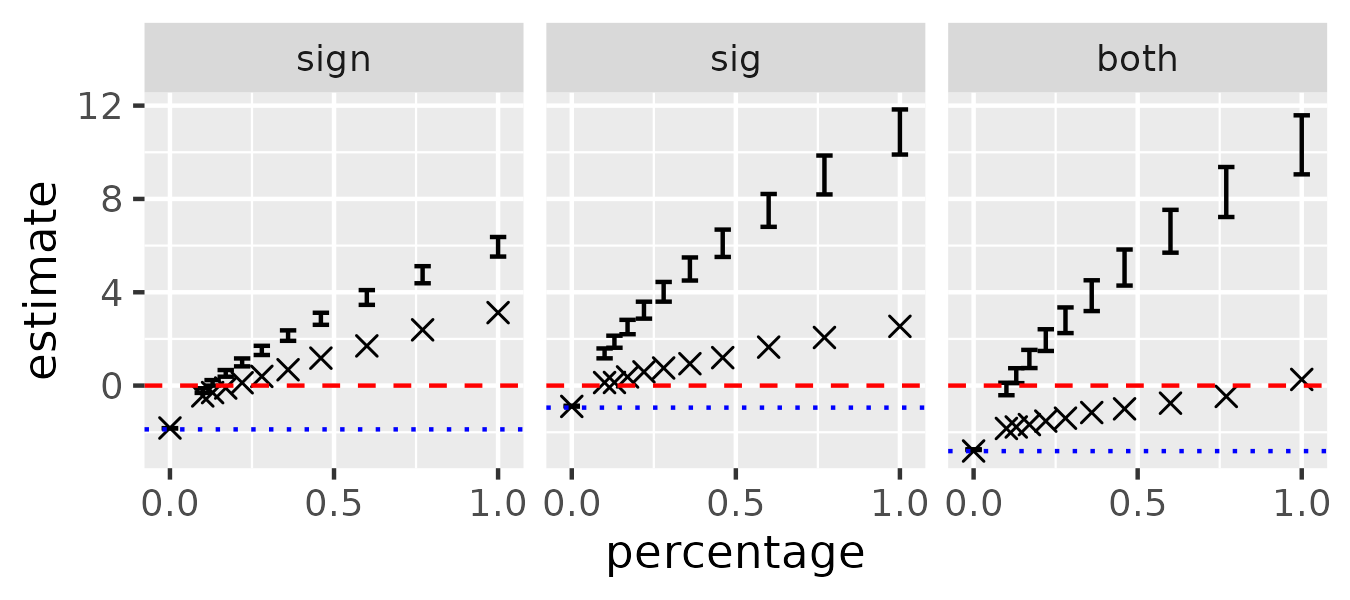}
    \caption{(Hierarchical model for tree mortality) Confidence interval and refit. See the caption of \cref{fig:Mexico_refit} for the meaning of the panels and the distinguished lines.}
    \label{fig:tree_mortality_refit}
\end{figure}

\Cref{fig:tree_mortality_refit} plots our confidence intervals and the result after removing the proposed data.
In general, our confidence interval predicts a more extreme change than realized by the refit: hence, our interval is not conservative. 
The overestimation is particularly severe for the `both' QoI and the `sig' QoI.
For changing sign, our method predicts there exists a data subset of relative size at most $0.17\%$ such that if we remove it, we change the posterior mean's sign; refitting does not confirm this prediction, however. 
The smallest $\dropout$ whose refit's posterior mean actually changes sign is $0.22\%$.
For changing significance, our method predicts there exists a data subset of relative size at most $0.10\%$ such that if we remove it, we change the sign of the right endpoint; refitting confirms this prediction.\footnote{The reason behind our correct prediction is likely the spacing between considered $\dropout$.
We expect the refit to be a continuous function of $\dropout$. Based on the scatter plot, it is likely that the refit's right endpoint changes sign at a $\dropout$ between $0.01\%$ and $0.10\%$. However, we do not evaluate the refit at any $\dropout$ in this interval.}
For generating a significant result of the opposite sign, our method predicts there exists a data subset of relative size at most $0.17\%$ such that if we remove it, we change the sign of the left endpoint; refitting does not confirm this prediction, however. 
The smallest $\dropout$ whose refit's left endpoint actually changes sign is $1.0\%$.

\subsubsection{Results on subsampled data} 

Running MCMC on the original dataset of size over $80{,}000$ took 12 hours.
In theory, we can spend time (on the order of thousands of hours) to run our quality checks, but we do not do so. 
Instead, we subsample $2{,}000$ observations at random from the original dataset.
Each MCMC on this subsample takes only $15$ minutes, making it possible to run quality checks in a few hours instead of weeks.
We hope that the subsampled data is representative enough of the original data that the quality checks on the subsampled data are indicative of the quality checks on the original data.

\begin{figure}[t]
    \centering 
    \includegraphics{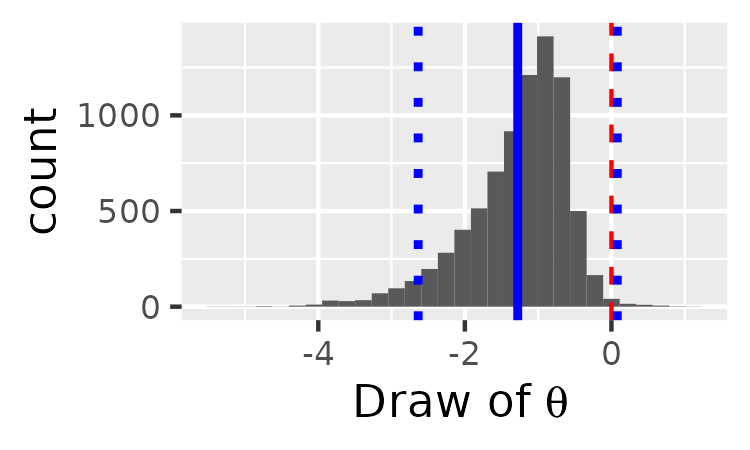}
    \caption{(Hierarchical model on subsampled tree mortality) Histogram of effect MCMC draws. See \cref{fig:mexico_hist} for the meaning of the distinguished lines.}
    \label{fig:tree_mortality_Subsample_hist}
\end{figure}

We use the same probabilistic model to analyze the subsampled data.
\Cref{fig:tree_mortality_Subsample_hist} plots the histogram of the association effect draws and sample summaries.
Based on the draws, a forest ecologist might tentatively say that drought is positively associated with canopy mortality if they relied on the posterior mean, but refrain from conclusively deciding, since the approximate credible interval contains zero.

\begin{figure}[t]
    \centering 
    \includegraphics{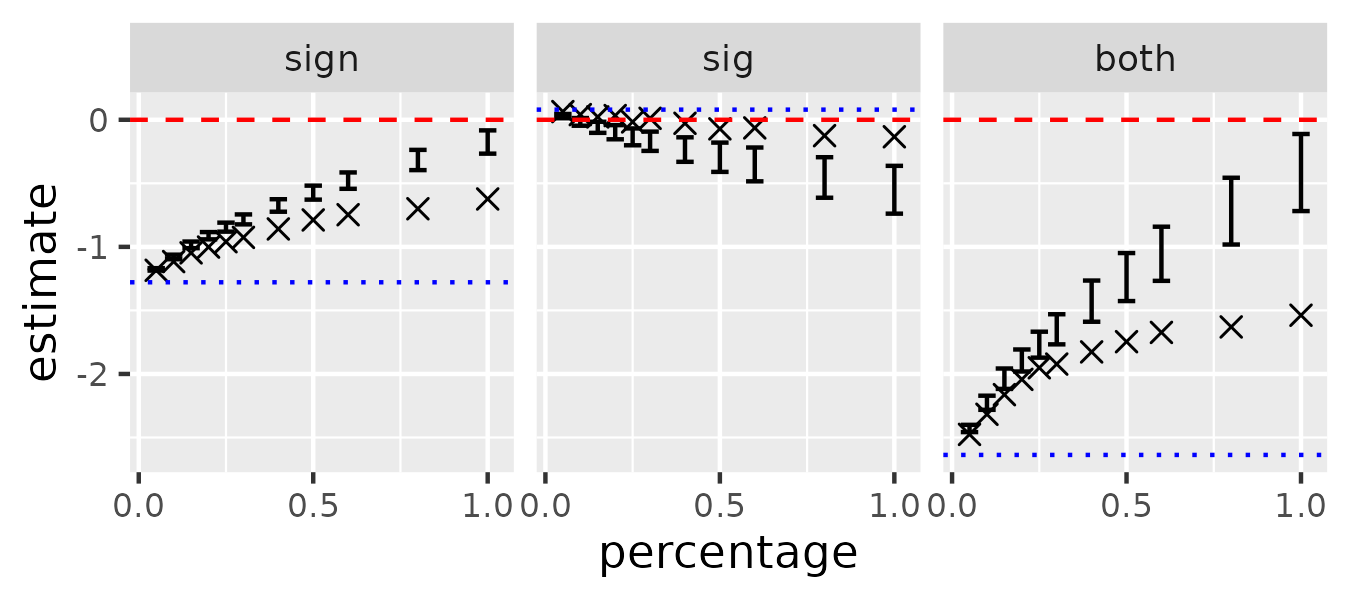}
    \caption{(Hierarchical model on subsampled tree mortality) Confidence interval and refit. See the caption of \cref{fig:Mexico_refit} for the meaning of the panels and the distinguished lines.}
    \label{fig:tree_mortality_Subsample_refit}
\end{figure}

\Cref{fig:tree_mortality_Subsample_refit} shows our confidence intervals and the actual refits.
Similar to \cref{fig:tree_mortality_refit}, our confidence intervals predict a more extreme change than realized by the refit.
The overestimation is most severe for `both' QoI.

\begin{figure}[t]
    \centering
    \includegraphics{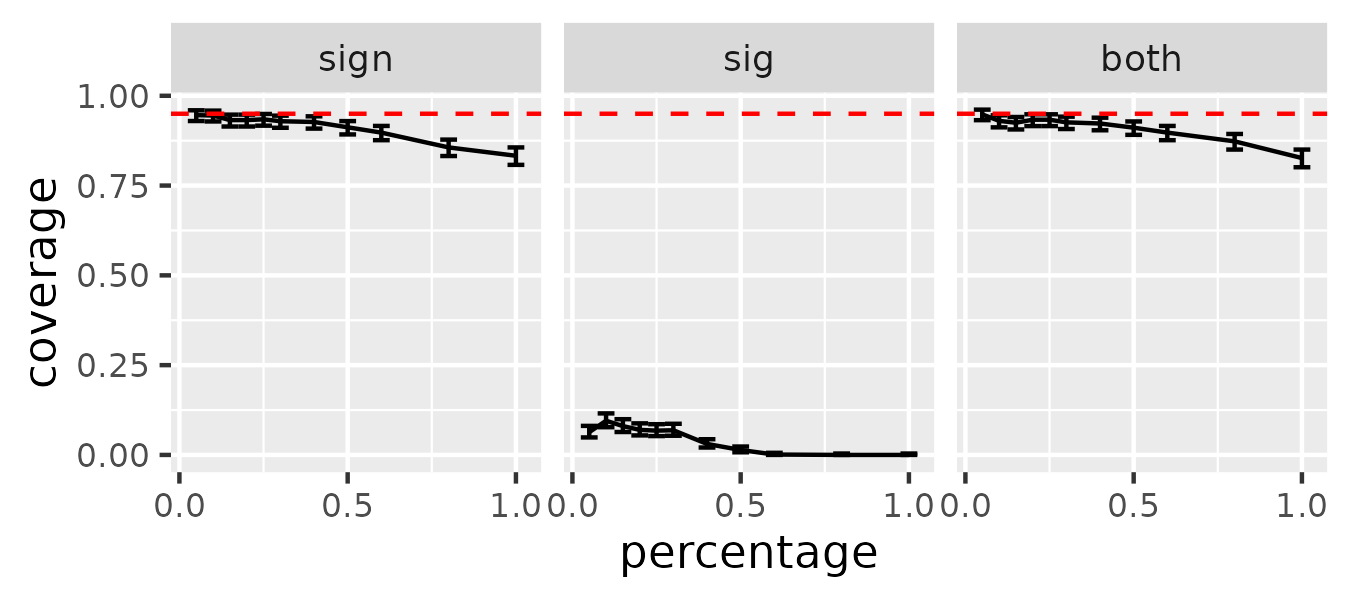}
    \caption{(Hierarchical model on subsampled tree mortality) Monte Carlo estimate of coverage of confidence interval for $\amip$. See \cref{fig:Mexico_amip_coverage} for the meaning of the panels and the distinguished lines.}
    \label{fig:tree_mortality_Subsample_amip_coverage}
\end{figure}

In \cref{fig:tree_mortality_Subsample_amip_coverage}, the confidence interval for AMIP undercovers for all quantities of interest.
The actual coverage decreases as $\dropout$ increases.
The undercoverage is most severe for `sig' QoI: while the nominal level is $\defaultnominal$, the confidence interval for the true coverage only contains values less than $0.15$.
This translates to a relative error of over $84\%$.
In other words, our confidence interval for significance change is too narrow, and rarely contains the AMIP.
For `both' QoI and `sig' QoI, the worst-case relative error between the nominal and the estimated coverage, which occurs under the largest $\dropout$, is $15.7\%$.

\begin{figure}[t]
    \centering
    \includegraphics{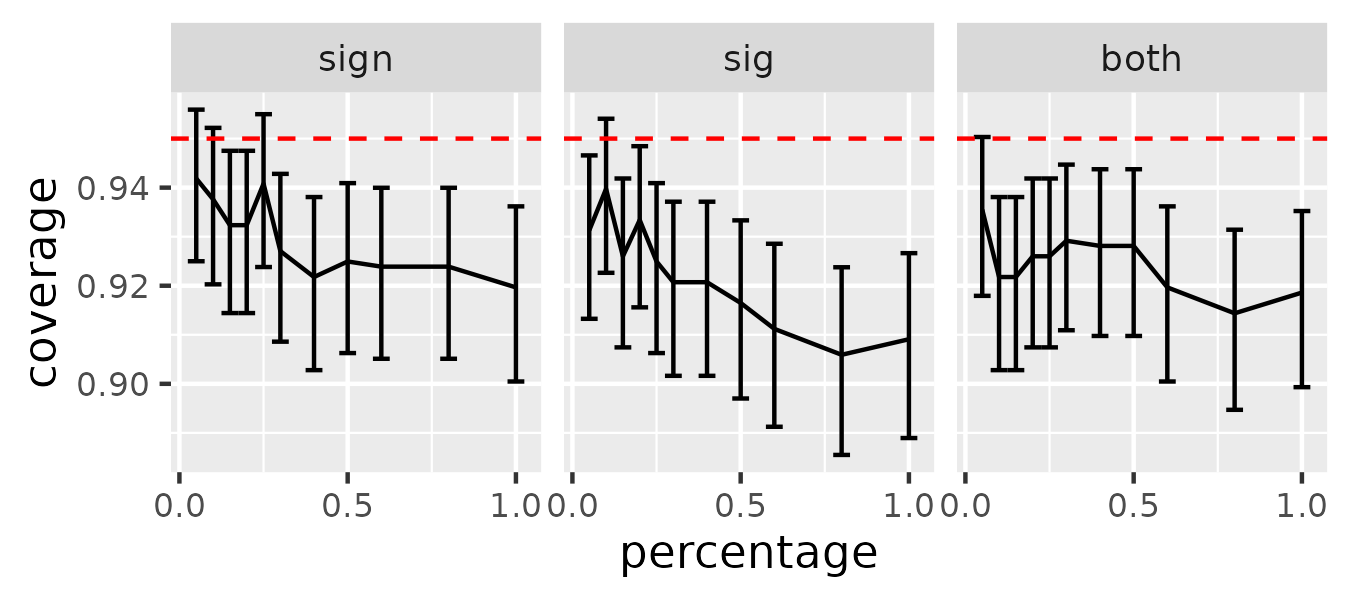}
    \caption{(Hierarchical model on subsampled tree mortality) Monte Carlo estimate of coverage of confidence interval for sum-of-influence. See \cref{fig:Mexico_soi_coverage} for the meaning of the panels and the distinguished lines.}
    \label{fig:tree_mortality_Subsample_soi_coverage}
\end{figure}

In \cref{fig:tree_mortality_Subsample_soi_coverage}, the estimated coverage of the confidence interval for sum-of-influence is close to the nominal coverage.
Note the stark contrast in the vertical scale of the `sig' panel in \cref{fig:tree_mortality_Subsample_amip_coverage} with that in \cref{fig:tree_mortality_Subsample_soi_coverage}.
At worst, our point estimate of the true coverage is $0.04$ less than the nominal level, which is only a $4.2\%$ relative error.
The success of the block bootstrap for the sum-of-influence (\cref{fig:tree_mortality_Subsample_soi_coverage}) indicates that the undercoverage observed in \cref{fig:tree_mortality_Subsample_amip_coverage} can be attributed to the sorting step involved in the construction of $\mcmcamip$.
We leave to future work to investigate why the interference caused by the sorting step is so much more severe for changing the significance than for changing sign or generating significant result of the opposite sign.

\begin{figure}[t]
    \centering
    \includegraphics{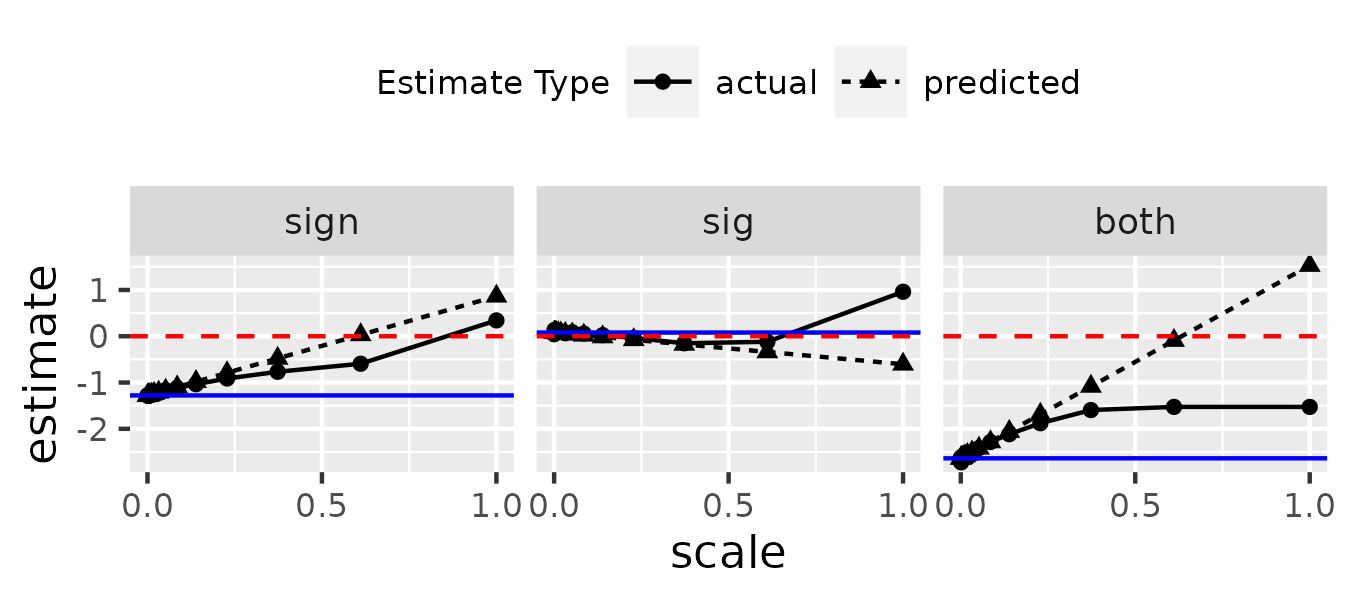}
    \caption{(Hierarchical model on subsampled tree mortality) Quality of linear approximation. See \cref{fig:Mexico_linear_approx} for the meaning of the panels and the distinguished lines.}
    \label{fig:tree_mortality_Subsample_linear_approx}
\end{figure}

\Cref{fig:tree_mortality_Subsample_linear_approx} shows that the linear approximation is good for the posterior mean (`sign' QoI) and the left credible endpoint (`both' QoI) up to $\zeta = 0.2276$; in data percentages, this is roughly $1.1\%$.
For larger $\zeta$, the refit for `both' QoI plateaus while the linear approximation continues to increase, and the linear approximation for the posterior mean slightly underestimates it.
For the left endpoint (`both' QoI), the linear approximation is close to the refit up to $\zeta = 0.6105$ (roughly $3\%$ of data); afterwards, the left endpoint increases while the linear approximation continues to decrease.

\section{Discussion}
We have provided a fast approximation to what happens to conclusions made with MCMC in Bayesian models when a small percentage of data is removed.
In real data experiments, our approximation is accurate in simple models, such as linear regression.
In complicated models, such as hierarchical ones with many random effects, our methods are less accurate.
A number of open questions remain.
We suspect that choosing the block length more carefully may improve performance; how to pick the block length in a data-driven way is an interesting question for future work.
Currently, we can assess sensitivity for quantities of interest based on posterior expectations and posterior standard deviations.
For analysts that use posterior quantiles to make decisions, we are not able to assess sensitivity.
To extend our work to quantiles, one would need to quantify how much a quantile changes under small perturbations of the total log likelihood.
Finally, we have not fully isolated the source of difficulty in complex models like those in \citet{Senf2020}.
In the analysis of tree mortality data, there are a number of conflating factors.
\begin{itemize}
   \item The model has a large number of parameters.
   \item The parameters are organized hierarchically.
   \item We use MCMC to approximate the posterior.
\end{itemize}
To determine if the difficulty comes from high dimensionality or if the error comes from hierarchical organization, future work might apply our approximation to a high-dimensional model without hierarchical structure.
For instance, one might use MCMC on a linear regression with many parameters and non-conjugate priors.
To check if MCMC is a cause of difficulty, one could experiment with variational inference (VI).
If we chose to approximate the posterior with VI, we can use the machinery developed for estimating equations \citep{Broderick2023} to assess small-data sensitivity.
If the dropping data approximation works well there, we have evidence that MCMC is part of the problem in complex models.

\section{Acknowledgments}

The authors are grateful to Hannah Diehl for useful discussions and comments. This work was supported in part by an ONR Early Career Grant (N000142012532), an NSF CAREER Award (1750286), the MIT-IBM Watson AI Lab, and
a SystemsThatLearn@CSAIL Ignite Grant.

\bibliography{export}
\bibliographystyle{plainnat}

\newpage
\appendix
\section{More Theory} \label{sec:more-theory}

In the normal means model, we show that the first-order error is smaller than the zeroth-order error under certain conditions.

We detail the data, prior, and likelihood for the normal means model.
Then, we specify a quantity of interest.
The $n$-th observation consists of real-valued measurement $\covarn$ and group assignment $\groupn$: $\obsn = (\covarn, \groupn)$.
The parameters of interest are the population mean $\globalmean$ and the group means $\groupmean = (\groupmean_1, \groupmean_2, \ldots, \groupmean_{\numgroup})$.
The measurements belonging to group $g$ are modeled as Gaussian centered at the group mean $\groupmean_g$ with a known standard deviation $\noise$.
In other words, the $n$-th log-likelihood is
$
\loglik{n}{\globalmean,\groupmean} = 
  \frac{1}{2}  \log \left( \frac{1}{2\pi \noise^2} \right) - \frac{1}{2\noise^2}[ (\covarn)^2 - 2 \covarn \groupmean_{\groupn}  + \groupmean_{\groupn}^2].
$
The prior over $(\globalmean, \groupmean)$ is the following.
We choose the uniform distribution over the real line as the prior for $\globalmean$.
Conditioned on $\globalmean$, the group means are Gaussian centered at $\globalmean$, with a known standard deviation $\dispersion$. 
The quantity of interest is the posterior mean of $\globalmean$.

This model, like the normal model, has closed-form posterior expectations. 
Before displaying the exact formula for the error $\firstOrderError$, we need to describe the weighted posterior in more detail. 
For each group $g$, we define three functions of $\weight$:
\begin{equation*}
    \nobsj{g}(\weight) := \sum_{n: \groupn = g } \weight_n , 
    \meanj{g}(\weight) := \frac{ \sum_{n: \groupn = g } \weight_n \covarn }{\nobsj{g}(\weight)}  , 
    \precj{g}(\weight) := \left( \frac{\noise^2}{\nobsj{g}(\weight)} + \dispersion^2 \right)^{-1}.
\end{equation*}
While $\nobsj{g}(\weight)$ sums up the weights of observations in group $g$, $\meanj{g}(\weight)$ is the weighted average of measurements in this group, and $\precj{g}(\weight)$ will be used to weight $\meanj{g}(\weight)$ in forming the posterior mean of $\globalmean$.
The \proofref{lem:normal-means-error} shows that $\Eposterior{\weight}\globalmean$ is equal to
\begin{equation*}
    \frac
    {\sum_{g=1}^{\numgroup} \precj{g}(\weight) \meanj{g}(\weight) }
    {\sum_{g=1}^{\numgroup} \precj{g}(\weight)  }.
\end{equation*}

To avoid writing $\sum_{g=1}^{\numgroup} \precj{g}(\weight)$, we define $\sumprec(\weight) := \sum_{g=1}^{\numgroup} \precj{g}(\weight)$.
To lighten notation, for expectations under the original posterior, we write $\fullmu$ instead of $\Eposterior{\allones}\globalmean$ and $\fullnobsj{g}$  instead of $\nobsj{g}(\allones)$.
An analogous shorthand applies to $\meanj{g}(\allones)$, $\precj{g}(\allones)$,s and $\sumprec(\allones)$.
In words, $\fullmu$ is the posterior mean of $\globalmean$ under the full-data posterior, $\fullnobsj{g}$ is the number of observations in group $g$ of the original dataset, and so on.

The first-order error in the normal means model is given in the following lemma.

\begin{mylemma}\label{lem:normal-means-error} 

    In the normal means model, let the index set $\indexset$ be such that there exists $k \in \{1,2,\ldots,\numgroup\}$ such that a) for all $n \in \indexset$, $\groupn = k$ and b) $|\indexset| < \fullnobsj{k}$.
    In other words, all observations dropped belong to the same group, but we do not drop all observations in this group.
    Define 
    \begin{equation*}
        \begin{aligned}
            F_1(\indexset) &:= \frac{ |\indexset|^2 } {\fullnobsj{k} [\fullnobsj{k}  - |\indexset| ] } 
                                (\fullmeanj{k}  - \meanOfIndexSet ) 
                                , \\
            F_2(\indexset) &:= 
            \frac{|\indexset| } { \fullnobsj{k} } 
            \frac{ \noise^2 \fullprecj{k} } {\fullnobsj{k} }
            (\fullmu - \fullmeanj{k}), \\
            E(\indexset) &:=  \frac{ |\indexset| }{ \fullnobsj{k}  [\fullnobsj{k} - |\indexset|] }  \sigma^2 \precj{k}(\wFromIndex(\indexset)) \fullprecj{k} .
        \end{aligned} 
    \end{equation*}
    Then, $\firstOrderError$ is equal to
    \begin{equation*} 
        \frac{
            \precj{k}(\wFromIndex(\indexset))  
            }
            {
                \fullprec 
            }  
        \left[ F_1(\indexset) + F_2(\indexset) \right] 
        + 
        \frac{
                \left(\sum_{g \neq k} \fullprecj{g}( \fullmeanj{g} - \meanj{k}(\wFromIndex(\indexset)))  \right)
            }
            {
                \fullprec \sumprec(\wFromIndex(\indexset))} 
        E(\indexset). 
    \end{equation*}
\end{mylemma} 

We prove \cref{lem:normal-means-error} in the \proofref{lem:normal-means-error}.
The constraint where all observations in $\indexset$ belong to the same group $k$ is made out of convenience.
We can derive the error without this constraint, but the formula will be much more complicated.
Because $|\indexset| < \fullnobsj{k}$, the denominators of $F_1(\indexset)$, $F_2(\indexset)$ and $E(\indexset)$ are non-zero.
So, the overall error is well-defined.

The zeroth-order error is as follows.

\begin{mylemma}\label{lem:normal-means-zeroth-order-error} 
    In the normal means model, let the index set $\indexset$ be such that there exists $k \in \{1,2,\ldots,\numgroup\}$ such that a) for all $n \in \indexset$, $\groupn = k$ and b) $|\indexset| < \fullnobsj{k}$.
    In other words, all observations dropped belong to the same group, but we do not drop all observations in this group.
    Let $F_1(\indexset)$, $F_2(\indexset)$, and $E(\indexset)$ be defined as in \cref{lem:normal-means-error}.
    Then, $\zerothOrderError$ is equal to 
    \begin{equation*}
        \frac{
            \precj{k}(\wFromIndex(\indexset))  
            }
            {
                \fullprec 
            }  
        \frac{\fullnobsj{k}}{|\indexset|}  F_1(\indexset) 
        + 
        \frac{
                \left(\sum_{g \neq k} \fullprecj{g}( \fullmeanj{g} - \meanj{k}(\wFromIndex(\indexset)))  \right)
            }
            {
                \fullprec \sumprec(\wFromIndex(\indexset))} 
        E(\indexset). 
    \end{equation*}
\end{mylemma}

We prove \cref{lem:normal-means-zeroth-order-error} in the \proofref{lem:normal-means-zeroth-order-error}.
Comparing the expression in \cref{lem:normal-means-zeroth-order-error} with the expression in \cref{lem:normal-means-error}, the (signed) difference between the zeroth-order error and the first-order error is
\begin{equation*}
    \frac{
        \precj{k}(\wFromIndex(\indexset))  
        }
        {
            \fullprec 
        }  \left[ \left(\frac{\fullnobsj{k}}{|\indexset|}  - 1  \right) F_1(\indexset) - F_2(\indexset) \right].
\end{equation*}
So, the first-order error is smaller than the zeroth-order error if and only if $\left(\frac{\fullnobsj{k}}{|\indexset|}  - 1  \right) F_1(\indexset) - F_2(\indexset)$ is positive.
This condition is equivalent to
\begin{equation} \label{eq:condition-normal-means}
    \fullmeanj{k} - \meanOfIndexSet > \noise^2 \frac{\fullprecj{k}}{\fullnobsj{k}} (\fullmu - \fullmeanj{k}).
\end{equation}
The left hand side of \cref{eq:condition-normal-means} is the difference between the sample mean of group $k$ and the sample mean of the measurements in $\indexset$.
The right hand side is a rescaled version of the difference between the global posterior mean and the sample mean of group $k$.

\section{Proofs}

\subsection{Taylor series proofs}

\begin{delayedproof}{thm:linear-map-exists}

At a high level, we rely on \citet[Chapter 5.12, Theorem 5.9]{Fleming1977} to  interchange integration and differentiation. 

Although the theorem statement does not explicitly mention the normalizer, to show that the quantity of interest is continuously differentiable and compute its partial derivatives, it is necessary to show that the normalizer is continuously differentiable and compute its partial derivatives.
To do so, we verify the following conditions on the integrand defining $\normalizer$:
\begin{enumerate}
    \item For any $\param$, the mapping 
        $
            \weight \mapsto \weightedlogprob
        $ 
        is continously differentiable.

    \item There exists a Lebesgue integrable function $\bound{1}$ such that for all $\weight \in \diffdomain$, 
        $
            \weightedlogprob \leq \bound{1}(\param).
        $

    \item For each $n$, there exists a Lebesgue integrable function $\bound{2}^{(n)}$ such that for all $\weight \in \diffdomain$,
        $
            \left| \frac{\partial}{\partial \weight_n} \weightedlogprob \right| \leq \bound{2}^{(n)}(\param).
        $
\end{enumerate}
The first condition is satisfied since the exponential function is continuously differentiable.
To construct $\bound{1}$ that satisfies the second condition, we partition the parameter space $\Rd{\paramdim}$ into a finite number of disjoint sets.
To index these sets, we use a subset of $\{1,2,\ldots,\nobs\}$.
If the indexing subset were $I = \{n_1,n_2,\ldots,n_M\}$, the corresponding element of the partition is 
\begin{equation} \label{eq:partition}
    B_{I} := \{ \param \in \Rd{\paramdim}: \forall n \in I, \loglik{n}{\param} \geq 0 \}.
\end{equation}
This partition allows us to upper bound the integrand with a function that is independent of $\weight$.
Suppose $\param \in B_{I}$ with $I \neq \emptyset$.
The maximum $\sum_{n=1}^{\nobs} \weight_n \loglik{n}{\param}$ is attained by setting $\weight_n = 1$ for all $n \in I$ and $\weight_n = 0$ for all $n \notin I$.
Suppose $\param \in B_{\emptyset}$.
As $\loglik{n}{\param} < 0$ for all $1 \leq n \leq \nobs$, and we are constrained by $\max_{n} \weight_n \geq \lowerbound$, the maximum of $\sum_{n=1}^{\nobs} \weight_n \loglik{n}{\param}$ is attained by setting $\weight_n = \lowerbound$ for $\arg\max_{n}\loglik{n}{\param}$ and $\weight_n = 0$ for all other $n$.
In short, our envelope function is
\begin{equation*}
    \bound{1}(\param) := 
        \begin{cases} 
            \prior(\param) \prod_{n \in I} \exp(\loglik{n}{\param}) \hspace{4pt} &\text{ if }  \param \in B_{I}, I \neq \emptyset. \\
            \prior(\param) \left( \max_{n=1}^{\nobs} \exp(\lowerbound \loglik{n}{\param})\right)  \hspace{4pt} &\text{ if }  \param \in B_{\emptyset}. \\
        \end{cases}
\end{equation*}
The last step is to show $\bound{1}$ is integrable.
It suffices to show that the integral of $\bound{1}$ on each $B_{I}$ is finite.
By \cref{assume:feasible-has-finite-normalization}, for any $n$, the integral of $\prior(\param) \exp(\lowerbound \loglik{n}{\param})$ over $\Rd{\paramdim}$ is finite, and $B_{\emptyset}$ is a subset of $\Rd{\paramdim}$.
As $\bound{1}(\param)$ is the maximum of a finite number of integrable functions, it is integrable.
Similarly, the integral of $\bound{1}$ over $B_{I}$ where $I \neq \emptyset$ is at most the integral of $\prior(\param) \prod_{n \in I} \exp(\loglik{n}{\param})$ over $\Rd{\paramdim}$, which is finite by \cref{assume:feasible-has-finite-normalization}.
To construct $\bound{2}^{(n)}$ that satisfies the third condition, we use the same partition of $\Rd{\paramdim}$, and the envelope function is 
$
    \bound{2}^{(n)}(\param) := \loglik{n}{\param} \bound{1}(\param),
$
since the partial derivative of the weighted log probability is clearly the product of the $n$-th log likelihood and the weighted log probability.
The integrability of $\bound{2}^{(n)}$ follows from \cref{assume:finiteness}'s guarantee that the expectation of $| \loglik{n}{\param} |$ is finite under different weighted posteriors.  
In all, we can interchange integration with differentiation, and the partial derivatives are 
\begin{equation*}
    \frac{\partial Z(w)}{\partial \weight_n} = \normalizer \times \Eposterior{\weight} \left[ \loglik{n}{\param} \right].
\end{equation*}

We move on to prove that $\Eposterior{\weight}\func{\param}$ is continuously differentiable and find its partial derivatives.
The conditions on $\func{\param} \frac{1}{\normalizer} \weightedlogprob$ that we wish to check are:
\begin{enumerate}
    \item For any $\param$, the mapping 
        $
            \weight \mapsto \func{\param} \frac{1}{\normalizer} \weightedlogprob
        $ 
        is continously differentiable. 
    \item There exists a Lebesgue integrable function $\bound{3}$ such that for all $\weight \in \diffdomain$, 
        $
             \left| \func{\param} \frac{1}{\normalizer} \weightedlogprob \right| \leq \bound{3}(\param).
        $
    \item For each $n$,  there exists a Lebesgue integrable function $\bound{4}^{(n)}$ such that for all $\weight \in \diffdomain$, \\
        $
            \left| \frac{\partial}{\partial \weight_n} \func{\param} \frac{1}{\normalizer} \weightedlogprob \right| \leq \bound{4}^{(n)}(\param).
        $
\end{enumerate}
We have already proven that $\normalizer$ is continuously differentiable: hence, there is nothing to do for the first condition.
It is straightforward to use \cref{assume:finiteness} and check that the second condition is satisfied by the function 
$
    \bound{3}(\param) := \frac{1}{\normalizer} \func{\param}  \bound{1}(\param),
$ and the third condition is satisfied by 
$
    \bound{4}^{(n)}(\param) := \frac{1}{\normalizer} \func{\param} \loglik{n}{\param}  \bound{1}(\param).
$
Hence, we can interchange integration with differentiation. 
The partial derivatives of $\Eposterior{\weight} \func{\param}$ is equal to tthe sume of two integrals.
The first part is 
\begin{equation*}
    \begin{aligned}
        &\int \left( \frac{\partial \normalizer^{-1}}{\partial \weight_n} \func{\param} \weightedlogprob   \right) d\param  \\
        &=-\left( \Eposterior{\weight} \left[ \loglik{n}{\param} \right] \right) \int \left( \frac{1}{Z(w)} \func{\param} \weightedlogprob   \right) d\param \\
        &= - \Eposterior{\weight} \left[ \loglik{n}{\param} \right]  \times \Eposterior{\weight} \left[ \func{\param} \right].
    \end{aligned}
\end{equation*}
The second part is 
\begin{equation*}
    \begin{aligned} 
        \int \left(\frac{1}{\normalizer} \func{\param} \loglik{n}{\param}  \weightedlogprob   \right) &d\param = \\
    &\Eposterior{\weight} \left[ \func{\param} \loglik{n}{\param} \right].
    \end{aligned}
\end{equation*}
Putting the two parts together, the partial derivative is equal to a covariance: 
\begin{equation*}
    \frac{\partial \Eposterior{\weight}\func{\param}}{\partial \weight_n} =  \Cov_{\weight} \left[ \func{\param}, \loglik{n}{\param} \right].
\end{equation*}

The proof that $\Eposterior{\weight}\func{\param}^2$ is continuously differentiable is similar to that for $\Eposterior{\weight}\func{\param}$.
The partial derivative is 
\begin{equation*}
    \frac{\partial [\Eposterior{\weight}\func{\param}^2]}{\partial \weight_n} =  \Cov_{\weight} \left[ \func{\param}^2, \loglik{n}{\param} \right].
\end{equation*}

Since the posterior standard deviation is a continuously differentiable function of the mean and second moment, it is also continuously differentiable.
The formula for the partial derivative of the posterior standard deviation is a simple application of the chain rule, and we omit the proof for brevity. 

\end{delayedproof}

\subsection{First-order accuracy proofs}

\begin{delayedproof}{lem:normal-error}

    Our proof finds exact formulas for the posterior mean and the partial derivatives of the posterior mean with respect to $\weight_n$. 
    Then, we take the difference between the posterior mean and its Taylor series.

    In the normal model, the total log probability at $\weight$ is equal to 
    \begin{equation*}
        \begin{aligned}
            &\sum_{n=1}^{\nobs} 
            \weight_n \left[ \frac{1}{2} \log \left( \frac{1}{2\pi \noise^2} \right) - 
                            \frac{1}{2\noise^2}[ (\covarn)^2 - 2 \covarn \globalmean + \globalmean^2] 
                        \right] \\
            &= - \left( \frac{\sum_{n=1}^{\nobs} \weight_n}{2\noise^2}  \right) \left( \globalmean - \frac{\sum_{n=1}^{\nobs} \weight_n \covarn }{\sum_{n=1}^{\nobs} \weight_n} \right)^2 + C,
        \end{aligned}
    \end{equation*}
    where $C$ is a constant that does not depend on $\globalmean$.
    Hence, the distribution of $\globalmean$ under weight $\weight$ is normal with mean $(\sum_{n=1}^{\nobs} \weight_n \covarn)/(\sum_{n=1}^{\nobs} \weight_n)$ and precision $(\sum_{n=1}^{\nobs} \weight_n)/(\noise^2)$.   
    The partial derivative of the posterior mean with respect to $\weight_n$ is 
    \begin{equation*}
        \frac{\covarn (\sum_{n=1}^{\nobs} \weight_n) - (\sum_{n=1}^{\nobs} \weight_n \covarn)   }{(\sum_{n=1}^{\nobs} \weight_n)^2}.
    \end{equation*}
    Plugging in $\weight = \allones$, we have that $\infln$ is equal to $(\covarn - \bar{\covar})/\nobs$.

    After removing the index set $\indexset$, the actual posterior mean is 
    \begin{equation*}
        \frac{\nobs \bar{\covar} - |\indexset| \meanOfIndexSet}{\nobs - |\indexset|},
    \end{equation*}
    while the Taylor series approximation is
    \begin{equation*}
        \bar{\covar} - \sum_{n \in \indexset} \frac{\covarn - \bar{\covar}}{\nobs}  = \frac{\nobs \bar{\covar} + |\indexset|(\bar{\covar} - \meanOfIndexSet)  }{\nobs}.
    \end{equation*}
    The difference between the actual posterior mean and its approximation is as in the statement of the lemma.

\end{delayedproof}

\begin{delayedproof}{lem:normal-zeroth-order-error}

    We reuse the calculations from \proofref{lem:normal-error}.
    Namely, the posterior mean as a function of $\weight$ is 
    \begin{equation*}
        \sum_{n=1}^{\nobs} \weight_n \covarn/(\sum_{n=1}^{\nobs} \weight_n),
    \end{equation*}
    and the posterior mean for the original analysis is $\bar{\covar}$.
    If we remove $\indexset$ from the analysis, the posterior mean is
    \begin{equation*}
        \frac{\nobs \bar{\covar} - |\indexset| \meanOfIndexSet }{ \nobs - |\indexset| }
    \end{equation*}
    So, the value of $\zerothOrderError$ is
    \begin{equation*}
        \frac{|\indexset|}{\nobs - |\indexset|} \left( \bar{\covar} - \meanOfIndexSet \right).
    \end{equation*}
\end{delayedproof}

\begin{delayedproof}{lem:normal-means-error}

    Similar to the proof of \cref{lem:normal-error}, we first find exact formulas for the posterior mean and its Taylor series.

    In the normal means model, the total log probability at $\weight$ is 
    \begin{equation*}
    \begin{aligned}
        &\sum_{g=1}^{\numgroup} \left[ \frac{1}{2} \log \left( \frac{1}{2\pi \dispersion^2} \right) -  \frac{1}{2 \dispersion^2} (\groupmean_g - \globalmean)^2 \right] \\
        &+ 
        \sum_{n=1}^{\nobs}  \weight_n \left\{ \frac{1}{2} \log \left( \frac{1}{2\pi \noise^2} \right) - \frac{1}{2\noise^2} \left[ (\covarn)^2 - 2 \covarn \groupmean_{\groupn}  + \groupmean_{\groupn}^2 \right]  \right\}.
    \end{aligned}
    \end{equation*}
    By completing the squares, we know that 
    \begin{itemize}
        \item The distribution of $\globalmean$ is normal:
        \begin{equation*}
            N \left( \frac{\sum_{g=1}^{\numgroup} \precj{g}(\weight) \meanj{g}(\weight)  }{\sum_{g=1}^{\numgroup} \precj{g}(\weight)} , \frac{1} {\sum_{g=1}^{\numgroup} \precj{g}(\weight)}  \right)
        \end{equation*}
        \item Condition on $\globalmean$, the group means are independent normals:
        \begin{equation*}
            \groupmean_{g} \mid \globalmean \sim N \left( \frac{\globalmean/\dispersion^2 + [\nobsj{g}(\weight) \meanj{g}(\weight)]/\noise^2 }{1/\dispersion^2 + \nobsj{g}(\weight)/\noise^2} , \frac{1}{1/\dispersion^2 +  \nobsj{g}(\weight)/\noise^2 }  \right).
        \end{equation*}
    \end{itemize}

    To express the partial derivative of the posterior mean of $\globalmean$ with respect to $\weight_n$, it is helpful to define 
    the following ``intermediate'' value between $\Eposterior{\weight} \globalmean$ and $\Eposterior{\weight} \groupmean_{g}$:
    \begin{equation*}
        \shrmeanj{g}(\weight) := \frac{ 
                                    \meanj{g}(\weight) \nobsj{g}(\weight)/\noise^2 + \Eposterior{\weight} \globalmean /\dispersion^2
                                    }
                                    {
                                        \nobsj{g}(\weight)/\noise^2 + 1/\dispersion^2
                                    }.
    \end{equation*}
    In addition, we need the partial derivatives of the functions $\nobsj{g}$, $\precj{g}$, and $\meanj{g}$. 
    They are as follows. 
    \begin{equation*}
        \begin{aligned}
            \frac{\partial \nobsj{g} }{\partial \weight_n} &= 
                \begin{cases}
                    0 &\text{ if } g \neq \groupn \\
                    1 &\text{ if } g = \groupn
                \end{cases}, \\
            \frac{\partial \meanj{g} }{\partial \weight_n} &= 
                \begin{cases}
                    0 &\text{ if } g \neq \groupn \\
                    \frac{\covarn - \meanj{g}(\weight) }{\nobsj{g}(\weight)}  &\text{ if } g = \groupn
                \end{cases}, \\
            \frac{\partial \precj{g} }{\partial \weight_n} &= 
                \begin{cases}
                     0 &\text{ if } g \neq \groupn \\
                    \noise^2 \frac{\precj{g}(\weight)^2}{\nobsj{g}(\weight)^2} &\text{ if } g = \groupn
                \end{cases}.
        \end{aligned}
    \end{equation*}
    
    If $n$ is in the $k$-th group, the partial derivative of the posterior mean with respect to $\weight_n$ is 
    \begin{equation*}
        \frac{1}{\sumprec(\weight)} \frac{1}{\noise^2 + \dispersion^2 \nobsj{k}(\weight) }\left(\covarn - \shrmeanj{k}(\weight) \right).
    \end{equation*}

    After removing only observations from the $k$-th group, the actual posterior mean is
    \begin{equation*}
        \frac{
            \precj{k}(\wFromIndex(\indexset)) \meanj{k}(\wFromIndex(\indexset)) +
            \sum_{g \neq k}  \precj{g}(\allones) \meanj{g}(\allones)
            }{
                \precj{k}(\wFromIndex(\indexset)) +
                \sum_{g \neq k}  \precj{g}(\allones)   
            }.
    \end{equation*}
    Between $\weight = \wFromIndex(\indexset)$ and $\weight = \allones$, the $\nobsj{g}, \meanj{g}$, and $\precj{g}$ functions do not change for $g \neq k$.
    The Taylor series approximation of the posterior mean is 
    \begin{equation*}
        \frac{
            \precj{k}(\allones) \left[ \meanj{k}(\allones) + \sum_{n \in \indexset} \left( \shrmeanj{k}(\allones) - \covarn  \right)/\nobsj{k}(\allones)  \right] +
            \sum_{g \neq k}  \precj{g}(\allones) \meanj{g}(\allones)
            }{
                \precj{k}(\allones) +
                \sum_{g \neq k}  \precj{g}(\allones)   
            }.
    \end{equation*}
    If we denote 
    \begin{equation*}
        \begin{aligned} 
            A_1 &:= \sum_{g \neq k}  \precj{g}(\allones) \meanj{g}(\allones), A_2 := \sum_{g \neq k}  \precj{g}(\allones) \\
            B_1 &:= \precj{k}(\wFromIndex(\indexset)) \meanj{k}(\wFromIndex(\indexset)) , B_2 := \precj{k}(\wFromIndex(\indexset)) \\
            C_1 &:= \precj{k}(\allones) \left[ \meanj{k}(\allones) + \sum_{n \in \indexset} \left( \shrmeanj{k}(\allones) - \covarn  \right)/\nobsj{k}(\allones)  \right], C_2 := \precj{k}(\allones) \\
        \end{aligned},
    \end{equation*}
    then $\firstOrderError$ is equal to $(A_1 + B_1)/(A_2 + B_2) - (A_1 + C_1)/(A_2 + C_2)$.
    The last equation is equal to 
    \begin{equation*}
        \frac{A_2(B_1-C_1) + A_1 (C_2 - B_2) + (B_1 C_2 - C_1 B_2)}{(A_2 + B_2)(A_2 + C_2)}. 
    \end{equation*}
    We analyze the differences $C_2 - B_2$, $B_1 C_2 - C_1 B_2$, and $B_1 - C_1$ separately.

    \textbf{$C_2 - B_2$.}
    This difference is 
    \begin{equation*} 
            \frac{1}{\noise^2/\nobsj{k}(\allones)  + \dispersion^2 }   - \frac{1}{\noise^2/\nobsj{k}(\wFromIndex(\indexset)) + \dispersion^2}.
    \end{equation*}
    Since we remove $|\indexset|$ from group $k$, $\nobsj{k}(\wFromIndex(\indexset)) = \nobsj{k}(\allones) - |\indexset|$.
    Hence, the difference $C_2 - B_2$ is 
    \begin{equation*} 
        \noise^2 \precj{k}(\allones) \precj{k}(\wFromIndex(\indexset)) \frac{|\indexset|}{\nobsj{k}(\allones)(\nobsj{k}(\allones) - |\indexset|)},
    \end{equation*}
    which is exactly the $E(\indexset)$ mentiond in the lemma statement.  

    \textbf{$B_1 C_2 - C_1 B_2$.}
    The difference is 
    \begin{equation*} 
        \precj{k}(\allones) \precj{k}(\wFromIndex(\indexset)) \left\{ \meanj{k}(\wFromIndex(\indexset)) - \meanj{k}(\allones) - \frac{\sum_{n \in \indexset} [\shrmeanj{k}(\allones) - \covarn ]  } {\nobsj{k}(\allones)}   \right\}.
    \end{equation*}
    We analyze the term in the curly brackets.
    It is equal to 
    \begin{equation*}
        \begin{aligned}
            \left\{\meanj{k}(\wFromIndex(\indexset)) - \meanj{k}(\allones) - \frac{\sum_{n \in \indexset} [\meanj{k}(\allones) - \covarn ]  } {\nobsj{k}(\allones)} \right\}
            + \sum_{n \in \indexset } \left( \frac{\meanj{k}(\allones) - \shrmeanj{k}(\allones)  }{\nobsj{k}(\allones)} \right)
         \end{aligned}
    \end{equation*}
    The left term is equal to 
    \begin{equation*}
        \frac{|\indexset|^2 (  \meanj{k}(\allones) - \meanOfIndexSet ) }{\nobsj{k}(\allones)[\nobsj{k}(\allones) - |\indexset|] }.
    \end{equation*}
    The right term is equal to 
    \begin{equation*}
        \frac{|\indexset| } { \nobsj{k}(\allones) } \frac{ \noise^2 \precj{k}(\allones) } {\nobsj{k}(\allones) }
                                (\Eposterior{\allones}\globalmean - \meanj{k} (\allones)  ).
    \end{equation*}
    The sum of the two terms is exactly $F(\indexset)$ mentioned in the lemma statement.
    Overall, the difference $B_1 C_2 - C_1 B_2$ is equal to $\precj{k}(\allones) \precj{k}(\wFromIndex(\indexset)) F(\indexset)$.

    \textbf{$B_1 - C_1$.}
    If we introduce $D :=  \precj{k}(\allones) \meanj{k}(\wFromIndex(\indexset))$, then the difference $B_1 - C_1$ is equal to $(B_1 - D) + (D - C_1)$.
    The former term is 
    \begin{equation*}
        \meanj{k}(\wFromIndex(\indexset))(B_2 - C_2) = -\meanj{k}(\wFromIndex(\indexset)) E(\indexset).
    \end{equation*}
    The later term is 
    \begin{equation*}
        \precj{k}(\allones) \left\{ \meanj{k}(\wFromIndex(\indexset)) - \meanj{k}(\allones) - \frac{\sum_{n \in \indexset} [\shrmeanj{k}(\allones) - \covarn ]  } {\nobsj{k}(\allones)}   \right\}.
    \end{equation*}
    We already know that the term in the curly brackets is equal to $F(\indexset)$.
    Hence $B_1 - C_1$ is equal to $ \precj{k}(\allones) F(\indexset) -\meanj{k}(\wFromIndex(\indexset)) E(\indexset)$. 

    With the differences $C_2 - B_2$, $B_1 C_2 - C_1 B_2$, and $B_1 - C_1$, we can now state the final form of $\firstOrderError$.
    The final numerator is 
    \begin{equation*}
        \begin{aligned}
            &\left[ \precj{k}(\wFromIndex(\indexset))  +  \sum_{g \neq k}  \precj{g}(\allones) \right] \precj{k}(\allones)  F(\indexset) \\
       &+ \left[ \sum_{g \neq k}  \precj{g}(\allones) \meanj{g}(\allones) - \meanj{k}(\wFromIndex(\indexset)) \sum_{g \neq k}  \precj{g}(\allones)  \right] E(\indexset)
        \end{aligned}.
    \end{equation*}
    Divide this by the denominator $\left[\sum_{g}  \precj{g}(\allones) \right] \left[ \sum_{g}  \precj{g}(\wFromIndex(\indexset))  \right]$, we have proven the lemma.  

\end{delayedproof}

\begin{delayedproof}{lem:normal-means-zeroth-order-error}

    We reuse many calculations from \proofref{lem:normal-means-error}.
    Recall that, after removing only observations from the $k$-th group, the actual posterior mean is
    \begin{equation*}
        \frac{
            \precj{k}(\wFromIndex(\indexset)) \meanj{k}(\wFromIndex(\indexset)) +
            \sum_{g \neq k}  \precj{g}(\allones) \meanj{g}(\allones)
            }{
                \precj{k}(\wFromIndex(\indexset)) +
                \sum_{g \neq k}  \precj{g}(\allones)   
            }.
    \end{equation*}
    The zeroth-order approximation of this posterior mean is
    \begin{equation*}
        \frac{
            \precj{k}(\allones) \meanj{k}(\allones) +
            \sum_{g \neq k}  \precj{g}(\allones) \meanj{g}(\allones)
            }{
                \precj{k}(\allones) +
                \sum_{g \neq k}  \precj{g}(\allones)   
            }.
    \end{equation*}

    If we denote 
    \begin{equation*}
        \begin{aligned} 
            A_1 &:= \sum_{g \neq k}  \precj{g}(\allones) \meanj{g}(\allones), A_2 := \sum_{g \neq k}  \precj{g}(\allones) \\
            B_1 &:= \precj{k}(\wFromIndex(\indexset)) \meanj{k}(\wFromIndex(\indexset)) , B_2 := \precj{k}(\wFromIndex(\indexset)) \\
            V_1 &:= \precj{k}(\allones) \meanj{k}(\allones), V_2 := \precj{k}(\allones) \\
        \end{aligned},
    \end{equation*}
    then $\zerothOrderError$ is equal to $(A_1 + B_1)/(A_2 + B_2) - (A_1 + V_1)/(A_2 + V_2)$.
    The last equation is equal to 
    \begin{equation*}
        \frac{A_2(B_1-V_1) + A_1 (V_2 - B_2) + (B_1 V_2 - V_1 B_2)}{(A_2 + B_2)(A_2 + C_2)}. 
    \end{equation*}
    We analyze the differences $V_2 - B_2$, $B_1 V_2 - V_1 B_2$, and $B_1 - V_1$ separately.
    Note that the quantities $A_1,A_2,B_1,B_2$ are the same as those defined in \proofref{lem:normal-means-error}.

    \textbf{$V_2 - B_2$.}
    $V_2$ is the same as $C_2$ from \proofref{lem:normal-means-error}.
    Hence, $V_2 - B_2$ is equal to $E(\indexset)$.

    \textbf{$B_1 - V_1$.}
    If we introduce $D :=  \meanj{k}(\allones) \fullprecj{k}$, then the difference $B_1 - V_1$ is equal to $(B_1 - D) + (D - C_1)$.
    The former term is 
    \begin{equation*}
        -\meanj{k}(\wFromIndex(\indexset)) E(\indexset).  
    \end{equation*}
    The later term is 
    \begin{equation*}
        \fullprecj{k} \frac{|\indexset|}{ \fullnobsj{k}  - |\indexset| } (\fullmeanj{k} - \meanOfIndexSet).
    \end{equation*}
    So, the total $B_1 - V_1$ is equal to 
    \begin{equation*}
        \fullprecj{k} \frac{|\indexset|}{ \fullnobsj{k} - |\indexset|} (\fullmeanj{k} - \meanOfIndexSet) - \meanj{k}(\wFromIndex(\indexset))   E(I).
    \end{equation*}

    \textbf{$B_1 V_2 - V_1 B_2$.}
    This is equal to $B_1 V_2  - B_1 B_2 + B_1 B_2 - V_1 B_2$, which is equal to $B_1(V_2 - B_2) + B_2(B_1 - V_1)$.
    The former term is 
    \begin{equation*}
        \precj{k}(\wFromIndex(\indexset)) \meanj{k}(\wFromIndex(\indexset)) E(\indexset).
    \end{equation*}
    The later term is 
    \begin{equation*}
        \precj{k}(\wFromIndex(\indexset)) \left[ \fullprecj{k} \frac{|\indexset|}{ \fullnobsj{k} - |\indexset|} (\fullmeanj{k} - \meanOfIndexSet) - \meanj{k}(\wFromIndex(\indexset))   E(I)   \right].
    \end{equation*}
    The sum of the two terms is equal to 
    \begin{equation*}
        \precj{k}(\wFromIndex(\indexset))  \fullprecj{k} \frac{|\indexset|}{ \fullnobsj{k} - |\indexset|} (\fullmeanj{k} - \meanOfIndexSet).
    \end{equation*}
        
    With the differences $V_2 - B_2$, $B_1 V_2 - V_1 B_2$, and $B_1 - V_1$, we can now state the final form of $\zerothOrderError$.
    The final numerator is 
    \begin{equation*}
        \begin{aligned}
            &\precj{k}(\wFromIndex(\indexset))  \fullprecj{k} \frac{|\indexset|}{ \fullnobsj{k} - |\indexset|} (\fullmeanj{k} - \meanOfIndexSet) \\
       &+ \left[ \sum_{g \neq k}  \precj{g}(\allones) \meanj{g}(\allones) - \meanj{k}(\wFromIndex(\indexset)) \sum_{g \neq k}  \precj{g}(\allones)  \right] E(\indexset)
        \end{aligned}.
    \end{equation*}
    Divide this by the denominator $\left[\sum_{g}  \precj{g}(\allones) \right] \left[ \sum_{g}  \precj{g}(\wFromIndex(\indexset))  \right]$, we have proven the lemma.  

\end{delayedproof}

A corrolary of \cref{lem:normal-means-error} is that the absolute value of the error behaves like $|\indexset|^2/(\numgroup  |\fullnobsj{k}|^2)$.

\begin{mycorollary}\label{cor:normal-means-bound}
    In the normal means model, for all groups $g$, assume that 
    $
        \fullnobsj{g} \geq \noise^2/\dispersion^2. 
    $
    Let the index set $\indexset$ be such that there exists $k \in \{1,2,\ldots,\numgroup\}$ such that $\groupn = k$ for all $n \in \indexset$.
    For this $k$, assume that 
    $
        \fullnobsj{k} - |\indexset| \geq \noise^2/\dispersion^2.
    $
    Then, 
    \begin{equation*}
            |\firstOrderError| \leq C(\|\covar\|_{\infty}, \noise, \dispersion)  \frac{1}{\numgroup}\frac{|\indexset|^2} {|\fullnobsj{k}|^2},
    \end{equation*}
    where $C(\|\covar\|_{\infty}, \noise, \dispersion)$ is a constant that only depends on $\|\covar\|_{\infty}$, $\noise$, and $\dispersion$.
\end{mycorollary}
We prove \cref{cor:normal-means-bound} in the \proofref{cor:normal-means-bound}.
In addition to the assumptions \cref{lem:normal-means-error}, the corrolary assumes that the number of observations in each group is not too small, and that after removing $\indexset$, group $k$ still has enough observations.
This condition allows us to approximate $\fullprecj{k}$ and $\precj{g}(\wFromIndex(\indexset))$ with a constant.
The factor $\|\covar\|_{\infty}$ in the bound comes from upper bounding 
$|\fullmeanj{g} - \meanj{k}(\wFromIndex(\indexset))|$ by $2 \max_{n=1}^{\nobs} |\covarn|$.

\begin{delayedproof}{cor:normal-means-bound}

    Under the assumption that $\fullnobsj{g} \geq \noise^2 / \dispersion^2$, we have that $\precj{g}(\allones) \in \left[ \frac{1}{2\dispersion^2}, \frac{1}{\dispersion^2} \right]$.
    Since $\fullmeanj{k} - |\indexset| \geq \noise^2 / \dispersion^2$, it is also true that $\precj{k}(\wFromIndex(\indexset)) \in \left[ \frac{1}{2\dispersion^2}, \frac{1}{\dispersion^2} \right]$.

    Because of \cref{lem:normal-means-error}, an upper bound on $\firstOrderError$ is 
    \begin{equation*}
        \frac{
            \precj{k}(\wFromIndex(\indexset))  
            }
            {
                \fullprec 
            }  
        \left| F(\indexset) \right|
        + 
        \left|
        \frac{
                \left(\sum_{g \neq k} \fullprecj{g}( \fullmeanj{g} - \meanj{k}(\wFromIndex(\indexset)))  \right)
            }
            {
                \fullprec \sumprec(\wFromIndex(\indexset))
            }
        \right|  
        \left| E(\indexset) \right|. 
    \end{equation*}
    The fraction $\precj{k}(\wFromIndex(\indexset))/\fullprec$ is at most $(\frac{1}{\dispersion^2})/\left( \numgroup \frac{1}{2\dispersion^2} \right)$ , which is equal to $2/\numgroup$.
    The absolute value $\left| F(\indexset) \right|$ is at most 
    \begin{equation*}
        \frac{2 |\indexset|^2 \|\covar\|_{\infty} }
                {(\fullnobsj{k})^2} 
        + 
        \frac{2 |\indexset|  \|\covar\|_{\infty} (\noise^2/\dispersion^2)}{(\fullnobsj{k})^2} \leq 
        \frac{2 |\indexset|^2 \|\covar\|_{\infty} (\noise^2/\dispersion^2 + 1) }{(\fullnobsj{k})^2}.
    \end{equation*}
    The absolute value 
    \begin{equation*}
        \left|
        \frac{
                \left(\sum_{g \neq k} \fullprecj{g}( \fullmeanj{g} - \meanj{k}(\wFromIndex(\indexset)))  \right)
            }
            {
                \fullprec \sumprec(\wFromIndex(\indexset))
            }
        \right|  
    \end{equation*}
    is at most 
    \begin{equation*}
        \frac{\numgroup (1/\dispersion^2) 2 \|\covar\|_{\infty} }{ \numgroup^2 (1/2\dispersion^2) } \leq \frac{4 \|\covar\|_{\infty} }{\numgroup}.
    \end{equation*}
    Finally, the absolute value $\left| E(\indexset) \right|$ is at most 
    \begin{equation*}
        \frac{|\indexset| (\noise^2/(4\dispersion^4))}{(\fullnobsj{k})^2} \leq \frac{|\indexset|^2 (\noise^2/(4\dispersion^4))}{(\fullnobsj{k})^2}.
    \end{equation*} 
    In all, the constant $C(\|\covar\|_{\infty}, \noise, \dispersion)$ in the corollary's statement is 
    \begin{equation*}
        \| \covar \|_{\infty} \left( 4(\noise^2/\dispersion^2 + 1) + \noise^2 / \dispersion^4  \right).
    \end{equation*}

\end{delayedproof}

\subsection{Consistency and asymptotic normality proofs} 

The following lemma on covariance between sample covariances under i.i.d.\ sampling will be useful for later proofs.

\begin{mylemma}\label{lem:cov-of-cov}
    Suppose we have $\samplesize$ i.i.d.\ draws $(\Adraw{s},\Bdraw{s},\Cdraw{s})_{s=1}^{S}$. 
    Let $f_1$ be the (biased) sample covariance between the $A$'s and the $B$'s.
    Let $f_2$ be the (biased) sample covariance between the $A$'s and $C$'s.
    In other words,
    \begin{equation*}
        \begin{aligned}
        f_1 &:= \left( \frac{1}{S} \sum_{s=1}^{S} \Adraw{s} \Bdraw{s} \right) - \left( \frac{1}{S} \sum_{s=1}^{S} \Adraw{s} \right) \left( \frac{1}{S} \sum_{s=1}^{S}\Bdraw{s} \right), \\
        f_2 &:= \left( \frac{1}{S} \sum_{s=1}^{S}\Adraw{s}  \Cdraw{s} \right) - \left( \frac{1}{S} \sum_{s=1}^{S}\Adraw{s}  \right) \left( \frac{1}{S} \sum_{s=1}^{S}\Cdraw{s} \right).
        \end{aligned}
    \end{equation*}
    Suppose that the following are finite: $\E [(A - \E [A])^2 (B - \E [B]) (C - \E [C]) ]$, $\Cov(B,C)$, $\Var(A)$, $\Cov(A,B)$, $\Cov(A,C)$.
    Then, the covariance of $f_1$ and $f_2$ is equal to
    \begin{equation*}
        \begin{aligned}
           &\frac{(S-1)^2 }{S^3} \E [(A - \E [A])^2 (B - \E [B]) (C - \E [C]) ] \\
            &+ \frac{S-1}{S^3} \Cov(B,C) \Var(A) - \frac{(S-1)(S-2)}{S^3} \Cov(A,B) \Cov(A,C).
        \end{aligned}
    \end{equation*}
\end{mylemma}

\begin{delayedproof}{lem:cov-of-cov}

    It suffices to prove the lemma in the case where $\E[A] = \E[B] = \E[C] = 0$.
    Otherwise, we can subtract the population mean from the random variable: the value of $f_1$ and $f_2$ would not change (since covariance is invariant to constant additive changes).
    In other words, we want to show that the covariance between $f_1$ and $f_2$ is equal to 
    \begin{equation} \label{eq:cov-rewrite}
        \frac{(S-1)^2 }{S^3} \E [A^2 B C ] + \frac{S-1}{S^3} \E[BC] \E[A^2] - \frac{(S-1)(S-2)}{S^3} \E[AB] \E[AC].
    \end{equation}

    Since $f_1$ is the biased sample covariance, $\E f_1 = \frac{S-1}{S} \E[AB]$.
    Similarly, $\E f_2 = \frac{S-1}{S} \E[AC]$.
    To compute $\Cov(f_1,f_2)$, we only need an expression for $\E[f_1f_2]$.
    The product $f_1f_2$ is equal to the sum of $D_1, D_2, D_3, D_4$ where:
    \begin{equation*}
        \begin{aligned}
            D_1 &:= -\left( \frac{1}{\samplesize}\sum_s \Adraw{s} \Bdraw{s} \right) 
            \left( \frac{1}{\samplesize}\sum_s \Adraw{s} \right) 
            \left( \frac{1}{\samplesize}\sum_s \Cdraw{s} \right), \\
            D_2 &:= \left( \frac{1}{\samplesize}\sum_s \Adraw{s}\right)^2  
            \left( \frac{1}{\samplesize}\sum_s \Bdraw{s} \right) 
            \left( \frac{1}{\samplesize}\sum_s \Cdraw{s} \right), \\ 
            D_3 &:= -\left( \frac{1}{\samplesize}\sum_s \Adraw{s} \Cdraw{s} \right)  
            \left( \frac{1}{\samplesize}\sum_s \Adraw{s} \right) 
            \left( \frac{1}{\samplesize}\sum_s \Bdraw{s} \right), \\
            D_4 &:= \left( \frac{1}{\samplesize}\sum_s \Adraw{s} \Bdraw{s} \right)  
            \left( \frac{1}{\samplesize}\sum_s \Adraw{s} \Cdraw{s} \right). 
        \end{aligned}
    \end{equation*}

    We compute the expectation of each $D_j$. 

    \textbf{$D_1$}.
    By expanding $D_1$, we know that $\E D_1 = \frac{1}{\samplesize^3} \sum_{i,j,k} \E[\Adraw{k} \Bdraw{k} \Adraw{i} \Cdraw{j}]$.
    The value of $\E[\Adraw{k} \Bdraw{k} \Adraw{i} \Cdraw{j}]$ depends on the triplet $(i,j,k)$ in the following way:
    \begin{equation*}
        \E[\Adraw{k} \Bdraw{k} \Adraw{i} \Cdraw{j}] = 
            \begin{cases}
             0  &\text{ if } i = k, j \neq k \\
             \E[A^2 BC]  &\text{ if } i = k, j = k \\ 
             0  &\text{ if } i \neq k, j = k \\ 
             \E[AB] \E[AC]  &\text{ if } i \neq k,  j \neq k, i = j \\
             0  &\text{ if } i \neq k,  j \neq k, i \neq j \\ 
            \end{cases}.
    \end{equation*}
    We have used the independence of $(\Adraw{s},\Bdraw{s},\Cdraw{s})_{s=1}^{S}$ to factorize the expectation $\E[\Adraw{k} \Bdraw{k} \Adraw{i} \Cdraw{j}]$.
    For certain triplets, the factorization reveals that the expectation is zero.
    By accounting for all triplets, the expectation of $D_1$ is 
    \begin{equation*}
        \frac{1}{\samplesize^3} \left[ \samplesize \E[A^2 BC] + \samplesize(\samplesize-1) \E[AB] \E[AC] \right].
    \end{equation*}

    \textbf{$D_2$}.
    By expanding $D_2$, we know that $\E D_2  = \frac{1}{\samplesize^4} \sum_{i,j,p,q} \E[\Adraw{i} \Adraw{i} \Bdraw{p} \Cdraw{q}]$.
    We can do a similar case-by-case analysis of how $\E[\Adraw{i} \Adraw{i} \Bdraw{p} \Cdraw{q}]$ depend on the quartet $(i,j,p,q)$.
    The outcome of this analysis is that, the expectation of $D_2$ is
    \begin{equation*}
        \frac{1}{\samplesize^3} \left[ \E[A^2BC] + (\samplesize-1) \E[A^2] \E[BC] + 2 (\samplesize-1) \E[AB] \E[AC] \right]. 
    \end{equation*}

    \textbf{$D_3$}.
    By symmetry between $D_1$ and $D_3$, the expectation of $D_3$ is also
    \begin{equation*}
        \frac{1}{\samplesize^3} \left[ \samplesize \E[A^2 BC] + \samplesize(\samplesize-1) \E[AB] \E[AC] \right].
    \end{equation*}

    \textbf{$D_4$}.
    By expanding $D_4$, we know that $\E D_4  = \frac{1}{\samplesize^2} \sum_{i,j} \E[\Adraw{i}  \Bdraw{i} \Adraw{j} \Cdraw{j}]$.
    The case-by-case analysis of $\E[\Adraw{i}  \Bdraw{i} \Adraw{j} \Cdraw{j}]$ for each $(i,j)$ is simple, and is omitted.
    The expectation of $D_4$ is 
    \begin{equation*}
        \frac{1}{\samplesize} \E[A^2 BC] + \frac{\samplesize-1}{\samplesize} \E[AB] \E[AC].
    \end{equation*}

    Some algebra shows that $\sum_{i=1}^{4} \E[D_i] - \frac{S-1}{S} \E[AB] \frac{S-1}{S} \E[AC] $ is equal to \cref{eq:cov-rewrite}.

\end{delayedproof}

\begin{delayedproof}{lem:mcmcinfln-variance}

    In this proof, we will only consider expectations under the full-data posterior.
    Hence, to alleviate notation, we shall write $\E$ instead of $\Eposterior{\allones}$: similarly, covariance and variance evaluations are understood to be at $\weight = \allones$.

    Applying \cref{lem:cov-of-cov}, the covariance of $\mcmcinfln$ and $\mcmcinfln$ i.e.\ the variance of $\mcmcinfln$ is equal to
    \begin{equation*}
        \begin{aligned}
        &\frac{(\samplesize-1)^2}{\samplesize^3} \E \{ ( \func{\param} - \E [\func{\param}]  )^2 ( \loglik{n}{\param} - \E [\loglik{n}{\param}]  )^2 \} \\
        &+  \frac{\samplesize-1}{\samplesize^3} \Var(\loglik{n}{\param}) \Var(\func{\param}) 
        - \frac{(\samplesize-1)(\samplesize-2)}{\samplesize^3} \Cov(\func{\param}, \loglik{n}{\param})^2.
        \end{aligned}
    \end{equation*}

    We define the constant $C$ to be the maximum over $n$ of
    \begin{equation*}
        \begin{aligned}
            &\Cov( \func{\param},  \loglik{n}{\param}  )^2 + \Var(\func{\param}) \Var(\loglik{n}{\param}) \\
            &+ \E \{ ( \func{\param} - \E [\func{\param}]  )^2 ( \loglik{n}{\param} - \E [\loglik{n}{\param}]  )^2 \}.
        \end{aligned}
    \end{equation*}
    Clearly, $\Var(\mcmcinfln) \leq \frac{C}{\samplesize}$.

\end{delayedproof}

\begin{delayedproof}{thm:mcmcinfl-mcmcamip-consistent}
    Similar to the proof of \cref{lem:mcmcinfln-variance}, expectations (and variances and covariances) are understood to be taken under the full-data posterior. 

    Since $\mcmcinfln$ is the biased sample covariance, we know that 
    \begin{equation*}
        \E \mcmcinfln = \frac{\samplesize-1}{\samplesize} \infln.
    \end{equation*}
    The bias of $\mcmcinfln$ goes to zero at rate $1/\samplesize$.
    Because of \cref{lem:mcmcinfln-variance}, the variance also goes to zero at rate $1/\samplesize$.
    Then, an application of Chebyshev's inquality shows that $\mcmcinfln \xrightarrow{p} \infln$.
    Since $\nobs$ is a constant, the pointwise convergence $|\mcmcinfln - \infln| \xrightarrow{p} 0$ implies the uniform convergence $\max_{n=1}^{\nobs} |\mcmcinfln - \infln| \xrightarrow{p} 0$.

    We now prove that $|\mcmcamip - \amip| \xrightarrow{p} 0$.
    We first recall some notation.
    The ranks $r_1, r_2,\ldots,r_N$ sort the influences $\infl_{r_1} \leq \infl_{r_2} \leq \ldots \leq \infl_{r_{\nobs}}$, and $\amip = -\sum_{m=1}^{\numdrop} \infl_{r_m} \indict{\infl_{r_m}  < 0} $.
    Similarly, $v_1, v_2,\ldots,v_N$ sort the estimates $\mcmcinfl_{v_1} \leq \mcmcinfl_{v_2} \leq \ldots \leq \mcmcinfl_{v_{\nobs}}$, and $\mcmcamip = -\sum_{m=1}^{\numdrop} \mcmcinfl_{v_m} \indict{\mcmcinfl_{v_m}  < 0}$.
    It suffices to prove the convergence when $\numdrop \geq 1$: in the case $\numdrop = 0$, both $\mcmcamip$ and $\amip$ are equal to zero, hence the distance between them is identically zero. 
    Denote the $T$ unique values among $\infln$ by $u_1 < u_2 < \ldots < u_T$. 
    If $T = 1$ i.e.\ there is only one value, let $\gap := 1$.
    Otherwise, let $\gap$ be the smallest gap between subsequent values: $\gap := \min_{t}  (u_{t+1} - u_{t})$.

    Suppose that $\max_{n=1}^{\nobs} |\mcmcinfln - \infln| \leq \gap/3$, and let $A$ be the indicator for this event.
    For any $n$, each $\mcmcinfln$ is in the interval $[\infln - \gap/3, \infln + \gap/3$].
    In the case $T = 1$, clearly all $k$ such that $\mcmcinfl_k$ is in $[\infln - \gap/3, \infln + \gap/3$] satisfy $\infl_k = \infl_n$. 
    In the case $T > 1$, since the unique values of $\infln$ are at least $\gap$ apart, all $k$ such that $\mcmcinfl_k$ is in $[\infln - \gap/3, \infln + \gap/3$] satisfy $\infl_k = \infl_n$. 
    This means that the ranks $v_1, v_2,\ldots, v_{\nobs}$, which sort the influence estimates, also sort the true influences in ascending order: $\infl_{v_1} \leq \infl_{v_2} \leq \ldots \leq \infl_{v_{\nobs}}$.
    Since the ranks $r_1, r_2, \ldots, r_{\nobs}$ also sort the true influences, it must be true that $\infl_{v_m} = \infl_{r_m}$ for all $m$. 
    Therefore, we can write 
    \begin{equation*}
        \begin{aligned}
            |\mcmcamip - \amip| &= \left| \sum_{m=1}^{\numdrop} \left( \infl_{v_m} \indict{\infl_{v_m} < 0} - \mcmcinfl_{v_m} \indict{\mcmcinfl_{v_m} < 0} \right) \right| \\
            &\leq \sum_{m=1}^{\numdrop} \left| \infl_{v_m} \indict{\infl_{v_m} < 0} - \mcmcinfl_{v_m} \indict{\mcmcinfl_{v_m} < 0} \right|.
        \end{aligned}
    \end{equation*}

    We control the absolute values $\left| \infl_{v_m} \indict{\infl_{v_m} < 0} - \mcmcinfl_{v_m} \indict{\mcmcinfl_{v_m} < 0} \right|$.
    For any index $n$, by triangle inequality, $\left| \infl_{n} \indict{\infl_{n} < 0} - \mcmcinfl_{n} \indict{\mcmcinfl_{n} < 0} \right|$ is at most
    \begin{equation*}
        \indict{\mcmcinfl_{n} < 0} |\infl_{n} - \mcmcinfl_{n}| + |\infln| | \indict{\mcmcinfl_{n} < 0} - \indict{\infl_{n} < 0} |.
    \end{equation*}
    The first term is at most $|\infl_{n} - \mcmcinfl_{n}|$.
    The second term is at most $\indict{|\infl_{n} - \mcmcinfl_{n}| \geq |\infln|, \infln \neq 0}$.
    We next prove a bound on $\left| \infl_{n} \indict{\infl_{n} < 0} - \mcmcinfl_{n} \indict{\mcmcinfl_{n} < 0} \right|$ that holds across $n$.
    Our analysis proceeds differently based on whether the set $\{n: \infl_n \neq 0\}$ is empty or not.
    \begin{itemize}
        \item $\{n: \infl_n \neq 0\}$ is empty. This means $\infln = 0$ for all $n$.
        Hence, $\indict{|\infl_{n} - \mcmcinfl_{n}| \geq |\infln|, \infln \neq 0}$ is identically zero. 

        \item $\{n: \infl_n \neq 0\}$ is not empty. We know that $\min_{n} |\infln| > 0$. 
        Hence, $\indict{|\infl_{n} - \mcmcinfl_{n}| \geq |\infln|, \infln \neq 0}$ is upper bounded by 
        $\indict{|\infl_{n} - \mcmcinfl_{n}| \geq \min_{n}  |\infln|}$. 
        Since $|\infl_{n} - \mcmcinfl_{n}| \leq \max_{n} |\infl_{n} - \mcmcinfl_{n}|$, this last indicator is at most 
        $\indict{\max_{n} |\infl_{n} - \mcmcinfl_{n}| \geq \min_{n} |\infln|}$. 
    \end{itemize}

    To summarize, we have proven the following upper bounds on $|\mcmcamip - \amip|$.
    When $\{n: \infl_n \neq 0\}$ is empty, on $A$, $|\mcmcamip - \amip|$ is upper bounded by 
    \begin{equation} \label{eq:empty-bound}
            \numdrop \max_{n=1}|\infl_{n} - \mcmcinfl_{n}|
    \end{equation}
    When $\{n: \infl_n \neq 0\}$ is not empty, on $A$, $|\mcmcamip - \amip|$ is upper bounded by 
    \begin{equation}\label{eq:non-empty-bound}
            \numdrop \max_{n=1}|\infl_{n} - \mcmcinfl_{n}| + \numdrop \indict{\max_{n} |\infl_{n} - \mcmcinfl_{n}| \geq \min_{n} |\infln|}.
    \end{equation}

    We are ready to show that $\Pr(|\mcmcamip - \amip| > \epsilon )$ converges to zero. 
    For any positive $\epsilon$, we know that 
    \begin{equation*}
        \Pr(|\mcmcamip - \amip| > \epsilon ) \leq \Pr(|\mcmcamip - \amip| > \epsilon, A) + \Pr(A^c).
    \end{equation*}
    The later probability goes to zero because $\max_{n=1}^{\nobs} |\mcmcinfln - \infln| \xrightarrow{p} 0$.

    Suppose that $\{n: \infl_n \neq 0\}$ is empty.
    Using the upper bound \cref{eq:empty-bound}, we know that event in the former probability implies that $\max_{n=1}^{\nobs} |\mcmcinfln - \infln| \geq \epsilon/\numdrop$.
    The probability of this event also goes to zero because $\max_{n=1}^{\nobs} |\mcmcinfln - \infln| \xrightarrow{p} 0$.
    
    Suppose that $\{n: \infl_n \neq 0\}$ is not empty.
    Using the upper bound \cref{eq:non-empty-bound}, we know that event in the former probability implies that $(\max_{n=1}^{\nobs} |\mcmcinfln - \infln| + \indict{\max_{n} |\infl_{n} - \mcmcinfl_{n}| \geq \min_{n} |\infln|}) \geq \epsilon/\numdrop$.
    Since $\max_{n=1}^{\nobs} |\mcmcinfln - \infln|$ converges to zero in probability,  $\indict{\max_{n} |\infl_{n} - \mcmcinfl_{n}| \geq \min_{n} |\infln|}$ also converges to zero in probability.
    Hence, the probability that $(\max_{n=1}^{\nobs} |\mcmcinfln - \infln| + \indict{\max_{n} |\infl_{n} - \mcmcinfl_{n}| \geq \min_{n} |\infln|}) \geq \epsilon/\numdrop$ converges to zero.

    In all, $\Pr(|\mcmcamip - \amip| > \epsilon )$ goes to zero in both the case where $\{n: \infl_n \neq 0\}$ is empty and the complement case. 
    As the choice of $\epsilon$ was arbitrary, we have shown $\mcmcamip \xrightarrow{p} \amip$.

\end{delayedproof}

\begin{delayedproof}{thm:mcmcinfl-asymptotic-normal}

    Similar to the proof of \cref{lem:cov-of-cov}, we only consider expectations under the full-data posterior.
    Hence, we will write $\E$ instead of $\Eposterior{\allones}$ to simplify notation.
    Variance and covariance operations are also understood to be taken under the full-data posteiror. 
    To lighten the dependence of the notation on the parameter $\param$, we will write $\func{\param}$ as $\funcname$ and $\loglik{n}{\param}$ as $\LL{n}$ when talking about the expectation of $\func{\param}$ and $\loglik{n}{\param}$.

    Define the the following multivariate function
    \begin{equation*}
        f(\param) := 
            \begin{bmatrix}
                \func{\param}, \loglik{1}{\param}, \func{\param} \loglik{1}{\param}, \ldots, \loglik{\nobs}{\param}, \func{\param} \loglik{\nobs}{\param}
            \end{bmatrix}^T.
    \end{equation*}
    As defined, $f(\cdot)$ is a mapping from $\paramdim$-dimensional space to $2\nobs + 1$-dimensional space.
    Since $\markovchain$ is an i.i.d.\ sample, $\left( f(\mcmcdraws{1}), f(\mcmcdraws{2}), \ldots, f(\mcmcdraws{\samplesize}) \right) $ is also an i.i.d.\ sample.
    Because of the moment conditions we have assumed, each $f(\param)$ has finite variance.
    We apply the Lindeberg-Feller multivariate central limit theorem \citep[Proposition 2.27]{vanderVaart1998}, and conclude that 
    \begin{equation*}
        \sqrt{\samplesize} \left( \frac{1}{\samplesize} \sum_{s} f(\mcmcdraws{s}) - \E f(\param)   \right) \xrightarrow{D} N(\mathbf{0}, \Xi)
    \end{equation*}
    where the limit is $\samplesize \to \infty$, and $\Xi$ is a symmetric $(2\nobs+1) \times (2\nobs + 1)$ dimensional matrix, which we specify next.
    It suffices to write down the formula for $(i,j)$ entry of $\Xi$ where $i \leq j$: 
    \begin{equation*}
        \Xi_{i,j} = 
            \begin{cases}
                \Var(\funcname) &\text{ if } i  = j = 1 \\
                \Cov(\funcname, \LL{n}) &\text{ if } i = 1, j > 1 \\
                \Cov(\LL{n}, \LL{m}) &\text{ if } i = 2n, j = 2m\\
                \Cov(\LL{n},  \funcname\LL{m}) &\text{ if } i = 2n, j = 2m + 1\\
                \Cov(\funcname\LL{n},  \LL{m}) &\text{ if } i = 2n+1, j = 2m \\
                \Cov(\funcname \LL{n}, \funcname\LL{m}) &\text{ if } i = 2n+1, j = 2m + 1 \\
            \end{cases}.
    \end{equation*}

    To relate the asymptotic distribution of $f(\param)$ to that of the vector $\mcmcinfl$, we now use the delta method.
    Define the following function which acts on $2\nobs+1$ dimensional vectors and returns $\nobs$ dimensional vectors:
    \begin{equation*}
        h([x_1,x_2,\ldots,x_{2\nobs + 1}]^T) := 
        \begin{bmatrix}
            x_3 - x_1x_2, x_5 - x_1x_4, x_7 - x_1x_6, \ldots, x_{2\nobs + 1} - x_1 x_{2\nobs}
        \end{bmatrix}^T.
    \end{equation*}
    Written this way, $h(\cdot)$ transform the sample mean $\frac{1}{\samplesize} \sum_{s} f(\mcmcdraws{s})$ into the estimated influences: $\mcmcinfl = h\left( \frac{1}{\samplesize} \sum_{s} f(\mcmcdraws{s}) \right)$.
    Furthermore, $h(\cdot)$ applied to $\E f(\param)$ yields the vector of true influences: $\infl = h\left(\E f(\param) \right)$.
    $h(\cdot)$ is continuously differentiable everywhere.
    Its Jacobian is the following $ \nobs \times (2\nobs+1)$ matrix
    \begin{equation*}
        \jacob{h} = 
            \begin{bmatrix}
                -x_2        & -x_1     & 1     & 0      & 0 & \hdots & 0 \\
                -x_4        & 0        & 0     & -x_1   & 1 & \ldots & 0 \\
                \vdots      & \vdots   &\vdots & \ddots & 0 & \ldots & 0  \\
                -x_{2\nobs} & 0        &  0    & \hdots & 0 & \ldots & 1  \\
            \end{bmatrix},
    \end{equation*}
    which is non-zero.
    Therefore, we apply the delta method \citep[Theorem 3.1]{vanderVaart1998} and conclude that 
    \begin{equation*}
        \sqrt{\samplesize} \left( \mcmcinfl - \infl  \right) \xrightarrow{D} N\left(\mathbf{0}, \jacob{h} \big|_{x=\E f(\param)} \Xi (\jacob{h} \big|_{x=\E f(\param)} )^T \right).
    \end{equation*}

    The $(i,j)$ entry of the asymptotic covariance matrix is the dot product between the $i$-th row of $\jacob{h} \big|_{x=\E f(\param)}$ and the $j$-th column of $\Xi (\jacob{h} \big|_{x=\E f(\param)} )^T$. 
    The former is 
    \begin{equation*}
        [-\E\LL{i}, 0, 0, \ldots, \underbrace{-\E \funcname}_{2i \text{ entry}}, \underbrace{1}_{(2i+1) \text{ entry}}, \ldots, 0].
    \end{equation*}
    The later is 
    \begin{equation*}
        \begin{bmatrix}
            (-\E\LL{j}) \Cov(\funcname, \funcname) - (\E \funcname) \Cov(\funcname,\LL{j} ) + \Cov(\funcname, \funcname \LL{j}) \\
            \vdots \\
             (-\E\LL{j}) \Cov(\funcname \LL{\nobs},\funcname)  - (\E \funcname) \Cov(\funcname \LL{\nobs},\LL{j} ) + \Cov(\funcname \LL{\nobs}, \funcname \LL{j})
        \end{bmatrix}.
    \end{equation*}
    Taking the dot product, the $(i,j)$ entry of the asymptotic covariance matrix is equal to
    \begin{equation*}
        \begin{aligned}
            &\Cov(\funcname \LL{i}, \funcname \LL{j}) - (\E g) \left[\Cov(\funcname \LL{i},\LL{j}) + \Cov(\funcname \LL{j},\LL{i})  \right] \\
            &- \left[ (\E  \LL{j}) \Cov(\funcname, \funcname \LL{i}) + (\E  \LL{i}) \Cov(\funcname, \funcname \LL{j})  \right] \\
            &+(\E  \LL{j})  (\E  \LL{i}) \Var(\funcname) \\
            &+(\E g)^2 \Cov(\LL{i}, \LL{j}) \\
            &+(\E g) \left[ (\E  \LL{j}) \Cov(\funcname, \LL{i}) + (\E  \LL{i}) \Cov(\funcname, \LL{j})  \right]
        \end{aligned}.
    \end{equation*}
    It is simple to check that the last display is equal to the covariance between $(\funcname - \E [\funcname])(\LL{j} - \E[\LL{j}])$ and $(\funcname - \E [\funcname])(\LL{i} - \E[\LL{i}])$.

\end{delayedproof}

\begin{delayedproof}{lem:normal-gamma-limitVarMat}

    We use the (shape, rate) parametrization of the gamma distribution.
    Let the prior over $\tau$ be $\text{Gamma}(\alpha, \beta)$ where $\alpha, \beta > 0$.
    Conditioned on observations, the posterior distribution of $(\mu,\tau)$ is normal-gamma:
    \begin{equation*} 
        \begin{aligned}
            \tau &\sim \text{Gamma}\left( \alpha +\frac{\nobs}{2} , \beta + \frac{\nobs}{2} \left[ \frac{1}{\nobs}\sum_{n=1}^{\nobs} (\covarn)^2 - \bar{\covar}^2 \right]  \right), \\
            \epsilon &\sim \nobs(0,1), \\ 
            \mu \mid \tau, \epsilon &= \bar{\covar} + \frac{\epsilon}{\sqrt{\nobs\tau}}.
        \end{aligned}
    \end{equation*}
    In this section, since we only take expectations under the original full-data posterior, we will lighten the notation's dependence on $\weight$, and write $\E$ instead of $\Eposterior{\allones}$.
    Similarly, covariance and variance operators are understood to be under the full-data posterior.

    For completeness, we compute $\Cov(\mu, \loglik{n}{\mu,\tau})$.
    We know that $\mu - \E \mu = \epsilon/\sqrt{\nobs\tau}$.
    The log likelihood, as a function of $\tau$ and $\epsilon$, is 
    \begin{equation*}
        \frac{1}{2} \log \left( \frac{\tau}{2\pi} \right) - \frac{1}{2}\tau (\covarn - \bar{\covar})^2 - \frac{1}{2\nobs} \epsilon^2 + \frac{\covarn - \bar{\covar}}{\sqrt{\nobs}} \epsilon \sqrt{\tau}.
    \end{equation*}
    The covariance of $\mu$ and $\loglik{n}{\mu,\tau}$ is equal to the covariance between $\epsilon/\sqrt{\nobs\tau}$ and $\loglik{n}{\mu,\tau}$.
    Since $\epsilon/\sqrt{\nobs\tau}$ is zero mean, the covariance is equal to the expectation of the product.
    Since $\epsilon$ is indedependent of $\tau$, many of the terms that form the expectation of the product is zero. 
    After some algebra, the only term that remains is 
    \begin{equation*}
        \E\left[ \frac{\covarn - \bar{\covar}}{\nobs} \epsilon^2 \right] = \frac{\covarn - \bar{\covar}}{\nobs}.
    \end{equation*}

    To compute the asymptotic variance of $\mcmcinfln$, it suffices to compute the expectation of $\frac{\epsilon^2}{\nobs \tau} \left( \loglik{n}{\mu,\tau}  -  \E \loglik{n}{\mu,\tau} \right)^2$.
    The calculations are simple, but tedious, and we omit them.
    We will only state the result.
    The expectation of $\frac{\epsilon^2}{\nobs \tau} \left( \loglik{n}{\mu,\tau}  -  \E \loglik{n}{\mu,\tau} \right)^2$ is 
    \begin{equation*}
        \begin{aligned}  
            &\left[ \frac{1}{4\nobs} \E [\tau^{-1}(\tau - \mathbb{E}\tau) ^2] \right] (\covarn-\bar{\covar})^4 \\
            &+ \left[\frac{3 + \E [\tau^{-1}(\tau - \mathbb{E}\tau)]}{\nobs^2} - \frac{\E [\tau^{-1} (\log \tau - \E \log \tau)]}{2\nobs}  \right] (\covarn-\bar{\covar})^2 \\
            &+ \frac{1}{2\nobs^3} \E [\tau^{-1}] + \frac{1}{2 \nobs} \E [\tau^{-1}(\log \tau - \E \log \tau)^2] -\frac{1}{\nobs^2} \E [\tau^{-1} (\log \tau - \E \log \tau)^2 ].
        \end{aligned}
    \end{equation*}
    Since the asymptotic variance is equal to this expectation minus the square of the covariance between $\loglik{n}{\mu,\tau}$ and $\mu$, our final expression for the asymptotic variance $\limitVarMat_{n,n}$ is 
    \begin{equation*}
        \begin{aligned}  
            &\left[ \frac{1}{4\nobs} \E [\tau^{-1}(\tau - \mathbb{E}\tau) ^2] \right] (\covarn-\bar{\covar})^4 \\
            &+ \left[\frac{2 + \E [\tau^{-1}(\tau - \mathbb{E}\tau)]}{\nobs^2} - \frac{\E [\tau^{-1} (\log \tau - \E \log \tau)]}{2\nobs}  \right] (\covarn-\bar{\covar})^2 \\
            &+ \frac{1}{2\nobs^3} \E [\tau^{-1}] + \frac{1}{2 \nobs} \E [\tau^{-1}(\log \tau - \E \log \tau)^2] -\frac{1}{\nobs^2} \E [\tau^{-1} (\log \tau - \E \log \tau)^2 ].
        \end{aligned}
    \end{equation*}

    The constants $D_1$, $D_2$, and $D_3$ mentioned in the lemma statement can be read off this last display.
    It is possible to replace the posterior functionals of $\tau$ with quantities that only depends on the prior $(\alpha,\beta)$ and the observed data.
    Such formulas might be helpful in studying the behavior of $\limitVarMat_{n,n}$ in the limit where some $\covarn$ becomes very large. 

\end{delayedproof}

\section{Additional Experimental Details} \label{sec:experimental_details}

\subsection{Linear model} \label{subsec:linear_model_details}

Recall that the \textit{t} location-scale distribution has three hyperparameters: $\nu,\mu, \sigma$.
$\nu$ is the degrees of freedom, $\mu$ is the location, and $\sigma$ is the scale.
The density at $y$ of this distribution is 
\begin{equation*}
    \frac{\Gamma( (\nu+1)/2 )}{\Gamma(\nu/2)} 
    \frac{1}{\sqrt{\pi \nu \sigma^2}} 
    \left( 1 + \frac{(y - \mu)^2}{\nu \sigma^2} \right)^{-(\nu+1)/2}.
\end{equation*}

Recall that the latent parameters of our model are the baseline $\intercept$, the treatment effect $\slope$, and the noise $\sigma$.   
We set the the prior over $\intercept$ to be \textit{t} location-scale with degrees of freedom $3$, location $0$, and scale $1000$.
We set the the prior over $\slope$ to be \textit{t} location-scale with degrees of freedom $3$, location $0$, and scale $1000$.
We set the the prior over $\sigma$ to be \textit{t} location-scale with degrees of freedom $3$, location $0$, and scale $1000$, and constrain $\sigma$ to be positive.

\subsection{Hierarchical model for microcredit data} \label{subsec:hierarchical_model_microcredit_details}

The entire generative process, from the top-down (observations to priors), is as follows.
\begin{equation*}
    \begin{aligned}
        |\responsen| &\sim \text{Log-Normal}\left( \lControlLocation_{\groupn} + \lTreatmentLocation_{\groupn} \covarn, \exp (\lControlScale_{\groupn}  + \lTreatmentScale_{\groupn} \covarn )  \right), \\
        \lControlLocation_{k} &\sim \text{Normal}(\gControlLocation, \sdControl^2) \quad \text{i.i.d. across k}, \\
        \lTreatmentLocation_{k} &\sim \text{Normal}(\gTreatmentLocation, \sdTreatment^2) \quad \text{i.i.d. across k}, \\
        \lControlScale_{k} &\sim \text{Normal}(\gControlScale, \sdControlScale^2) \quad \text{i.i.d. across k}, \\
        \lTreatmentScale_{k} &\sim \text{Normal}(\gTreatmentScale, \sdTreatmentScale^2) \quad \text{i.i.d. across k}, \\
        \gControlLocation &\sim \text{Normal}(0, 10^2), \\
        \gTreatmentLocation &\sim \text{Normal}(0, 10^2), \\
        \sdControl &\sim \text{Cauchy}(0,2), \\
        \sdTreatment &\sim \text{Cauchy}(0,2), \\
        \gControlScale &\sim \text{Normal}(0, 10^2), \\
        \gTreatmentScale &\sim \text{Normal}(0, 10^2), \\
        \sdControlScale &\sim \text{Cauchy}(0,2), \\
        \sdTreatmentScale &\sim \text{Cauchy}(0,2). \\
    \end{aligned}
\end{equation*}
The observed data are $\covarn, \groupn, \responsen$.
All other quantities are latent, and estimated by MCMC.

\subsection{Hierarchical model for tree mortality data}\label{subsec:hierarchical_model_ecological_details}

The likelihood for the $n$-th observation is exponentially modified Gaussian with standard deviation $\sigma$, scale $\lambda$ and mean 
\begin{equation*}
     \left(\tIntercept_{\timeVarn} + \lIntercept_{\locationVarn} + \intercept \right) + \left(\tSlope_{\timeVarn} + \lSlope_{\locationVarn} + \slope \right) \covarn + f(\covarn),
\end{equation*}
with $f(x) := \sum_{i=1}^{10} B_i(x) \gamma_i$ where $B_i$'s are fixed thin plate spline basis functions \citep{Wood2003} and $\gamma_i$'s are random: $\gamma_i \sim \text{Normal}(0, \sigma_{(\text{smooth})}^2)$. 
In all, the parameters of interest are 
\begin{itemize}
    \item Fixed effects: $\intercept$ and $\slope$.
    \item Random effects:  time ($\tIntercept_{\timeVarn}, \tSlope_{\timeVarn}$) and location ($\lIntercept_{\locationVarn}, \lSlope_{\locationVarn}$).
    \item Degree of smoothing: $\sigma_{(\text{smooth})}$.
\end{itemize}
Since there are many regions (nearly $3{,}000$) and periods of time (30), the number of random effects is large. 
\citet{Senf2020} uses brms()'s default priors for all parameters: in this default, the fixed effects are given improper uniform priors over the real line.
To work with proper distributions, we set the priors for the random effects and degree of smoothing in the same way set by \citet{Senf2020}.
For fixed effects, we use $t$ location-scale distributions with degrees of freedom $3$, location $0$, and scale $1000$.

\end{document}